\documentclass[a4paper,11pt]{article}
\pdfoutput=1 
\usepackage{jcappub} 
\makeatletter
\gdef\@fpheader{\footnotesize
Published 30 January 2025 • © 2025 The Author(s)\\
Citation: U.\ Andrade \textit{et al.}, \textbf{JCAP} \textbf{01} (2025) 128\\
doi:\href{https://doi.org/10.1088/1475-7516/2025/01/128}{10.1088/1475-7516/2025/01/128}}
\makeatother
\usepackage{lineno}
\usepackage{graphicx}	
\usepackage{amsmath}	
\usepackage{hyphenat}
\usepackage{booktabs,caption}
\usepackage{pdflscape}
\usepackage[export]{adjustbox}
\usepackage{tensor}
\usepackage{xcolor}
\usepackage{CJK}
\usepackage{aas_macros}
\usepackage[T1]{fontenc} 
\usepackage{enumerate}
\usepackage{float}
\usepackage{multirow}

\DeclareRobustCommand{\VAN}[3]{#2}
\let\VANthebibliography\thebibliography
\def\thebibliography{\DeclareRobustCommand{\VAN}[3]{##3}\VANthebibliography}

\usepackage{makecell}
\usepackage{verbatim}
\usepackage{comment}
\usepackage{soul}
\usepackage{adjustbox}
\usepackage{enumitem}
\usepackage{placeins}
\usepackage{afterpage}

\usepackage{orcidlink} 

\usepackage{hanging}

\usepackage{tabularx}

\definecolor{WildStrawberry}{HTML}{EE2967}

\newcommand{\abacus}[0]{{\sc AbacusSummit}}
\newcommand{\alphaiso}[0]{$\alpha_{\mathrm{iso}}$}
\newcommand{\alphaap}[0]{$\alpha_{\mathrm{AP}}$}
\newcommand{\dm}[0]{$m$}

\newcommand{\wowa}[0]{$(w_0,w_a)$}
\newcommand{\wo}[0]{$w_0$}
\newcommand{\wa}[0]{$w_a$}
\newcommand{\fnl}[0]{$f_\mathrm{{NL}}$}
\newcommand{\chitwo}[0]{$\chi^2$}
\newcommand{\recsym}{{\tt Rec-Sym}}

\renewcommand{\sectionautorefname}{\S\kern-4pt}
\renewcommand{\subsectionautorefname}{\S\kern-4pt}
\renewcommand{\subsubsectionautorefname}{\S\kern-4pt}
\renewcommand{\appendixautorefname}{\S\kern-4pt}

\usepackage[nameinlink,noabbrev]{cleveref}
\crefname{equation}{Eq.}{Eqs.}
\crefname{section}{Section}{Sections}
\crefname{figure}{Figure}{Figures}
\crefname{table}{Table}{Tables}
\crefname{appendix}{Appendix}{Appendices}

\title{Validating the Galaxy and Quasar Catalog-Level Blinding Scheme for the DESI 2024 analysis}

\affiliation{Affiliations are in \cref{sec:affiliations}}

\author[1,2]{{U.~Andrade}\orcidlink{0000-0002-4118-8236},}
\author[3]{{J.~Mena-Fern\'andez}\orcidlink{0000-0001-9497-7266},}
\author[1]{{H.~Awan}\orcidlink{0000-0003-2296-7717},}
\author[4,5,6]{{A.~J.~Ross}\orcidlink{0000-0002-7522-9083},}
\author[7]{{S.~Brieden}\orcidlink{0000-0003-3896-9215},}
\author[2]{{J.~Pan}\orcidlink{0000-0001-9685-5756},}
\author[8]{{A.~de~Mattia},}
\author[9]{{J.~Aguilar},}
\author[10]{{S.~Ahlen}\orcidlink{0000-0001-6098-7247},}
\author[2]{{O.~Alves},}
\author[11]{{D.~Brooks},}
\author[12,13]{{E.~Buckley-Geer},}
\author[9]{{E.~Chaussidon}\orcidlink{0000-0001-8996-4874},}
\author[9]{{T.~Claybaugh},}
\author[14]{{S.~Cole}\orcidlink{0000-0002-5954-7903},}
\author[15]{{A.~de la Macorra}\orcidlink{0000-0002-1769-1640},}
\author[16]{{A.~Dey}\orcidlink{0000-0002-4928-4003},}
\author[11]{{P.~Doel},}
\author[17,18]{{K.~Fanning}\orcidlink{0000-0003-2371-3356},}
\author[19,20]{{J.~E.~Forero-Romero}\orcidlink{0000-0002-2890-3725},}
\author[21,22,23]{{E.~Gaztañaga},}
\author[24,21,25]{{H.~Gil-Mar\'in}\orcidlink{0000-0003-0265-6217},}
\author[9]{{S.~Gontcho A Gontcho}\orcidlink{0000-0003-3142-233X},}
\author[9]{{J.~Guy}\orcidlink{0000-0001-9822-6793},}
\author[26]{{C.~Hahn}\orcidlink{0000-0003-1197-0902},}
\author[2]{{M.~M.~S~Hanif}\orcidlink{0009-0006-2583-5006},}
\author[4,27,6]{{K.~Honscheid},}
\author[28]{{C.~Howlett}\orcidlink{0000-0002-1081-9410},}
\author[2]{{D.~Huterer}\orcidlink{0000-0001-6558-0112},}
\author[16]{{S.~Juneau},}
\author[9]{{A.~Kremin}\orcidlink{0000-0001-6356-7424},}
\author[9]{{M.~Landriau}\orcidlink{0000-0003-1838-8528},}
\author[29]{{L.~Le~Guillou}\orcidlink{0000-0001-7178-8868},}
\author[9]{{M.~E.~Levi}\orcidlink{0000-0003-1887-1018},}
\author[30,31]{{M.~Manera}\orcidlink{0000-0003-4962-8934},}
\author[4,5,6]{{P.~Martini}\orcidlink{0000-0002-4279-4182},}
\author[16]{{A.~Meisner}\orcidlink{0000-0002-1125-7384},}
\author[32,31]{{R.~Miquel},}
\author[33]{{J.~Moustakas}\orcidlink{0000-0002-2733-4559},}
\author[34]{{E.~Mueller},}
\author[15]{{A.~Muñoz-Gutiérrez},}
\author[35]{{A.~D.~Myers},}
\author[22]{{S.~Nadathur}\orcidlink{0000-0001-9070-3102},}
\author[36]{{J.~ A.~Newman}\orcidlink{0000-0001-8684-2222},}
\author[37]{{J.~Nie}\orcidlink{0000-0001-6590-8122},}
\author[38,39]{{G.~Niz}\orcidlink{0000-0002-1544-8946},}
\author[40,42]{{E.~Paillas}\orcidlink{0000-0002-4637-2868},}
\author[8,9]{{N.~Palanque-Delabrouille}\orcidlink{0000-0003-3188-784X},}
\author[40,41,42]{{W.~J.~Percival}\orcidlink{0000-0002-0644-5727},}
\author[8]{{M.~Pinon},}
\author[9,43,44]{{C.~Poppett},}
\author[45]{{F.~Prada}\orcidlink{0000-0001-7145-8674},}
\author[15]{{A.~P\'{e}rez-Fern\'{a}ndez}\orcidlink{0009-0006-1331-4035},}
\author[46]{{M.~Rashkovetskyi}\orcidlink{0000-0001-7144-2349},}
\author[47]{{M.~Rezaie}\orcidlink{0000-0001-5589-7116},}
\author[48]{{G.~Rossi},}
\author[49]{{E.~Sanchez}\orcidlink{0000-0002-9646-8198},}
\author[9]{{D.~Schlegel},}
\author[2]{{M.~Schubnell},}
\author[50]{{H.~Seo}\orcidlink{0000-0002-6588-3508},}
\author[16]{{D.~Sprayberry},}
\author[2]{{G.~Tarl\'{e}}\orcidlink{0000-0003-1704-0781},}
\author[15]{{M.~Vargas-Maga\~na}\orcidlink{0000-0003-3841-1836},}
\author[32,25]{{L.~Verde}\orcidlink{0000-0003-2601-8770},}
\author[16]{and {B.~A.~Weaver}}

\emailAdd{uendsa@umich.edu}
\abstract{

In the era of precision cosmology, ensuring the integrity of data analysis through blinding techniques is paramount -- a challenge particularly relevant for the Dark Energy Spectroscopic Instrument (DESI). DESI represents a monumental effort to map the cosmic web, with the goal to measure the redshifts of tens of millions of galaxies and quasars. Given the data volume and the impact of the findings, the potential for confirmation bias poses a significant challenge. To address this, we implement and validate a comprehensive blind analysis strategy for DESI Data Release 1 (DR1), tailored to the specific observables DESI is most sensitive to: Baryonic Acoustic Oscillations (BAO), Redshift-Space Distortion (RSD) and primordial non-Gaussianities (PNG). We carry out the blinding at the catalog level, implementing shifts in the redshifts of the observed galaxies to blind for BAO and RSD signals and weights to blind for PNG through a scale-dependent bias. We validate the blinding technique on mocks as well as on data by applying a second blinding layer to perform a series of sanity checks; the latter allows probing complexities in real data not captured in mocks. 
We find that the blinding strategy alters the data vector in a controlled way, and the BAO and RSD analysis choices are robust to blinding. The successful validation of the blinding strategy paves the way for the unblinded DESI DR1 analysis, alongside future blind analyses with DESI and other surveys.
}
\begin{document}
\maketitle
\flushbottom

\section{INTRODUCTION}

Cosmology has entered a precision era, where experiments are designed to measure key parameters of the Universe to unprecedented levels of accuracy. One of the most robust methodologies employed to understand the cosmic landscape is the two-point clustering statistics of 3D galaxy distributions. These statistics, the power spectrum \( P(k) \) in harmonic space and the correlation function \( \xi(r) \) in configuration space, provide vital clues about the underlying cosmological model and the nature of dark energy and constraints on primordial non-Gaussianities \citep{Peebles1980, Eisenstein1998}.

However, as we refine our methods and aim for increasingly precise results, the risk of confirmation bias becomes more of a concern. These biases can arise during the data analysis process and may lead to misleading conclusions, thereby affecting the veracity of the findings. It is in this context that the concept of blind analysis becomes critically important.  At the heart of it is the “blinding”, which involves the deliberate concealment or modification of key analysis outcomes, thereby ensuring that researchers' subsequent choices and interpretations remain unbiased \citep{Klein2005, Smith2012, Croft2011}. In other words, only after the full pipeline is frozen -- all the choices are made --, the unaltered results are unveiled in a step we refer to as ``unblinding''. Establishing clear criteria for when unbinding happens is a key part of the blind analysis procedure, which will be detailed later.

The primary goal of this paper is to present and validate the blinding technique applied to Data Release 1 (DR1; \cite{DESI2024.I.DR1}) of the Dark Energy Spectroscopic Instrument (DESI) \cite{DESI2016a.Science,DESI2022.KP1.Instr,levi2019dark,DESI2022.KP1.Instr}. With its ability to collect high-quality spectroscopic data, DESI enables a range of cosmological analyses, including constraints from Baryon Acoustic Oscillations (BAO), Redshift-Space Distortions (RSD), and the investigation of scale-dependent bias due to primordial non-Gaussianities (PNG) \citep{Levi2013, DESI2016a.Science}.

The scope of this paper is to validate the catalog-level blinding technique for galaxy and quasar samples, specifically focusing on the signals extracted from BAO and RSD analyses. While we apply the blinding scheme to PNG analyses as well, the detailed validation of the PNG-related results lies outside the current scope and will be addressed in a follow-up paper. Similarly, further optimization and validation of the BAO and RSD analyses are beyond the scope of this paper and are discussed in detail in the dedicated BAO and RSD publications \cite{DESI2024.III.KP4, DESI2024.V.KP5}.

DESI has been designed to perform a galaxy survey spanning approximately 14,000 square degrees of the sky, encompassing regions in both the southern and northern galactic caps \cite{dey2019overview}, over a period of five years. During its operation, DESI aims to determine the redshifts of around 40 million galaxies, ranging from redshifts 0.05 to 3.5. The survey has successfully completed its validation stage \cite{collaboration2024validation} and made its early data publicly available \cite{adame2023early}, while the analysis of DR1 is underway (for which this work is a supporting paper; more details below). DESI's target selection program classifies its tracers into four distinct types: Bright Galaxy Survey (BGS), Luminous Red Galaxy (LRG), Emission Line Galaxy (ELG), and Quasars (QSO), in increasing order of redshift. Moreover, DESI also probes the Universe using Lyman-$\alpha$, for which the blind analysis will follow a different type of blinding scheme; we refer the reader to \cite{DESI2024.IV.KP6}.

We begin by giving an overview of blinding in cosmology in \cref{sec:blinding-in-general}, followed by a description of the DESI DR1 blinding scheme in \cref{sec:blinding-scheme_}, expanding first on the DESI observables to motivate the parameters subjected to blinding, followed by the details of the blinding strategy; we also discuss when and under what conditions the blinding was planned to be unblinded. Next, we detail the analysis framework in \cref{sec:analysis-framework}, discussing the data vector and covariance used, the theory model, as well as the inference framework. Then, in \cref{sec:validation-wmocks}, we validate the blinding strategy using mock datasets and the analysis framework, demonstrating the blinding technique is robust. We then validate the strategy on blinded data in \cref{sec:validation-wdata}, delving into the statistical tests and analyses to ensure that the blinding process does not introduce any spurious features or artifacts in the data. We conclude in \cref{sec:conclusion}.

\section{BLINDING IN COSMOLOGY\label{sec:blinding-in-general}}

The practice of blind analysis is not new to cosmology. In fact, different blinding strategies have been adopted for various cosmology analyses, e.g., the Supernovae analysis presented in \cite{Zhang2017} and for weak lensing surveys such as the Kilo-degree Survey (KiDS) \cite{Kuijken2015,Hildebrandt2020,Sellentin2020} and the Dark Energy Survey (DES) \cite{Muir2020}. In these applications, the blinding strategy was carefully tailored to the unique requirements and complexities of each survey and analysis. For example, the KiDS collaboration focused on the gravitational lensing signal and hence blinded their main observable, galaxy ellipticities at the catalog level \cite{Kuijken2015}. On the other hand, the DES collaboration, carrying out a multi-probe experiment, employed a blinding scheme at the data-vector level ensuring internal model consistency between the galaxy clustering and weak lensing signal.

Considering the various uses of blinding for cosmology, we can distill several key criteria that a successful blinding scheme must satisfy:

\begin{enumerate}[label=\roman*),noitemsep,topsep=1pt]
    \item \textbf{Preservation of data quality:} The blinding scheme should maintain the statistical properties of the data to permit accurate validation tests.
    \item \textbf{Difficult reversibility:} Blinding should not be easily reversible by those conducting the analysis, avoiding accidental unblinding.
    \item \textbf{Parameter specificity:} Blinding should be specific to the cosmological parameters of interest, without affecting other variables and diagnostics used in the analysis. Note that the cosmological parameters of interest are defined by the actual observables the survey is most sensitive to.
\end{enumerate}

These principles serve as a guideline for the choices made while developing a catalog-level blinding scheme for spectroscopic galaxy surveys in general, which we can then tailor to DESI in particular (as done in \cref{sec:blinding-scheme}). 

The choice for catalog-level blinding is particularly motivated in order to satisfy criterion ii). The relevant quantities of a (galaxy) catalog here include two angular coordinates (right ascension, RA, and declination, DEC), the measured redshift $z$, and a set of weights $w_x$  to correct for the variations in completeness. A given spectroscopic survey yields redshifts and the weights for the corresponding, pre-existing angular coordinates from the photometric catalogs from which the galaxy targets are selected. Therefore, one only needs to perturb redshifts and weights for catalog-level blinding, while leaving angular positions unchanged as these are already ``unblinded'' via the photometric catalog.

This procedure complies with criterion i) when it comes to the treatment of survey systematics. Due to the unchanged angular positions, the determination of systematic weights impacts the blinded and unblinded catalogs in the same fashion, hence allowing for an effective treatment of angular systematics.

Finally, to satisfy criterion iii), the exact per-object shifts in redshift and weight are not chosen to be random, and instead to distort the primary observables of a spectroscopic survey, i.e., the BAO, RSD and PNG signals. In order to ensure that the validation tests on the blinded analysis lead to insights valid on the unblinded analysis, the catalog-level blinding strategy needs to ensure that the blinded data can be represented by a viable underlying cosmological model.

\section{DESI DR1 BLINDING SCHEME}
\label{sec:blinding-scheme_}

For DESI DR1, we develop a comprehensive blinding strategy based on foundational criteria described in \cref{sec:blinding-in-general} and methodologies described in the literature to blind for BAO and RSD \cite{Brieden2020} as well as PNG \cite{KP3s10-Chaussidon}. The procedure ensures that neither individual scientists nor the collective team can inadvertently unblind the data or induce experimenter biases based on intermediate results.

\subsection{DESI Observables\label{sec:desi-observables}}

The main capacity of DESI relies on examining the full 3D (along and across the line of sight) clustering of galaxies over a wide redshift range. The precise map of galaxy positions allows us to identify the cosmic web and accurately constrain the expansion history (via BAO) and growth history (via RSD) of the universe. Furthermore, the high number of large-scale modes arising from the full 3D information provides us with a powerful window to investigate the presence of PNG. 

\subsubsection{Summary Statistics}
To create a 3D map of galaxies, we first convert measured galaxy redshifts $z$ to comoving distances using a fiducial cosmological model $\mathbf{\Omega}_{\rm fid}$; see \cref{eq:fiducial-cosmo} for the fiducial model used in this paper. From the resulting catalog, we infer the galaxy redshift-space overdensity field, $\delta_\mathrm{g}^\mathrm{red}(\vec{r})$; this field depends on the comoving coordinate $\vec{r}$. Then, we calculate the galaxy two-point clustering statistics: the correlation function $\xi_\mathrm{g}(s)$, which depends on pair separation $s$, and its Fourier analog, the power spectrum $P_\mathrm{g}(k)$, which depends on the wavevector $k$.

For exact implementation details of the $\xi_\mathrm{g}(s)$ and $P_\mathrm{g}(k)$ measurement, see \cref{sec:analysis-framework}. These statistics exhibit a few distinct cosmological features described below.

\subsubsection{Probing the Expansion History}
The expansion history of the universe is encoded in the Baryonic Acoustic Oscillations (BAO) signal observed within the two-point clustering statistics. In the early universe, pressure waves, driven by the interplay between radiation and matter, propagated through the primordial plasma. This propagation continued until the universe cooled enough for protons and electrons to combine into neutral hydrogen, an epoch known as recombination. Shortly after recombination, at a redshift denoted by $z_d$ (the drag epoch), the decoupling of baryons from photons occurred. This decoupling effectively ``froze'' the acoustic waves in space, marking the maximum extent to which these pressure-driven waves could travel. This maximum extent is referred to as the sound horizon and is mathematically expressed as

\begin{align}
    r_d = \int_{\infty}^{z_d} \! \frac{c_s(z)}{H(z)} \, dz,
\end{align}
where $c_s(z)$ is the sound speed and $H(z)$ the Hubble expansion rate. At late times, we can still see the impact of these waves as an overabundance of galaxy pairs and separation of the sound horizon, the BAO feature. Given that the sound horizon in comoving coordinates is fixed, it represents a \textit{standard ruler}. Hence, by measuring the angular and parallel position of the BAO feature at different redshift bins, we can exquisitely map the angular diameter distance $D_A(z)$ and the  Hubble distance $c/H(z)$ in units of the sound horizon. This is often parameterized via the scaling parameters: 
\begin{equation}
    \alpha_\perp(z)\equiv \frac{D_{A}(z)r_{d,\text{fid}}}{D_{A,\text{fid}}(z)r_d},\quad
    \alpha_\parallel(z)\equiv \frac{H_\text{fid}(z) r_{d,\text{fid}}}{H(z) r_d}\, ,
\label{eq:aperp_apar}
\end{equation}
defined with respect to a fiducial template, which is fitted to the data as detailed in \cref{sec:standard-bao}. 
The notion that the sound horizon is isotropic, i.e., its size perpendicular and parallel to the line of sight is the same, allows us to perform the so-called Alcock-Paczynski (AP) test \cite{Alcock:1979mp}. If the fiducial cosmology chosen to transform redshifts to distances does not correspond to the underlying distance-redshift relation, this manifests itself as an anisotropy between the BAO distance perpendicular and parallel the line of sight. For this test, it is useful to combine the scaling parameters of \cref{eq:aperp_apar} into the isotropic (``iso'') and anisotropic (``AP'') components given as 
\begin{equation}\label{eq:aisoaap}
    \alpha_{\text{iso}} = (\alpha_\perp^2 \alpha_\parallel)^{\frac{1}{3}},\quad \alpha_{\text{AP}} = \frac{\alpha_{\parallel}}{\alpha_{\perp}}~.
\end{equation}

The BAO analysis hence represents a powerful tool to reconstruct the expansion history of the universe. In what follows, we consider the \wo\wa CDM model with varying dark energy equation of state 
\begin{equation}
   w(a) = w_0 + (1-a)w_a\, , 
\end{equation}
where $a$ is the scale factor; see \cite{2003PhRvL..90i1301L} for a review. This parameterization allows for describing a much richer range of dynamical behavior of dark energy than a constant equation of state (i.e., the cosmological constant $\Lambda$), allowing for a test of the $\Lambda$CDM model.

Within the flat \wo\wa CDM model, the cosmological expansion law at late times is given by

\begin{equation}
    H(z) = H_0 \sqrt{\Omega_\mathrm{m}(1+z)^3  + (1-\Omega_\mathrm{m})(1+z)^{3(1+w_0+\frac{z}{1+z}w_a)}} \, , 
\end{equation}
with the present time Hubble constant $H_0$ and the relative matter energy density $\Omega_\mathrm{m}$.

\subsubsection{Probing the Growth History\label{sec:growth-history}}

The growth history of the universe is mapped via the Redshift-Space Distortions (RSD) signal. By measuring galaxy redshifts, we measure galaxy velocities along the line of sight, each consisting of two contributions: the Hubble flow velocity (recession) and their own velocity (peculiar). Therefore, by converting redshifts to distances, the true real-space positions are contaminated by the peculiar velocities, giving rise to RSD. On large scales, the galaxy bulk flows can be characterized by the so-called displacement field, $\boldsymbol{\Psi}=\nabla \phi$, the gradient of the gravitational potential $\phi$ sourced by the real-space matter density field $\delta_\mathrm{m}$. This leads to
\begin{align}
    \nabla \cdot \boldsymbol{\Psi} = - \frac{\delta_g}{b_1}~,
\end{align}
where $b_1=\delta_\mathrm{g}/\delta_\mathrm{m}$ is the scale independent linear galaxy bias. The mapping between real-space coordinate $\vec{x}$ and redshift-space coordinate $\vec{r}$ on large scales induced by the line-of-sight component of the displacement field $\boldsymbol{\Psi}\cdot \hat{r}$ is given, following \cite{Nusser:1994}, by
\begin{align}
    \vec{r} = \vec{x} + f\left( \boldsymbol{\Psi} \cdot \mathbf{\hat{r}} \right) \mathbf{\hat{r}} 
\end{align}
with the so-called growth rate $f \equiv {d \ln D(a)} / {d \ln a}
$, where $D(a)$ is the linear growth function, defined as $D(a)=\delta(a)/\delta (a=1)$~\cite{Huterer:2013xky}. As a consequence, the redshift-space galaxy power spectrum $P_g(k, \mu)$ experiences an anisotropy as a function of the angle $\mu=\cos\theta$ between the galaxy pair and the line of sight given by the Kaiser formula,\cite{Kaiser:1987qv},
\begin{equation}\label{eq:kaiserformula}
    P_g(k, \mu) \approx \left(b_1+f \mu^2 \right)^2 P_\mathrm{lin}(k),
\end{equation}
where $P_\mathrm{lin}(k)$ is the linear matter power spectrum. Note that this approximation is only valid on large scales and we adopt a more sophisticated RSD model in \cref{sec:analysis-framework}.

\subsubsection{Probing the Primordial Non-Gaussianity}
As for the local primordial non-Gaussianity (PNG), the primordial gravitational potential \( \phi(x) \) is approximated, as in \cite{Komatsu2001}, by a quadratic contribution,
\begin{equation}
\phi_G(x) + f_{\text{NL}} \left(\phi_G(x)^2 - \langle \phi_G(x)^2 \rangle \right),
\end{equation}
where \( \phi_G \) is a Gaussian distributed random field and \fnl\ is the amplitude of the quadratic correction to the potential. The parameter \fnl\ in effect parameterizes the PNG. In the standard single-field slow-roll inflation model, the value of \fnl\ is approximately zero (it is roughly $(n_s-1) \simeq 0.03$). However, if \fnl\ is found to be substantially larger than zero, it would suggest interesting possibilities, such as the presence of multiple interacting scalar fields during the inflationary period.

\subsection{Blinding Strategy\label{sec:blinding-scheme}}

To blind for the observables described before, we start with shifting galaxy redshifts along the line of sight, mimicking a universe with a different underlying cosmological model without changing the galaxies' angular positions. This entails two kinds of shifts: one to blind the cosmological background evolution (i.e., BAO) and another to blind the growth of structures (i.e., RSD); the first mimics the Alcock-Paczynski (AP) effect while the second mimics RSD. Both use two cosmologies: one fiducial one and one we pick for our blinding scheme\footnote{Note that the fiducial cosmology is referred to as the ``reference'' cosmology in \cite{Brieden2020}  while the what we call blinding cosmology is termed as ``shifted'' cosmology. Also, note that the arbitrary choice of fiducial cosmology does not have a sizable impact on the blinding shifts, which primarily depend on the relative difference between the fiducial and the blind cosmology.}. We refer the reader to  \cite{Brieden2020} for details, but summarize the two shifts briefly for completeness:

\begin{enumerate}
\item AP-like shift: This shift emulates the Alcock-Paczynski effect by altering galaxy positions along the line of sight. This procedure is executed by first transforming the measured redshifts \(z_i\) into comoving distances \(D_{\mathrm{M}}(z_i, \mathbf{\Omega}_{\text{blind}})\) using the cosmology chosen for blinding, which are then transformed back to redshift space \(z_i\) using the fiducial cosmology, as captured in Equation in 3.10 \cite{Brieden2020}, reproduced here for completeness:
\begin{equation} \label{eqn:AP-shift}
z_i \, (\mathbf{\Omega}_{\text{true}}) \overset{\mathbf{\Omega}_{\text{blind}}}{\longrightarrow} D_{\mathrm{M}}(z_i, \mathbf{\Omega}_{\text{blind}}) = D_{\mathrm{M}}(z_i^\prime, \mathbf{\Omega}_{\text{fid}}) \overset{\mathbf{\Omega}_{\text{fid}}}{\longrightarrow} z_i^\prime \, (\mathbf{\Omega}_{\text{blind}}),
\end{equation}
where $\mathbf{\Omega}_{\text{true}}$ is the cosmology underlying the observed data. As a result, once the blinded galaxy redshift catalog is converted to distances via $\mathbf{\Omega}_{\text{fid}}$, the scaling parameters in \cref{eq:aperp_apar} are shifted with respect to $\mathbf{\Omega}_{\text{true}}$ as predicted by $\mathbf{\Omega}_{\text{blind}}$. Note that the AP-like shift is the same for all galaxies that have the same redshift.

\item RSD shift: This perturbative shift mimics the redshift-space distortions by adjusting galaxy positions based on the local galaxy density and the peculiar velocity field. The shifts in redshifts are calculated by first transforming the observed redshifts into distances using the fiducial cosmology. From the resulting galaxy redshift-space positions $\mathbf{r}$, and a fiducial value of the growth rate $f^\mathrm{fid}$ as input, an approximation of the real-space density and its gradient, the displacement field, are derived. Then, the line-of-sight component of the latter is used to transform each galaxy to a new, blinded redshift-space position $\mathbf{r^\prime}$ according to the blinding value of the growth rate $f^\mathrm{blind}$. This is encapsulated in Equation 3.18 in \cite{Brieden2020}: 
\begin{equation}\label{eq:RSD-shift}
\begin{aligned}
\mathbf{r^\prime} = \mathbf{r} - f^{\text{fid}} (\mathbf{\Psi} \cdot \mathbf{\hat{r}})\mathbf{\hat{r}} + f^\mathrm{blind}  (\mathbf{\Psi} \cdot \mathbf{\hat{r}})\mathbf{\hat{r}} ~.
\end{aligned}
\end{equation}
As a result, the galaxy power spectrum measured from the catalog after the blinding transformation of \cref{eq:RSD-shift} exhibits an RSD anisotropy reminiscent to \cref{eq:kaiserformula} with $f=f^\mathrm{true}$ being altered to $f=f^\mathrm{true}-f^\mathrm{fid}+f^\mathrm{blind}$.
Note that given the impact of local galaxy density and peculiar velocity field, the shift is different for each galaxy.

\end{enumerate}

The procedure described above blinds only two of the three observables that we are interested in, i.e.  BAO and RSD. For the third observable,  PNG, we follow the strategy in \cite{KP3s10-Chaussidon}, which entails blinding the large scales of the power spectrum by adding weights to the data, using a blinding \fnl\ value. Here we aim to mimic the scale-dependent bias signature of PNG on large scales, which impacts the theoretical real-space galaxy power spectrum as:
\begin{equation} \label{eqn:scale_depedent_bias}
    P(k) = \left(b_{1} + \dfrac{b_{\phi}}{\alpha(k)} f_{\mathrm{NL}} \right)^2 \times P_{\rm lin}(k),
\end{equation}
where $P_{\rm lin}(k)$ is the linear matter power spectrum, $\alpha(k)$ is a transfer function connecting the primordial gravitational field to the matter density perturbation, $b_{1}$ is the linear bias and $b_{\phi}$ is the bias quantifying the response of the tracer to local PNG.

To implement the scale-dependent part $b_{\phi} f_{\mathrm{NL}} / \alpha(k)$ of \cref{eqn:scale_depedent_bias} at the catalog level, we first approximately move galaxies to real-space by subtracting the RSD displacements estimated in the same manner as for the RSD shift mentioned above from the observed galaxy positions. The obtained shifted galaxies are`painted on a grid'\footnote{\textit{Painting galaxies on a grid} is a term commonly used to describe the process of mapping discrete galaxy positions onto a continuous grid, so that their spatial distribution can be converted into a density field.} to estimate the underlying matter density field in real space $\hat{\delta}_{r}$ (assuming a fiducial linear bias). The scale-dependent bias contribution $b_{\phi} f_{\mathrm{NL}} \hat{\delta}_{r}(\Vec{k}) / \alpha(k)$ is computed in Fourier space for a blinded value of $b_{\phi} f_{\mathrm{NL}}$, then transformed back to configuration space and read off at each galaxy position, thereby providing a weight to be applied to each galaxy to mimic the scale-dependent bias. 

For our blinding scheme, we pick \( w_0 \), \( w_a \), and \fnl\ values, as described in \cref{sec:blinding-pipeline}. As for the fiducial cosmology, throughout this work we use Planck-2018 results \cite{aghanim2020planck}. The cosmological parameters are, explicitly, given by:
\begin{equation}
\begin{split}
\omega_b &= 0.02237, \quad \omega_{cdm} = 0.12, \quad h = 0.6736, \\
A_s &= 2.083 \times 10^{-9}, \quad n_s = 0.9649, \quad \quad N_{ur} = 2.0328, \\
N_{ncdm} &= 1.0, \quad \omega_{ncdm} = 0.0006442, \quad w_0 = -1, \quad w_a = 0.
\end{split}
\label{eq:fiducial-cosmo}
\end{equation}
$\omega_b$ and $\omega_{cdm}$ denote the densities for baryons and cold dark matter, respectively, both scaled by $h^2$, where $h$ denotes the reduced Hubble constant. $A_s$ and $n_s$ characterize the amplitude and spectral index of primordial scalar perturbations. Further, $N_{ur}$ and $N_{ncdm}$ denote the effective number of ultra-relativistic and non-cold dark matter species, with $\omega_{ncdm}$ indicating the density of the latter. Finally, $w_0$ and $w_a$ are the dark energy equation of state and its evolution, as explained in \cref{sec:growth-history}.

\subsection{Blinding Pipeline\label{sec:blinding-pipeline}}
Taking into account the DESI observables discussed in \cref{sec:desi-observables}, we focus our blinding efforts on three key parameters, \( w_0 \), \( w_a \) and \fnl, as they are central to the primary science goals of the DESI experiment and, thus, are highly susceptible to experimenter bias during the data validation/interpretation stage. As discussed in \cref{sec:desi-observables}, all these parameters are constrained using two-point clustering statistics from large-scale structure observations, with details of the analysis framework in \cref{sec:analysis-framework}. While it would be possible to further extend the blinding parameter basis, for example, by adding non-zero curvature, we decided to limit the AP blinding to the flat \wo\wa CDM model introduced in \cref{sec:desi-observables}, since with DESI alone we do not expect to constrain the dark energy equation of state jointly with curvature, due to the strong degeneracy of these parameters. This means, that given the DESI precision, blinding for flat \wo\wa CDM model imprints sufficient freedom to the $H(z)$ function that makes it barely indistinguishable from a $H(z)$ function of a $k$-CDM model. 

\begin{figure}[hbt!]
    \centering 
    \includegraphics[width=.75\linewidth]{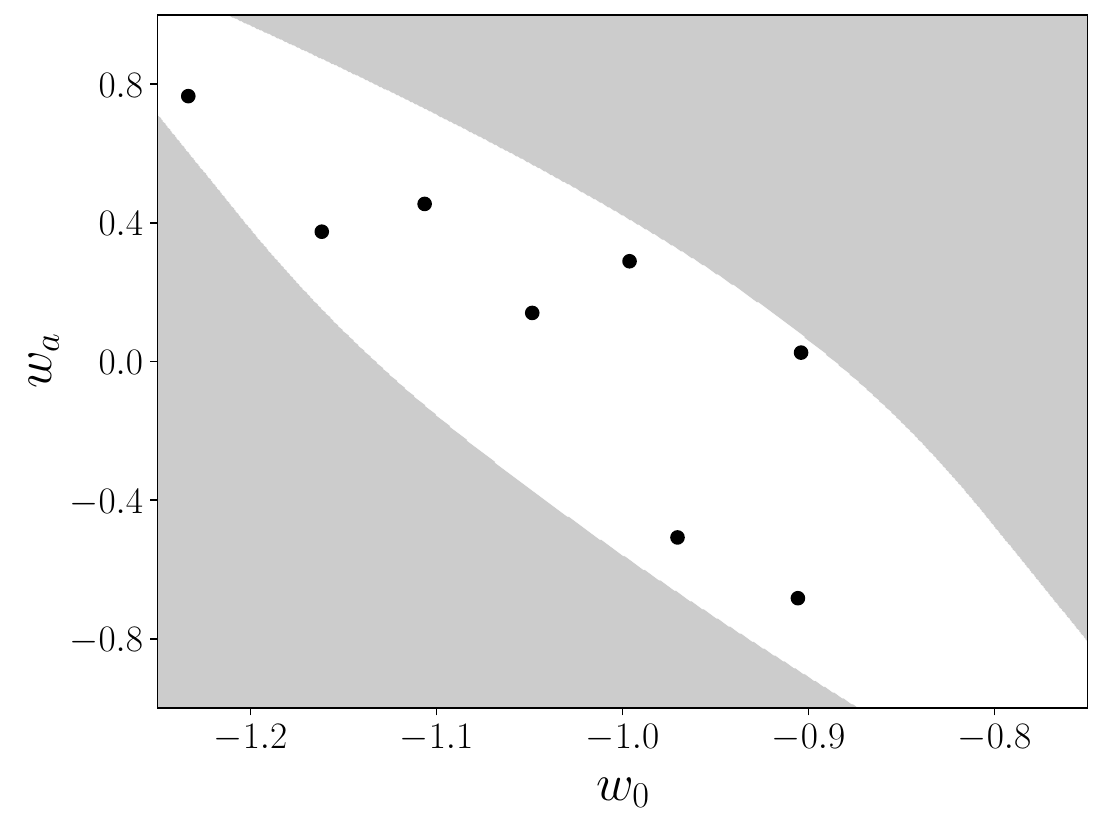}
    \caption{
    Parameter space of interest for $(w_0,w_a)$ under the DESI DR1 blinding scheme. The white region represents the parameter region that allows for changes in $\alpha_\parallel$ and $\alpha_\perp$ of less than 3\% with respect to a fiducial chosen value of 1 in the redshift range $0.4 < z < 2.1$. The black points are 8 random selections used to blind our mock catalogs, which we use to validate our methodology.
    }
    \label{fig:w0-wa_plane_zeffcombined_8realisations_4paper_0.4-2.1}
\end{figure}

To ensure the robustness of our blinding scheme, we confine the shifts in the blinded cosmology to specific regions within the \wowa\ parameter space. This allows us to ensure that these shifts can be accurately translated into galaxy redshift changes. In particular, we dictate that the shifts for the BAO-scaling parameters \( \alpha_{\perp} \) and \( \alpha_{\parallel} \), defined in \cref{eq:aperp_apar}, should be kept within a maximum deviation of \(3\%\) from their fiducial value of unity, i.e., \( |\alpha_{\perp} - 1| < 0.03 \) and \( |\alpha_{\parallel} - 1| < 0.03 \), respectively. \cref{fig:w0-wa_plane_zeffcombined_8realisations_4paper_0.4-2.1} illustrates these constraints, showcasing a \wowa\ region permissible within the redshift range \( 0.4 < z < 2.1 \) (white region), from which we pick \wowa\ values for blinding; details of the redshift range used for this selection are given in \cref{appendix:zrange_blinding}. Also, we require that the amplitude of the monopole of the clustering signal does not change significantly, i.e., we aim to keep it as close as reasonably possible to the true one\footnote{For a definition of multiples moments see \cref{sec:datavector}.}. We compute\footnote{\cref{eq:ffromw0wa} is derived by requiring the change in power spectrum monopole amplitude due to RSD from \cref{eq:kaiserformula} to compensate the volume dilation factor proportional to $\alpha_\mathrm{iso}^{-3}$ arising from the AP blinding.} the impact of the blinded cosmology on the growth factor \( f \) as
\begin{equation} \label{eq:ffromw0wa}
    f_\mathrm{blind}(z) = b_1(z) \left( \sqrt{\frac{D_{A,\mathrm{fid}}^2(z) H_\mathrm{blind}(z,w_0,w_a)}{D_{A,\mathrm{blind}}^2(z,w_0,w_a) H_\mathrm{fid}(z)} \left( \frac{f_\mathrm{fid}^2(z)}{b_1^2(z)} + \frac{10}{3} \frac{f_\mathrm{fid}(z)}{b_1(z)} \right) + \frac{25}{9}}-\frac{5}{3} \right)\, ,
\end{equation}
and require that the shifts in \( f \) do not exceed \(10\%\) of the fiducial value, \( f_{\text{fid}} = 0.8 \). These ranges (3\% for $\alpha_{\perp}$ and $\alpha_{\parallel}$, 10\% for $f$) were roughly based on the precision of such measurements before DESI DR1.

For validating the blinding scheme, we randomly select 8 pairs of \wowa, shown as black dots in \cref{fig:w0-wa_plane_zeffcombined_8realisations_4paper_0.4-2.1}, as well as two \fnl\ values ($\pm$20), to blind our mock catalogs. This validation is described in \cref{sec:validation-wmocks}.

To blind the data catalogs, we generate a list of 1,000 random combinations of \wowa, all within the white region of \cref{fig:w0-wa_plane_zeffcombined_8realisations_4paper_0.4-2.1}. Then, from the list of 1,000 pairs of \wowa, we randomly select one to blind our data catalog (using the same pair of values for all our tracers), following the prescription in \cref{sec:blinding-scheme}. We do not disclose the parameters used for blinding the data given that not all the DESI DR1 papers are unblinded as of this writing; note that the same blinding is used for BAO, RSD, and \fnl\ DESI analyses. Importantly, even the `blinding team' did not have access to the specific blinding parameters. This ensures that ongoing projects, such as the \fnl\ and RSD validations, remain securely blinded despite the unblinding of BAO results. 

By adhering to these principles, our DESI DR1 blinding scheme aims to provide a reliable and effective means to mitigate the confirmation bias effect, thereby ensuring the integrity of the DESI DR1 cosmology analysis.

\subsection{When to Unblind: Criteria and  Tests\label{sec:when-to-unblind}}
The decisions on the blinding pipeline were coordinated closely with the DR1 DESI collaboration 2-point clustering cosmological analysis team. Once milestones for analysis validation on blinded catalogs were reached\footnote{For a detailed list of milestones reached in the BAO analysis, see Section 6 of \cite{DESI2024.III.KP4}. A similar list of unblinding criteria and tests for the full-shape analysis will be available in the forthcoming full-shape paper \cite{DESI2024.V.KP5}.}; the results were unblinded. The results particularly relevant to this work are presented in:
\begin{enumerate}
    \item DESI 2024 II: Sample definitions, characteristics, and two-point clustering statistics \cite{DESI2024.II.KP3}
    \item DESI 2024 III: Baryon Acoustic Oscillations from Galaxies and Quasars \cite{DESI2024.III.KP4}
    \item DESI 2024 V: Analysis of the full shape of two-point clustering statistics from galaxies and quasars \cite{DESI2024.V.KP5}
\end{enumerate}

We emphasize once more that the scope of the present paper is to ensure that the blinding pipeline works as expected (as detailed in the next section), and we refer the reader to \cite{DESI2024.III.KP4, DESI2024.V.KP5} for details regarding the respective analyses,including optimizations and unblinding requirement tests, which are beyond the scope of the present paper. We also note that DESI  2024 IV focuses on BAO measurements Lyman-$\alpha$ forest; blinding, validation, and analysis for IV were distinct from those described above; we refer the reader to \cite{DESI2024.IV.KP6}. Finally, DESI 2024 VI focuses on likelihoods and combining the various probes; see \cite{DESI2024.VI.KP7A, DESI2024.VII.KP7B} for further details.

\subsection{Blinding Pipeline Validation}

As previously mentioned, the primary objective of this paper is to demonstrate that the blinding pipeline performs as intended. Specifically, we validate the pipeline by applying it to both mock and real (blinded) data (i.e., \cref{sec:validation-wmocks}, \cref{sec:validation-wdata}), 
\textit{ensuring that we observe only the expected shifts deliberately introduced by the blinding procedure, and no other remaining artifact}. These checks confirm that the blinding process is effective, without unintentionally revealing any cosmological information prior to unblinding, and that no uncontrolled deviations occur. Given that we understand how the blinding affects our measurements, we can safely utilize the catalogs that underwent through the pipeline and then study systematic error budgets. After those errors are set, one can finally unblind.

\section{ANALYSIS FRAMEWORK\label{sec:analysis-framework}}

Here we elaborate on how we calculate the two-point statistics from the catalog (our data vectors). We also explain the theoretical model as well as the compression employed to extract information from the two-point statistics, followed by the details of the inference framework that we adopted.

\subsection{Data Vector and Covariance}\label{sec:datavector}
\subsubsection{Correlation Function Estimator}

In order to measure the two-point correlation function, which measures the excess probability of finding two galaxies at a specific separation $s$ and angle relative to the line-of-sight $\mu$, we use the Landy-Szalay estimator~\cite{Landy:1993yu},
\begin{equation}
\widehat{\xi}(s, \mu) = \frac{DD(s, \mu) - DR(s, \mu) - RD(s, \mu) + RR(s, \mu)}{RR(s, \mu)},
\end{equation}
where $DD$, $DR$, $RD$, and $RR$ represent the weighted number of data ($D$ and random ($R$) pairs in specific distance and angle bins. From this, we calculate the multipole moments (monopole given by $\ell = 0$, quadrupole by $\ell = 2$, and hexadecapole by $\ell = 4$) using Legendre polynomials,
\begin{equation}
\widehat{\xi}_{\ell}(s) = \frac{2\ell + 1}{2} \int d\mu \, \widehat{\xi}(s, \mu) \mathcal{L}_{\ell}(\mu)\,.
\end{equation}

\subsubsection{Power Spectrum Estimator}

The power spectrum estimator, on the other hand, leverages the Feldman-Kaiser-Peacock (FKP) estimator ~\cite{yamamoto2006, Feldman:1993ky}, which combines galaxy and random field densities to calculate the weighted galaxy fluctuation field,
\begin{equation}
F(\Vec{r}) = n_{d}(\Vec{r}) - \alpha n_{r}(\Vec{r}),
\end{equation}
where $n_{d}(\Vec{r})$ and $n_{d}(\Vec{r})$ are the weighted galaxy and random number densities, the latter having a total weighted number $1/\alpha$ times the one of the data catalog. For a detailed discussion of the weighting scheme, we refer to \cite{DESI2024.II.KP3}.

Power spectrum multipoles are then calculated as an average over all Fourier modes \(\Vec{k}\) within a bin centered on the magnitude \(k\). The sum over \(\Vec{k}\) corresponds to summing over the different wavevectors that fall within the bin. The estimator is given by:
\begin{align}
\hat{P}_{\ell}(k) = \frac{2 \ell + 1}{A N_{k}} \sum_{\Vec{k} \in k} \sum_{\Vec{r}_{1}} \sum_{\Vec{r}_{2}} F(\Vec{r}_{1}) F(\Vec{r}_{2}) \mathcal{L}_{\ell}(\hat{k} \cdot \hat{\eta}) e^{i \Vec{k} \cdot (\Vec{r}_{2} - \Vec{r}_{1})} - \mathcal{N}_{\ell},
\end{align}
where the sums are performed over galaxy pairs with positions \(\Vec{r}_{1}, \Vec{r}_{2}\) and line-of-sight \(\hat{\eta}\), and over wavevectors \(\Vec{k}\) within the bin of magnitude \(k\). \(\mathcal{N}_{\ell}\) denotes the shot noise correction applied to the monopole term, and \(A\) is the normalization factor. \(N_k\) is the number of modes in the \(k\)-bin.

\subsubsection{Measurements}

We use \texttt{pycorr}\footnote{\url{https://github.com/cosmodesi/pycorr}: \texttt{pycorr} is essentially a wrapper of a modified version of the \textsc{Corrfunc} package~\cite{Sinha:2019reo}} and \texttt{pypower}\footnote{\url{https://github.com/cosmodesi/pypower}: \texttt{pypower} is based on the original \textsc{nbodykit}~\cite{Hand:2017pqn} implementation} to execute the two estimators above. As for covariances, we utilize those generated with the \texttt{RascalC}\footnote{\url{https://github.com/oliverphilcox/RascalC}} \cite{KP4s7-Rashkovetskyi} code for configuration space, and those from \texttt{TheCov}\footnote{\url{https://github.com/cosmodesi/thecov}} \cite{Wadekar:2019rdu,KP4s8-Alves} code for Fourier space.

\subsection{Compression Approaches and Theory Models}

We compress the information encoded in the two-point statistics, allowing fitting the two-point function to a \textit{template} with only a limited set of parameters. Before we delve into the fitting method, we summarize three approaches to compress information from the two-point clustering statistics, in increasing complexity:
\begin{enumerate}[noitemsep,topsep=1pt]
    \item \textbf{Standard BAO approach:} This method focuses on extracting the cosmological distance scale from the BAO observed in both pre- and post-reconstructed correlation functions and power spectra; see e.g. for reference~\cite{Seo:2015eyw}. Specifically, it utilizes the isotropic (\alphaiso) and anisotropic (\alphaap) dilation scales to infer the Hubble parameter and angular diameter distance relative to the sound horizon at the drag epoch. While extracting the BAO feature from clustering statistics, a polynomial expansion is often used to parameterize the broadband, allowing us to marginalize over non-BAO peak information. This approach provides a robust means to measure the expansion history of the Universe~\cite{BOSS:2012tck, BOSS:2013rlg, Xu:2012fw}.

    \item \textbf{Standard BAO+RSD approach:} This approach extends the standard BAO analysis by incorporating measurements of RSD, enabling the extraction of the growth rate of structure parameter (\(f\)) alongside the geometric BAO signals in the pre-reconstructed catalogs.
    The combined analysis not only enhances the constraining power on cosmological parameters, particularly those related to dark energy and gravity theories, but also measures the rate of gravitational clustering through the \(df\) parameter, providing a direct probe of the theory of gravity on cosmic scales~\cite{Percival:2009}. The set of parameters constrained then is \{\alphaiso, \alphaap, \(df\)\}, where \(df\) captures the rate of structure formation. (In our framework, $df$ is defined as the ratio $f/f^\mathrm{fid}$, and is calculated using the \texttt{desilike} package for model predictions\footnote{\url{https://github.com/cosmodesi/desilike}}.)

    \item \textbf{ShapeFit:} Applied exclusively to pre-reconstructed power spectra, ShapeFit goes beyond BAO and RSD by incorporating an additional parameter \dm\ to model the broadband shape of the power spectrum
    . The expanded set of parameters: \{\alphaiso, \alphaap, \dm\, \(df\)\} allows for a more comprehensive analysis of the cosmic expansion history and growth of structure, leveraging the complementary information encoded in the shape of the power spectrum and the amplitude of RSD~\cite{Brieden:2021edu, Ramirez:2023ads}.
\end{enumerate}

\noindent In this paper, we utilize the standard BAO approach and ShapeFit to capture the two ends of complexity for our validation tests. Now we turn to explaining how exactly the two compression analyses work.

\subsubsection{Standard BAO Compression \label{sec:standard-bao}}

The standard approach utilizes a pre-defined template based on theoretical predictions that capture the effects of BAO and a broadband term that marginalizes over RSD in the clustering of galaxies. By adjusting the template’s amplitude, scale, and shape to best match the observed data, we can infer distances. The fitting template for the power spectrum is defined, as in \cite{KP4s2-Chen}, as
\begin{equation} \label{eq:generic_model}
    P(k, \mu) = \mathcal{B}(k, \mu) P_{\rm nw}(k) + \mathcal{C}(k, \mu)P_{\rm w}(k) + \mathcal{D}(k)\,,
\end{equation}
where $P_{\rm nw}(k)$ is the smooth (no-wiggle) component of the linear power spectrum and $P_{\rm w}(k)$ is the BAO (wiggle) component. Both components are obtained using the \textit{peak average} method from \cite{Brieden2022:2204.11868}.

The term $\mathcal{B}(k, \mu)$ incorporates both the RSD and linear bias, while $\mathcal{C}(k, \mu)$ extends this by including an anisotropic damping factor that describes the broadening of the BAO signal due to large-scale bulk flows. This damping is characterized by the parameter $\Sigma$, which accounts for the scale-dependent distortion of the BAO signal. The $\Sigma$ parameters include contributions from both the real-space distortions and the effects of small-scale velocities, commonly known as the Fingers-of-God (FoG) effect. Recent work by \cite{KP4s2-Chen} suggests that the FoG damping should be applied exclusively to the smooth component, $P_{\rm nw}(k)$, which is the approach adopted here and in \cite{DESI2024.III.KP4}.

Finally, $\mathcal{D}(k)$ accounts for any residual deviations from linear theory in the broadband shape of the power spectrum multipoles. The model in \cref{eq:generic_model} is then integrated over $\mu$ to produce predictions for the power spectrum multipoles:

\begin{align}
    P_{\ell}(k) = \frac{2\ell+1}{2} \int_{-1}^1 &d\mu \mathcal{L}_\ell(\mu) \big[\mathcal{B}(k, \mu) P_{\rm nw}(k) \nonumber \\
    &+ \mathcal{C}(k^\prime(k,\mu), \mu^\prime(k,\mu))P_{\rm w}(k^\prime(k,\mu))\big] + \mathcal{D}_\ell(k).
    \label{eq:pow_spec_multipoles}
\end{align}
The term involving BAO wiggles is evaluated at $k^\prime$ and $\mu^\prime$, which are given by
\begin{equation}
    k^\prime(k,\mu)= \frac{k}{\alpha_\perp}\sqrt{1+\mu^2\left(\frac{\alpha_\perp^2}{\alpha_\parallel^2}-1\right)}
\end{equation}
and
\begin{equation}
    \mu^\prime(\mu)= \frac{\mu}{\frac{\alpha_\parallel}{\alpha_\perp}\sqrt{1+\mu^2\left(\frac{\alpha_\perp^2}{\alpha_\parallel^2}-1\right)}},
\end{equation}
where $\alpha_\perp$ and $\alpha_\parallel$ are the BAO scaling parameters across and along the line of sight, respectively, defined in \cref{eq:aperp_apar}. The measured $\alpha_\perp$ and $\alpha_\parallel$ can be 
transformed into the isotropic and anisotropic BAO dilations $\alpha_{\text{iso}}$ and  $\alpha_{\text{AP}}$ provided in \cref{eq:aisoaap}.
The latter represents the parameter basis we use throughout the rest of this work.

\subsubsection{ShapeFit Compression\label{sec:shapefit}}

This approach incorporates additional shape information from the galaxy power spectrum, while also fitting for BAO and RSD features.

Within the ShapeFit formalism the scale dependence of the linear power spectrum $P_\mathrm{lin}$ is represented by the following modification of the fiducial template $P_\mathrm{lin}^\mathrm{fid}$ via the shape parameter \dm\ 
\begin{equation}
    P_\mathrm{lin}(k) = P_\mathrm{lin}^\mathrm{fid}(k)\ \exp\left\{ \frac{m}{a_m}\tanh \left[a_m\ln\left(\frac{k}{k_p}\right) \right] \right\}\, .
\end{equation}
Here, \(k_{\text{p}}=\pi/r_d\) is the pivot scale and $a_m=0.6$ is tuned to fit the full numerical calculation of the linear power spectrum with a Boltzmann code such as CLASS~\cite{Blas:2011rf} or CAMB~\cite{Lewis:1999bs} over a wide model parameter space; see \cite{Brieden:2021edu} and in particular Figure 4 therein for reference.

Underpinning our ShapeFit analysis is the Lagrangian Perturbation Theory (LPT) approach to large-scale structure, provided by the \texttt{velocileptors}\footnote{\url{https://github.com/sfschen/velocileptors}} code~\cite{KP5s1-Maus}. This tool computes the redshift-space distortions and clustering statistics using perturbation theory, including non-linearities crucial for accurate modeling at scales smaller than $k \geq 0.07\,h\,\mathrm{Mpc}^{-1}$. We refer the reader to \cite{KP5s1-Maus} for the detailed modeling.

\subsection{Inference Framework}
We apply our analysis pipeline to extract key cosmological information from the two-point statistics, using theoretical models and data explained earlier, summarizing the complex data into a few interpretable parameters. We implement a series of scale cuts as detailed in \cref{tab:tracers-redshift-scale-cut} for analyses both in Fourier and configuration space, specifically, the correlation function multipoles are obtained as a Hankel-transform of the power spectrum multipoles. Finally, the measured power spectrum in relation to theoretical models is mediated through the window function, incorporating survey geometry and selection effects, using the formalism detailed in \cite{KP3s5-Pinon}.

We employ a Bayesian inference framework to extract the compressed parameters from galaxy correlation and power spectrum measurements, implemented in the \texttt{desilike} framework. To sample the posteriors, we utilize the Markov Chain Monte Carlo (MCMC) method implemented in the \texttt{emcee} package, with Gelman–Rubin convergence diagnostic of $R-1 < 0.02$. We also require the effective sample size of the chains\footnote{Defined as the maximum of the weighted chain length divided by the autocorrelation length for all parameters.} to be $\gtrsim 10^3$. Additionally, we do profile likelihood with \texttt{iminuit} package\footnote{\url{https://github.com/scikit-hep/iminuit}}. The priors used are included in \cref{tab:priors}.

\begin{table}[!htb]
\centering
\begin{tabular}{|c|c|c|c|c|}
\hline
Tracer & Redshift & Analysis Type & \(k_{\text{lim}}\) ($h$/Mpc) & \(s_{\text{lim}}\) (Mpc/$h$) \\
\hline
\multirow{2}{*}{BGS} & \multirow{2}{*}{\([0.1, 0.4]\)} & BAO & [0.02, 0.3] &  [50, 150]  \\
\cline{3-5}
& & ShapeFit &  [0.02, 0.2] & [32, 150]  \\ \hline
\multirow{2}{*}{LRG} & \multirow{2}{*}{\([0.4, 0.6], [0.6, 0.8], [0.8, 1.1]\)} & BAO & [0.02, 0.3] &  [50, 150]  \\
\cline{3-5}
& & ShapeFit &  [0.02, 0.2] & [30, 150]  \\ \hline
\multirow{2}{*}{ELG} & \multirow{2}{*}{\([0.8, 1.1], [1.1, 1.6]\)} & BAO & [0.02, 0.3] &  [50, 150]  \\
\cline{3-5}
& & ShapeFit &  [0.02, 0.2] & [27, 150]  \\ \hline
\multirow{2}{*}{QSO} & \multirow{2}{*}{\([0.8, 2.1]\)} & BAO & [0.02, 0.3] &  [50, 150]  \\
\cline{3-5}
& & ShapeFit &  [0.02, 0.2] & [25, 150]  \\ \hline
\end{tabular}
\caption{Summary of galaxy tracers, their redshift ranges, and the applied scale cuts for BAO analysis (\(\ell = 0, 2\)) and ShapeFit (\(\ell = 0, 2, 4\)).}
\label{tab:tracers-redshift-scale-cut}
\end{table}

\begin{table}[!htb]
\centering
\begin{tabular}{lll}
\hline
\textbf{Parameter} & \textbf{Prior} & \textbf{Description} \\
\hline
\\[-1em]
\multicolumn{3}{l}{BAO Template} \\
\hline
\alphaiso & $\mathcal{U}(0.8, 1.2)$ & Isotropic distortion parameter \\
\alphaap & $\mathcal{U}(0.8, 1.2)$ & Alcock-Paczynski distortion parameter \\
\hline
\\[-1em]
\multicolumn{3}{l}{ShapeFit Template} \\
\hline
\alphaiso & $\mathcal{U}(0.8, 1.2)$ & Isotropic distortion parameter \\
\alphaap & $\mathcal{U}(0.8, 1.2)$ & Alcock-Paczynski distortion parameter \\
\dm\ & $\mathcal{U}(-3, 3)$ & Shape parameter \\
$df$ & $\mathcal{U}(0, 2)$ & Growth rate parameter\\

\hline
\\[-1em]
\multicolumn{3}{l}{\texttt{velocileptors} Theory } \\
\hline
$b_{1}$ & $\mathcal{U}(-1, 10)$ & Linear bias, density relation. \\
$b_{2}$ & $\mathcal{N}(0, 10^2)$ & Second-order bias, non-linear effects. \\
$b_{s}$ & $\mathcal{N}(0, 5^2)$ & Tidal bias, anisotropic clustering. \\
$\alpha_{0}$ & $\mathcal{N}(0, 30^2)$ & Monopole shot noise. \\
$\alpha_{2(4)}$ & $\mathcal{N}(0, 50^2)$ & Quadrupole (and hexadecapole) shot noise. \\
$s_{n, 0}$ & $\mathcal{N}(0, 4^2)$ & Monopole stochastic term. \\
$s_{n, 2}$ & $\mathcal{N}(0, 100^2)$ & Quadrupole stochastic term. \\
$s_{n, 4}$ & $\mathcal{N}(0, 500^2)$ & Hexadecapole stochastic term. \\
\hline

\end{tabular}
\caption{Parameter priors and descriptions for the BAO and ShapeFit templates as well as \texttt{velocileptors} used in our analysis. We note that when excluding the hexadecapole from our analysis, we set $\{\alpha_{4}, s_{n,4}\}=0$. Moreover, following the \texttt{velocileptors} paper~\cite{KP5s1-Maus}, the $b_3$ parameter, representing third-order bias, is set to zero throughout this work.}

\label{tab:priors}
\end{table}

\subsection{Blinding Validation Metrics: Definition of $\Gamma$}
In this section, we describe how we quantify any net residual (i.e., bias) in the measured dilation and ShapeFit parameters after accounting for the expected effect of the shift introduced by the blind cosmologies. We first rescale the measured values with a given cosmology in terms of the baseline cosmology distance ratios as

\begin{equation}
\label{eq:rescale}
    \alpha^{\rm measured} \rightarrow \frac{\alpha^{\rm measured}}{\alpha^{\rm rescaling
    }}
\end{equation}
with
\begin{align}
\label{eq:alpha_rescale_factor}
    \alpha_{{\rm iso}}^{\rm rescaling} = \frac{D_{V}^{\rm baseline}}{D_{V}^{\rm fid}} \frac{r_d^{\rm fid}}{r_d^{\rm baseline}}, \qquad  \alpha_{\rm AP}^{\rm rescaling} = \frac{D_{\rm M}^{\rm baseline}/D_{\rm H}^{\rm baseline}}{D_{\rm M}^{\rm fid}/D_H^{\rm fid}} ,
\end{align}
\noindent and where $D_\mathrm{V}\equiv (D_H(z)D_M(z)^2)^{1/3}$ is the spherically averaged distance. The baseline here will be \textbf{a} blind cosmology; thus, the quantity $\alpha_{{\rm iso}}^{\rm rescaling}$ is telling you how much shift is expected due to \textbf{a} blind cosmology. Rescaling the measure values by it, therefore, tells you how far (deflated) you are from \textbf{a} blinded cosmology. Moreover, to reduce the sample variance, we focus on the ratio,
\begin{equation}
\label{eq:gamma_alpha}
    \Gamma_{\alpha} = \frac{\alpha^{\rm measured}}{\alpha^{\rm rescaling}} \times \frac{1}{\alpha_{\rm baseline}^{\rm measured}}\,.
\end{equation}
\noindent The first term accounts for how blinding shifts the measure parameters, while the second one is merely to reduce sample variance. $\sigma_{\Gamma}$ are obtained by standard error propagation.

Finally, we generalize  the notation defining $\Gamma_i$, where $i$ range in (\alphaiso, \alphaap, $df$), i.e., $\Gamma_i = \{\Gamma_{\alpha_{\mathrm{iso}}}, \Gamma_{\alpha_{\mathrm{AP}}}, \Gamma_{df} \}$. \textit{The ratios {\rm{($\Gamma_i$)}} is expected to be unity in an ideal scenario where no systematic errors were introduced}. Similarly,  $\tilde{\Gamma}_m$ is defined in terms of differences,

\begin{equation}
\label{eq:gamma_m}
    \tilde{\Gamma}_{m} = ({m^{\rm measured}} - {m^{\rm rescaling}}) - {m_{\rm baseline}^{\rm measured}}\,.
\end{equation}
\textit{The differences {\rm{($\tilde{\Gamma}_m$)}} is expected to be zero in the presence of no systematic errors.}

\section{VALIDATION WITH MOCKS\label{sec:validation-wmocks}}

\subsection{Mock Data and Preliminary Checks}

To rigorously validate the blinding scheme developed for DESI DR1, we utilize mock catalogs produced from the {\sc AbacusSummit}\footnote{The {\sc AbacusSummit} suite contains 25 different boxes at Planck-2018 cosmology with different phases in the initial conditions, which we refer to as base boxes.} $N$-body simulations \citep{AbacusSummit}. These mocks were produced by fitting the galaxy two-point correlation function at small scales using {\tt Abacus} halos and a halo occupation 
distribution model \citep{2022AbacusHOD}, in order to populate the dark matter halos with galaxies\footnote{The mocks used in this work are referred to as {\tt Abacus-1} in other DESI papers, i.e., they were produced with the fiducial cosmological parameters of \cref{eq:fiducial-cosmo}.}. Each tracer at each redshift is populated over all 25 base boxes, giving a total volume of 200$h^{-3}$Gpc$^{3}$. These mock datasets aim to comprehensively mimic the characteristics of the actual DESI DR1 data. Specifically, they include features like target galaxy distributions and redshift bins, observational systematics, and have a window footprint applied, thus providing an ideal dataset for testing our analysis pipeline. 

We applied our blinding pipeline to the mock catalogs, using the same procedure as for the actual data, but with a distinct realization of the blinding parameters.

\begin{figure}[!htb]
\centering
\includegraphics[width=0.495\textwidth]{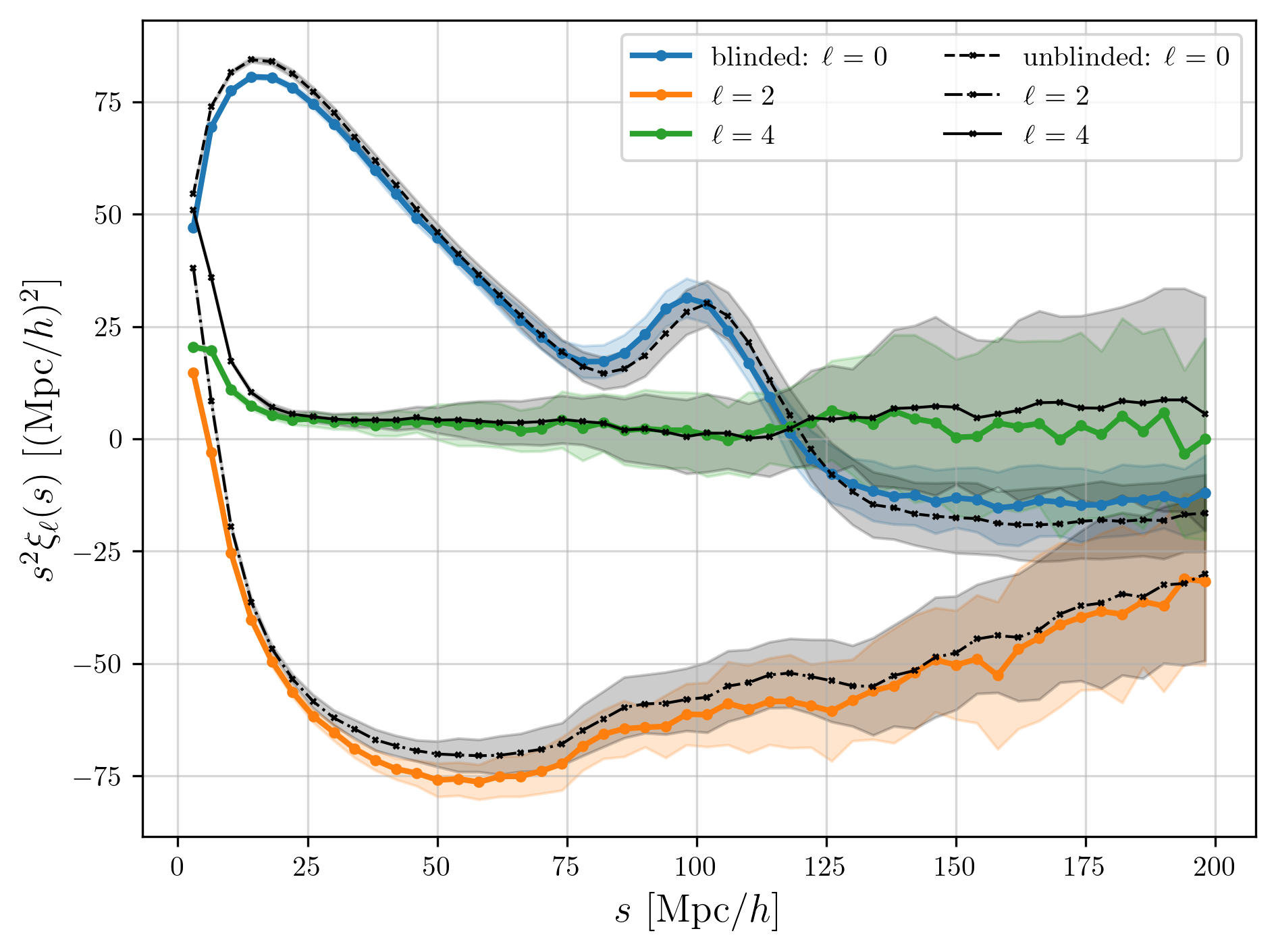}
\includegraphics[width=0.495\textwidth]{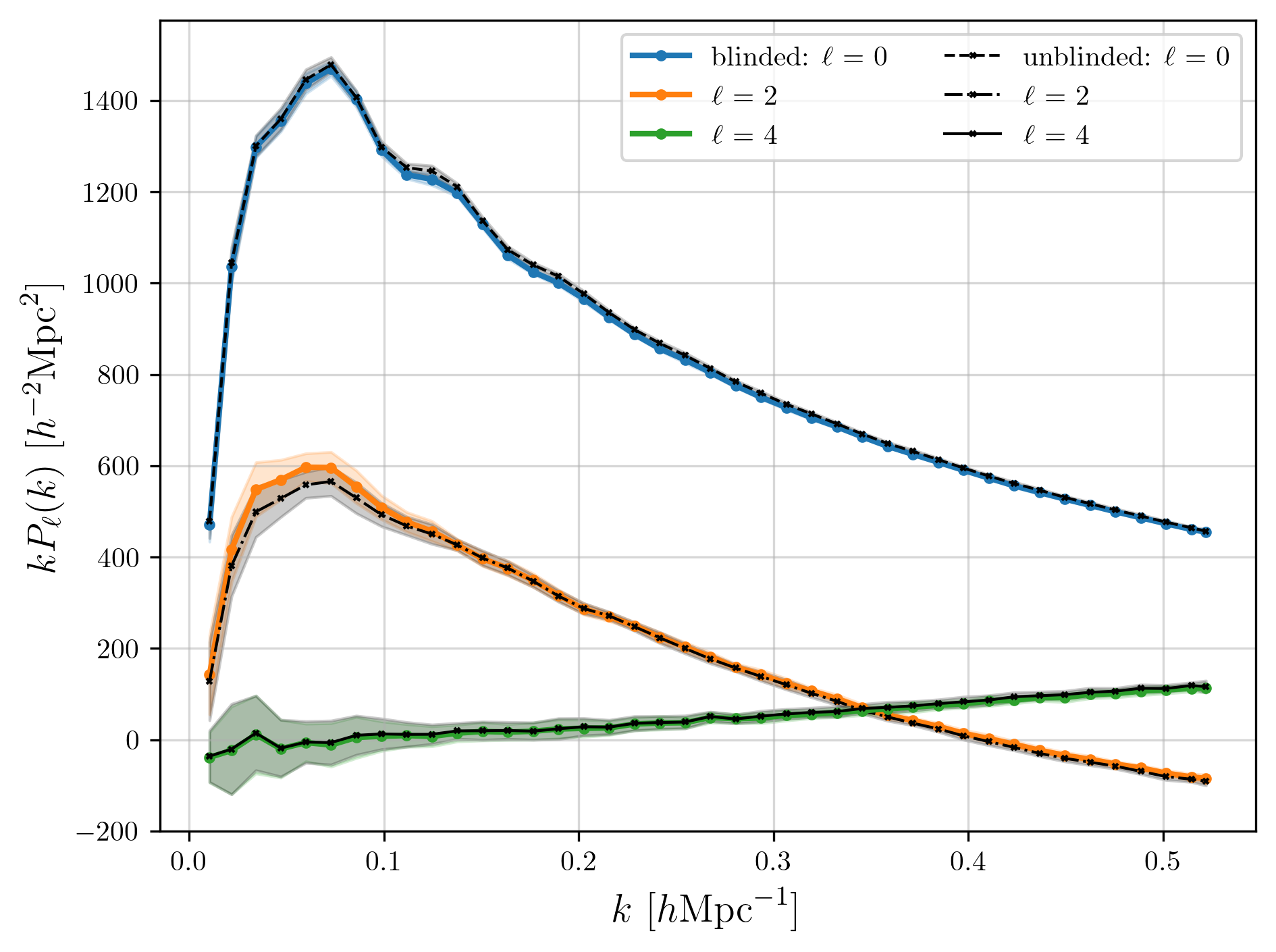}
\caption{Comparison of blinded and unblinded mocks for multipoles $\ell = 0$, $\ell = 2$, and $\ell = 4$, for the correlation function (left column) and power spectrum (right column). The curves show the mean is across 25 \abacus\ catalogs which are blinded with the same blinding parameters.
}
\label{fig:25mocks_multipoles}
\end{figure}

To assess the impact of these blinding parameters, we tested them across 25 \abacus\ mock catalogs. We aimed to confirm that the blinding did not inadvertently introduce any distortions or biases in the clustering signal, such as spurious changes to the correlation function or power spectrum. As depicted in \cref{fig:25mocks_multipoles}, examining both the correlation function and power spectrum multipoles reveals that the blinded and unblinded monopole amplitudes remain largely consistent, meaning the blinded catalogs retain the same overall shape and comparable amplitudes as the unblinded catalogs, while successfully altering the BAO position and quadrupole amplitude on large scales. This validation serves as a sanity check to confirm that the blinding procedure does not introduce any unwanted artifacts, such as excess clustering on any scale. Therefore, the blinding procedure changes the clustering signal as expected without introducing unexpected distortions or biases.

Second, we select \textit{one} of the 25 \abacus\ mock catalogs for more detailed testing, applying 8 $\times$ 2  blindings (following the 8 randomly selected \wowa\ pairs shown as black points in \cref{fig:w0-wa_plane_zeffcombined_8realisations_4paper_0.4-2.1} and 2 \fnl\ values, $\pm 20$) to this specific catalog. Note that all our fitting analyses in this section refer to this \textit{one} out of 25 \abacus\ mock catalogs we mention earlier. Clustering measurements from this catalog are displayed for completeness in the~\cref{appendix:sup}. 

\subsection{Testing and Evaluation}
We carry out several tests on LRGs, ELGs and QSO; see \cref{tab:tracers-redshift-scale-cut} for relevant details. These tracers are vital for testing the robustness and applicability of our blinding scheme for both BAO and ShapeFit analyses, as these are the samples for which the analysis is planned to be carried out with real data\footnote{Note that we are not including BGS and Ly$\alpha$ tracers in our tests; we do, however, have the sample \texttt{BGS\_BRIGHT-21.5} in the validation tests with real data, discussed in \cref{sec:validation-wdata}.}.

Next, we detail the results from the anisotropic pre- and post-reconstruction BAO fitting analysis in both configuration and Fourier spaces, as well as the ShapeFit pre-reconstruction power spectrum analysis. 

\subsubsection{Validation Results for BAO Analysis\label{sec:validation-bao}}

Starting with our first tracer, LRGs, we carry out anisotropic BAO fitting for LRG samples in the redshift ranges of $0.4 < z < 0.6$, $0.6 < z < 0.8$, and $0.8 < z < 1.1$.

\begin{figure}[hbt!]
    \centering
    \vspace*{-1em}
    \includegraphics[width=0.75\textwidth, clip=True, trim={0 25 0 0}]{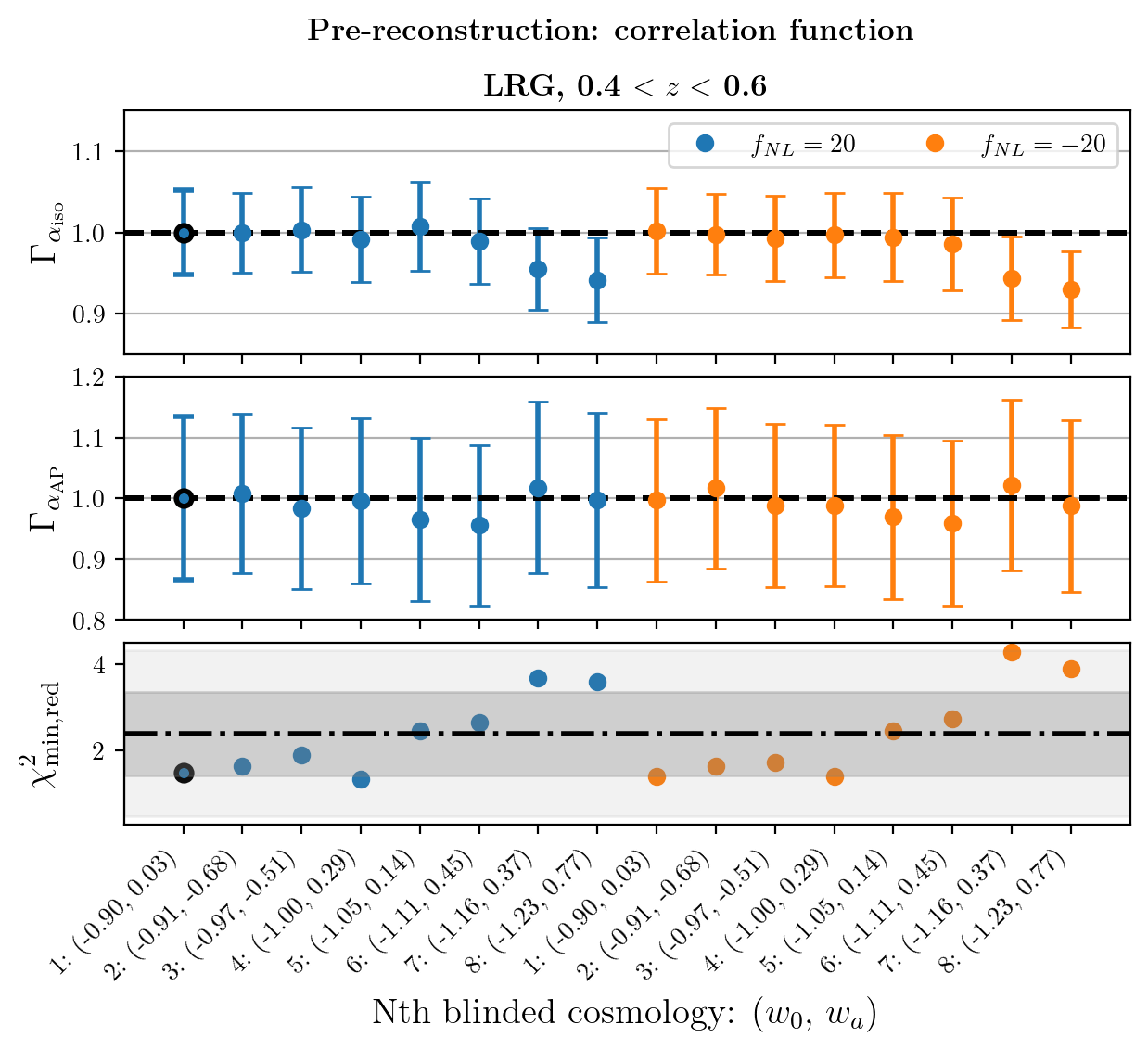}
    \includegraphics[width=0.75\textwidth, clip=True, trim={0 25 0 0}]{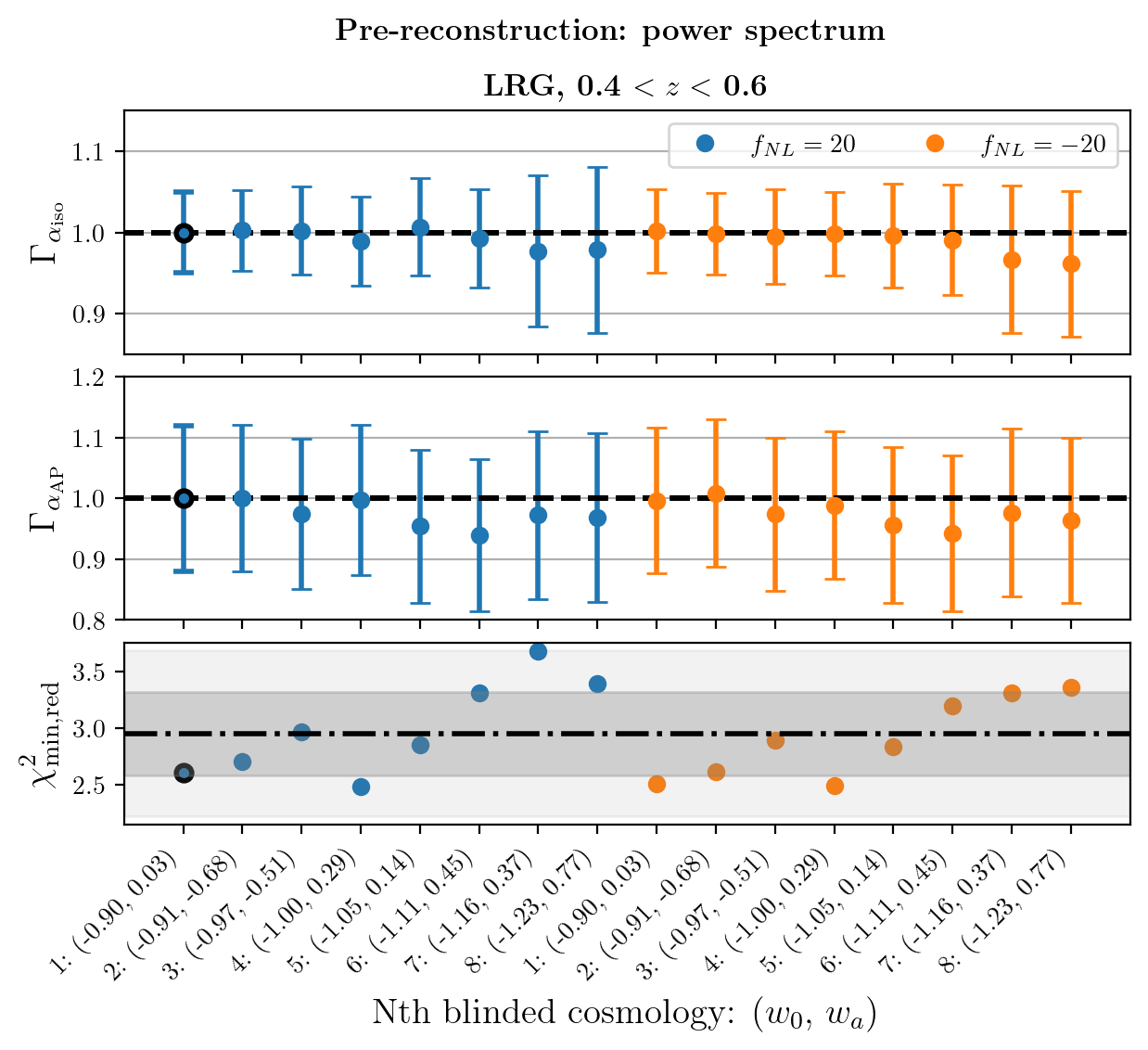}
    \caption{
    \footnotesize{
    Pre-reconstruction anisotropic BAO fits using the correlation function (top) and the power spectrum (bottom) for LRG samples for the first redshift bin (each row) from 16 different blinded mock catalogs with \wowa\ choices identified by indices 1-8 and two \fnl\ values by blue and orange, respectively. The top two subplots in each panel plot $\Gamma_i$, defined as the ratio of measured vs expected ratios of the $i$th parameter from each \texttt{sim} vs a reference \texttt{sim} (identified with black marker-edge); here $i$ = \alphaiso, \alphaap, where measured values are from the analysis pipeline while expected ones are from the theoretical connection with the respective \wowa; error bars capture the measurement uncertainties while propagating the errors. This statistic allows comparing all the sims against a reference sim. The bottom subplot in each panel displays the reduced \chitwo\ values, with shaded areas representing $1\sigma$ and $2\sigma$ regions;
    the $\sigma$-limits are obtained as the standard deviation from the mean of the \chitwo\ distribution of the 16 \chitwo\ values. This confirms the consistency and reliability of BAO measurements under various blinding shifts given the small variations.
    }
    }
    \label{fig:pre_recon_bao_fits_LRG}

\end{figure}

\cref{fig:pre_recon_bao_fits_LRG} shows the anisotropic BAO fitting results, for both pre- and post- reconstruction. We focus on three metrics: the ratios of measured to expected values for BAO fitting parameters (\alphaiso, \alphaap), as well as the reduced \chitwo\ of the fit\footnote{We emphasize that the $\chi^2$ values correspond to the inference of each blinded cosmology, not normalized by the first \texttt{sim}.}; error bars show measurement uncertainties. Here the measured values are using the analysis pipeline, while the expected values
are based on the true and blinded cosmology. As we see in the figure, while there is some variation across the 16 blinding catalogs in the measured vs. expected ratios of the BAO fitting parameters, \chitwo\ is within $1\sigma$ for $\sim 80\%$ of the cases and within $2\sigma$ for the rest, as represented by the light-gray and dark-gray areas, respectively\footnote{There are a few cases near the $2\sigma$ boundary, but they are not concerning as they remain close. These cases represent only about $3\%$ of the total.}. Note that these $\sigma$-limits are obtained as the standard deviation from the mean of the \chitwo\ distribution of the 16 blinding cases. This demonstrates that the blinding preserves the signal we aim to measure, but maintains the variances of the sample \chitwo\ low.

Also, it is interesting to note that, as expected, the BAO analysis is not sensitive to the $f_\mathrm{NL}$-blinding values, i.e., the scatter in (\alphaiso, \alphaap) follows the same trend for $f_\mathrm{NL}=20$ (blue points) and $f_\mathrm{NL}=-20$ (orange points).

The results of the post-reconstruction anisotropic BAO fitting for the LRG samples, including the quantitative assessments of \alphaiso, \alphaap\ across the 8 $\times$ 2 studied blinded cosmologies, are summarized in \cref{tab:test-metrics}. This table provides a comprehensive overview of the fitting accuracy in configuration space, emphasizing that \chitwo\ variation is within 1-2$\sigma$.

\begin{figure}[hbt!]
    \centering
    \includegraphics[width=0.75\textwidth, clip=True, trim={0 25 0 0}]{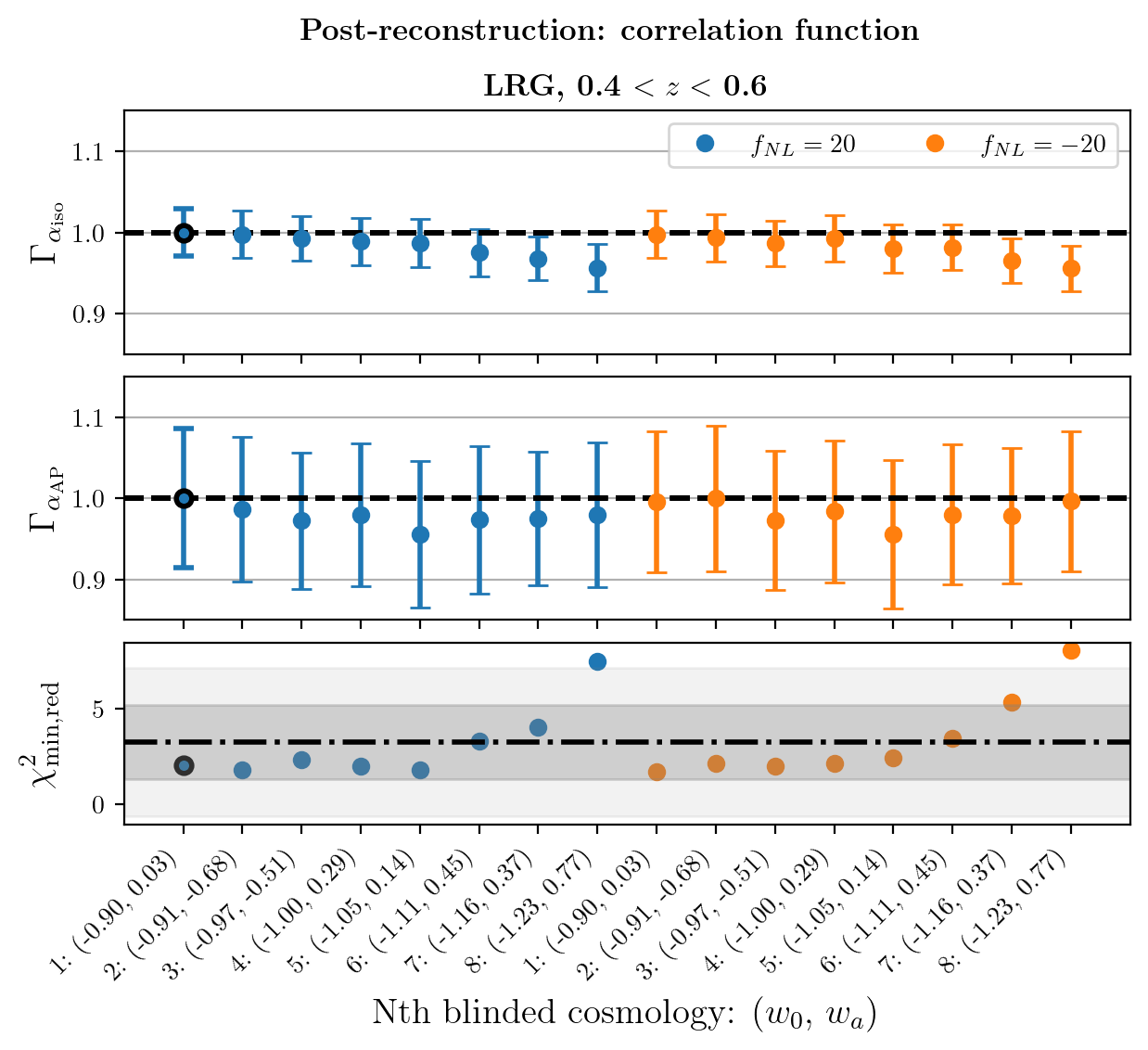}
    \includegraphics[width=0.75\textwidth, clip=True, trim={0 25 0 0}]{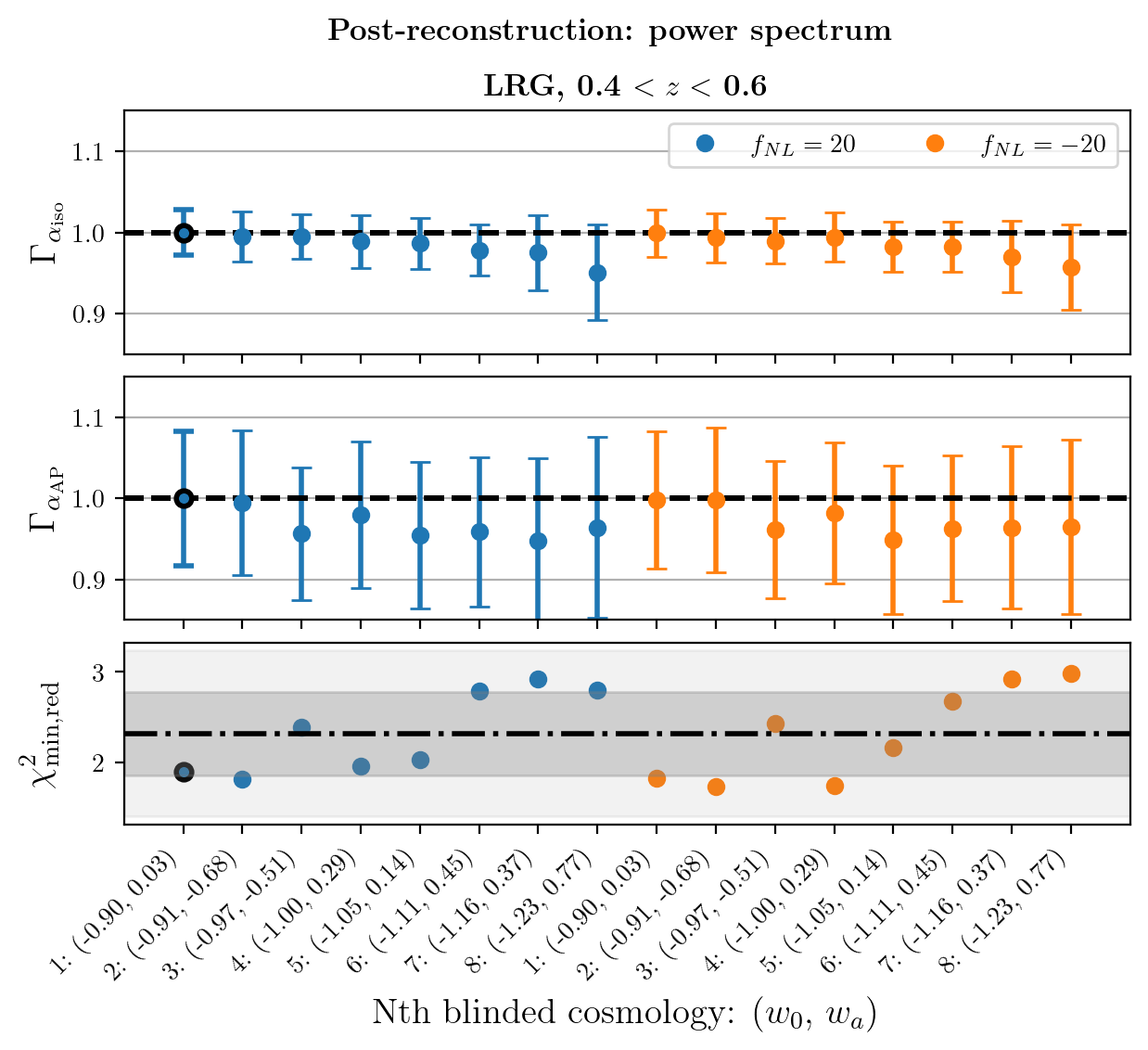}
    \caption{Post-reconstruction anisotropic BAO fits for LRG samples for the first redshift bin (each row) following the structure in \cref{fig:pre_recon_bao_fits_LRG}. Here, too, we see that while our $\Gamma$ statistic varies around the expected value of unity, the reduced \chitwo\ indicates good fits.}
    \label{fig:post_recon_bao_fits_LRG}
\end{figure}

We repeat the same for our other two tracers, ELGs and QSO, arriving at the same results: while we see some variation in measured vs expected ratios of the BAO fitting parameters across the 16 blinding catalogs, we see that \chitwo\ variation is always within 1-2$\sigma$; \cref{appendix:sup} shows the figures (\cref{fig:pre_recon_bao_fits_ELG} for pre-construction results for ELGs, \cref{fig:post_recon_bao_fits_ELG} for post-construction ones; and \cref{fig:pre_recon_bao_fits_QSO} for pre-reconstruction results for QSOs and \cref{fig:post_recon_bao_fits_QSO} for post-construction ones).

The consistency of our results in both real and configuration spaces underscores the robustness of the blinding technique, which is vital for ensuring unbiased cosmological parameter estimation, as well as the reconstruction pipeline to extract the BAO features. We note that there are a
few cases where \alphaiso\ uncertainties is $\sim 1\sigma$ away from the expected (e.g., pre-recon middle panels of \cref{fig:pre_recon_bao_fits_LRG}), but they still consistent within $2\sigma$.
\begin{table}
\centering
\begin{adjustbox}{width=0.55\paperwidth,center}
\begin{tabular}{|c|ccc|cc|c|}
\toprule
\multicolumn{1}{|c|}{Case} &
\multicolumn{3}{c|}{Blinding Parameters} & \multicolumn{2}{c|}{Test Metrics} & \multicolumn{1}{c|}{Fit $\chi^2$} \\
\midrule
\# & $w_0$ & $w_a$ & \fnl & $\Gamma_{\alpha_\mathrm{iso}}$
& $\Gamma_{\alpha_\mathrm{AP}}$
&
$\chi^{2} / (52 - 13)$ \\
\toprule\hline
1 & -0.90 & 0.03 & 20 & 1.000  $\pm$ 0.0128 & 1.000 $\pm$ 0.0346 & 1.56 \\
\hline
2 & -0.91 & -0.68 & 20 & 1.005 $\pm$ 0.0131 & 0.996 $\pm$ 0.0383 & 1.26 \\
3 & -0.97 & -0.51 & 20 & 1.002 $\pm$ 0.0126 & 0.999 $\pm$ 0.0376 & 0.90 \\
4 & -1.00 & 0.29 & 20 & 1.003 $\pm$ 0.0128 & 0.986 $\pm$ 0.0358 & 2.38 \\
5 & -1.05 & 0.14 & 20 & 1.002 $\pm$ 0.0126 & 1.009 $\pm$ 0.0385 & 1.21 \\
6 & -1.11 & 0.45 & 20 & 1.002 $\pm$ 0.0129 & 1.011 $\pm$ 0.0385 & 1.22 \\
7 & -1.16 & 0.37 & 20 & 1.005 $\pm$ 0.0135 & 1.003 $\pm$ 0.0408 & 0.93 \\
8 & -1.23 & 0.77 & 20 & 1.008 $\pm$ 0.0130 & 1.003 $\pm$ 0.0384 & 0.95 \\
\hline
9 & -0.90 & 0.03 & -20 & 1.000 $\pm$ 0.0125 & 0.993 $\pm$ 0.0345 & 1.85 \\
10 & -0.91 & -0.68 & -20 & 1.003 $\pm$ 0.0128 & 0.996 $\pm$ 0.0375 & 1.38 \\
11 & -0.97 & -0.51 & -20 & 1.005 $\pm$ 0.0129 & 0.987 $\pm$ 0.0382 & 1.17 \\
12 & -1.00 & 0.29 & -20 & 1.005 $\pm$ 0.0126 & 0.990 $\pm$ 0.0359 & 1.48 \\
13 & -1.05 & 0.14 & -20 & 1.004 $\pm$ 0.0130 & 0.987 $\pm$ 0.0382 & 0.89 \\
14 & -1.11 & 0.45 & -20 & 1.005 $\pm$ 0.0129 & 0.998 $\pm$ 0.0383 & 1.04 \\
15 & -1.16 & 0.37 & -20 & 1.006 $\pm$ 0.0134 & 1.008 $\pm$ 0.0409 & 1.08 \\
16 & -1.23 & 0.77 & -20 & 1.012 $\pm$ 0.0133 & 0.991 $\pm$ 0.0398 & 0.96 \\
\bottomrule
\end{tabular}
\end{adjustbox}

\caption{Configuration space post-reconstruction anisotropic BAO fitting parameters for the LRG $0.8 < z < 1.1$ sample from the \abacus\ mock catalog, blinded using various blinding cosmologies with varying $w_0$, $w_a$, and \fnl. This table presents two statistics: (1) $\Gamma_i$, defined as the ratio of the measured vs. expected values of the $i$th parameter (i.e., the two BAO fitting parameters) from each simulation compared to a reference simulation, and (2) the reduced \chitwo\ from the fit, comparing the theoretical model to the data. \autoref{tab:hyperparams} presents the numbers for other analysis parameters.
\label{tab:test-metrics}
}
\end{table}

\begin{table}
\centering
\begin{adjustbox}{width=\textwidth,center}
\begin{tabular}{|c|ccccccccccc|}
\toprule
\multicolumn{1}{|c}{Case} & \multicolumn{11}{|c|}{Analysis Hyperparameters}\\
\midrule
\# &
$b$ & $d\beta$ & $\Sigma_{s}$ &
$\Sigma_{\parallel}$ & $\Sigma_{\perp}$ &
$a_{2, 0}$ & $a_{2, 1}$ &
$b_{0, 0}$ & $b_{0, 2}$ & $b_{2, 0}$ & $b_{2, 2}$
\\
\toprule\hline
1 & 2.103 $\pm$ 0.078 & 1.30 $\pm$ 0.39 & 1.7 $\pm$ 1.8 & 3.4 $\pm$ 1.4 & 2.98 $\pm$ 0.92 & -110 & 0.13 & -0.0004 & 0.0025 & -0.0037 & 0.0087 \\
2 & 2.05 $\pm$ 0.23 & 1.14 $\pm$ 0.49 & 1.9 $\pm$ 1.9 & 5.1 $\pm$ 1.7 & 2.79 $\pm$ 0.93 & -65 & 0.13 & 0.0010 & -0.0051 & -0.0011 & -0.0035\\
3 & 2.17 $\pm$ 0.22 & 0.87 $\pm$ 0.49 & 1.9 $\pm$ 1.9 & 4.7 $\pm$ 1.7 & 2.70 $\pm$ 0.91 & 1.3 & 0.43 & 0.00066 & -0.0028 & -0.00053 & -0.0044 \\
4 & 2.093 $\pm$ 0.080 & 1.30 $\pm$ 0.44 & 1.8 $\pm$ 1.8 & 4.5 $\pm$ 1.5 & 2.90 $\pm$ 0.91 & -90. & -0.093 & -0.0001 & 0.0014 & -0.0039 & 0.010 \\
5 & 2.18 $\pm$ 0.22 & 0.80 $\pm$ 0.50 & 1.9 $\pm$ 1.9 & 4.9 $\pm$ 1.7 & 2.70 $\pm$ 0.91 & 4.9 & 0.73 & 0.0005 & -0.0022 & -0.0018 & -0.0004 \\
6 & 2.05 $\pm$ 0.22 & 1.23 $\pm$ 0.49 & 1.8 $\pm$ 1.9 & 4.9 $\pm$ 1.7 & 2.79 $\pm$ 0.92 & -75 & 0.062 & 0.00066 & -0.0034 & -0.0017 & 0.0011 \\
7 & 2.01 $\pm$ 0.27 & 1.29 $\pm$ 0.33 & 1.9 $\pm$ 1.9 & 5.8 $\pm$ 1.7 & 2.85 $\pm$ 0.94 & -72 & -0.22 & 0.00059 & -0.0027 & -0.0004 & -0.0014 \\
8 & 2.03 $\pm$ 0.23 & 1.25 $\pm$ 0.56 & 1.9 $\pm$ 1.9 & 5.1 $\pm$ 1.6 & 2.78 $\pm$ 0.93 & -48 & -0.29 & 0.0004 & -0.0019 & 0.0001 & -0.0034 \\
\hline
9 & 2.049 $\pm$ 0.080 & 1.30 $\pm$ 0.44 & 1.7 $\pm$ 1.8 & 3.6 $\pm$ 1.4 & 2.69 $\pm$ 0.90 & -130 & 0.21 & -0.0011 & 0.0038 & -0.0056 & 0.018 \\
10 & 2.15 $\pm$ 0.22 & 0.76 $\pm$ 0.57 & 1.9 $\pm$ 1.9 & 4.6 $\pm$ 1.7 & 2.70 $\pm$ 0.92 & 37 & 0.50 & -0.0001 & -0.00098 & -0.0002 & -0.0078 \\
11 & 2.186 $\pm$ 0.082 & 0.70 $\pm$ 0.37 & 2.0 $\pm$ 2.0 & 4.9 $\pm$ 1.7 & 2.77 $\pm$ 0.92 & 40. & 0.68 & -0.0002 & -0.00078 & -0.0005 & -0.0050 \\
12 & 2.075 $\pm$ 0.080 & 1.30 $\pm$ 0.48 & 1.7 $\pm$ 1.8 & 4.7 $\pm$ 1.5 & 2.69 $\pm$ 0.91 & -90. & -0.11 & -0.0011 & 0.0049 & -0.0038 & 0.010 \\
13 & 2.08 $\pm$ 0.23 & 1.05 $\pm$ 0.44 & 1.9 $\pm$ 1.9 & 4.9 $\pm$ 1.6 & 2.78 $\pm$ 0.93 & -48 & 0.43 & -0.0004 & 0.00064 & -0.0020 & 0.0012 \\
14 & 2.10 $\pm$ 0.22 & 0.97 $\pm$ 0.39 & 1.9 $\pm$ 1.9 & 4.9 $\pm$ 1.7 & 2.73 $\pm$ 0.92 & -37 & 0.56 & -0.00057 & 0.0011 & -0.0024 & 0.0037 \\
15 & 2.03 $\pm$ 0.24 & 1.16 $\pm$ 0.42 & 1.9 $\pm$ 1.9 & 5.6 $\pm$ 1.7 & 2.85 $\pm$ 0.94 & -25 & -0.12 & -0.0004 & 0.0001 & 0.0003 & -0.0046 \\
16 & 2.00 $\pm$ 0.24 & 1.26 $\pm$ 0.45 & 1.9 $\pm$ 1.9 & 5.5 $\pm$ 1.6 & 2.79 $\pm$ 0.93 & -77 & -0.18 & -0.0004 & -0.00057 & -0.0013 & 0.0024 \\
\bottomrule
\end{tabular}
\end{adjustbox}
\caption{Values of the various analysis parameters for the blinding cosmologies in \autoref{tab:test-metrics} with matching case number \#. From left to right: linear galaxy bias ($b$), linear RSD nuisance parameter  accounting for the anisotropy of the signal amplitude ($d\beta = \beta / \beta_\mathrm{fid}$ with $\beta=f/b$), Fingers of God damping ($\Sigma_{s}$), line-of-sight BAO damping ($\Sigma_{\parallel}$), transverse BAO damping ($\Sigma_{\perp}$), with the remaining parameters ($a_{n, n}$ -- $b_{n, n}$) being the DESI baseline parametrization for broadband term, capturing any deviation from the linear theory~\cite{KP4s2-Chen}.
\label{tab:hyperparams}
}
\end{table}

\afterpage{\FloatBarrier}
\subsubsection{Validation Results for ShapeFit Analysis\label{sec:validation-shapefit}}

We repeat the framework in \cref{sec:validation-bao}, but now with ShapeFit. We carry out ShapeFit analysis for LRGs in the three redshift ranges of $0.4 < z < 0.6$, $0.6 < z < 0.8$, and $0.8 < z < 1.1$, with the first redshift shown in \cref{fig:shapefit_fits_LRG1}, and other two in the \cref{appendix:sup} (\cref{fig:shapefit_fits_LRG23}). In addition to \alphaiso, \alphaap, we now have two additional parameters \dm\ and \(df\), where \dm\ is the additional parameter due to ShapeFit while \(df\) comes from RSD. We see that the measured values of the parameters are close to the expected ones, although with some variation. Again, we note that as in the BAO case $\alpha_{\rm iso}$ and $\alpha_{\rm AP}$ are insensitive to the $f_\mathrm{NL}$-blinding choice. However, the ShapeFit parameter \dm\ shows a systematic offset between the $f_\mathrm{NL}=20$ and $f_\mathrm{NL}=-20$ choices. This is expected, given the degeneracy between the scale-dependent bias and the power spectrum slope on large scales captured by $f_\mathrm{NL}$ and \dm\, respectively \cite{Brieden:2021cfg}. Also, the growth rate $df$ exhibits a very mild $f_\mathrm{NL}$ dependence, which can be explained by the small correlation of $df$ with \dm\. Finally, the \chitwo\ of the fits is always within 1-2$\sigma$, indicating robustness.

We repeat the same for ELGs, arriving at the same results: while we see some variation in measured versus expected parameters across the 16 blinding catalogs, we see that \chitwo\ variation is always within 1-2$\sigma$; \cref{fig:shapefit_fits_ELG} shows the results.

By expanding our tests to different tracers and using both BAO and ShapeFit in Fourier and configuration spaces, we have substantially validated the robustness and applicability of our blinding scheme. We refer the reader to~ \cite{KP5s4-Lai} for a complete list of unblinding tests and optimizations for the full shape of the power spectrum.

\subsection{Concluding Remarks on Validation on Mocks}

The suite of tests conducted on mock datasets confirms the robustness and efficacy of the blinding scheme developed for DESI DR1. These validation efforts provide strong evidence that our blinding scheme can be reliably used for DR1 and beyond, both for DESI and other large-scale galaxy surveys, mitigating the potential risks of experimenter bias in cosmological parameter inference. It serves as a foundational step toward more complex, multi-probe cosmological analyses that may require intricate blinding techniques.

Specifically, we applied 16 possible blinding configurations to one \textsc{Abacus} simulation, resulting in 16 new blinded catalogs from which we performed our tests. Based on these tests, we are confident that our blinding scheme does not introduce any unwanted artifacts. While the blinding procedure does alter summary statistics and shift parameter inference, this is a core feature of the blinding process, not a flaw. Importantly, the results demonstrate that the blinding does not degrade the model's ability to fit the measurem evenents.

While our tests on mock datasets were comprehensive, it is crucial to note that real-world data might present complexities not accounted for in our mock datasets; we probe these in \cref{sec:validation-wdata}, where we carry out tests on blinded real data. Future work may include updating the blinding scheme and continually validating it against more complex and realistic mock datasets.

\section{VALIDATION WITH REAL DATA\label{sec:validation-wdata}}

So far, we have only discussed validation in the realm of mock datasets. In this section, however, we dive into the validation using real (blinded) data, focusing on only BAO given that constraints from RSD are not unblinded at the time of this writing, as explained in \cref{sec:when-to-unblind}. In the following, we detail the methodology we use to further validate the blinding scheme but first, we explain the data we work with.

After the validation tests passed on mocks (as detailed in \cref{sec:validation-wmocks}), we blinded DESI DR1 using the blinding pipeline described in \cref{sec:blinding-pipeline}, implementing blinding as detailed in \cref{sec:blinding-scheme}. The resulting catalog is the \textit{blinded catalog}, consisting of real DR1 data that is blinded. To validate the blinding scheme with real data, we carry out tests on this blinded catalog. As mentioned in \cref{sec:blinding-scheme}, we do not yet disclose the parameters used to blind the data.

\begin{figure}[hbt!]
    \vspace*{-0.5em}
    \centering
    \includegraphics[width=0.5\paperwidth]{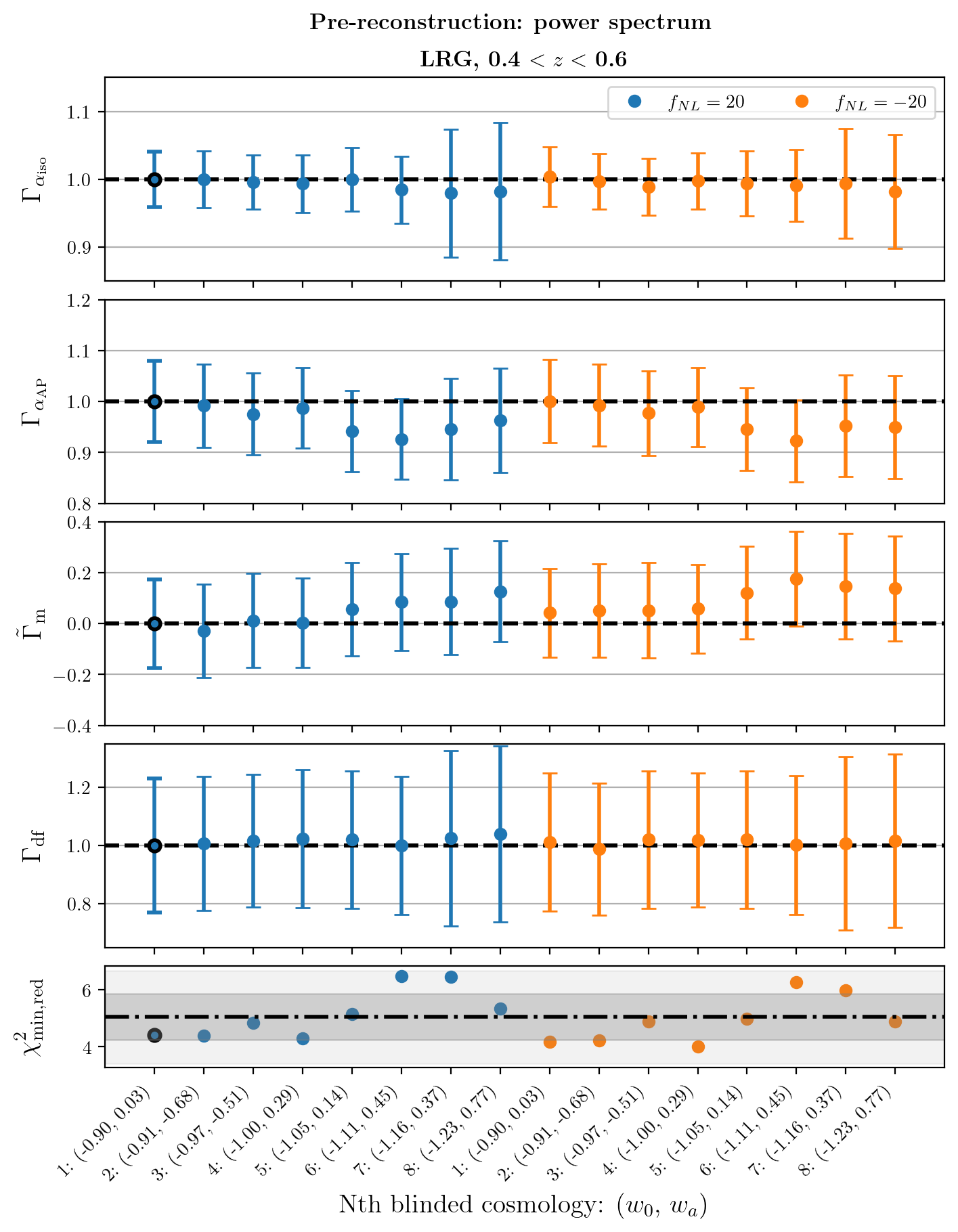}
    %
    \caption{ShapeFit fits using LRG samples for the first redshift bin (each column) from 16 different blinded mock catalogs. Various details here are the same as in \cref{fig:pre_recon_bao_fits_LRG}, except that $i$ = \alphaiso, \alphaap, $df$, \dm\ in $\Gamma_i$ while $\tilde\Gamma_i$ is the same as $\Gamma_i$ but comparing differences as opposed to ratios between measured and expected (since expected is 0). As for BAO fits, we see that the ratios (differences) are close to 1 (0) and the \chitwo\ variations are within 1-2$\sigma$, demonstrating the robustness of the fits.
    }
    \label{fig:shapefit_fits_LRG1}
\end{figure}

To probe the efficacy of our blinding scheme, we apply a second layer of blinding to the blinded catalog, using a \wowa\ pair that is at the edge of the allowed region in order to stress-test the scheme while arbitrarily choosing \fnl\ = 20. This second layer of blinding enables us to directly check that our blinding strategy does not introduce any unintended biases and that our analysis pipelines are robust, as we can apply and then remove this second layer at will.

For clarity, we refer to the original blinded catalog as the \textit{fiducial blinded catalog} while the doubly-blinded data is referred to as the \textit{double-blinded catalog}. Comparing analysis results from the fiducial blinded and double-blinded catalogs allows us to check the impact of blinding while preserving characteristics of real data that may not have been captured fully in the mocks.

\subsection{Validation Tests and Results}
\subsubsection{Varying Analysis Choices} 

As a first test, we run the baseline analysis pipeline, defined below, on the double-blinded catalogs, alongside a suite of alternative fitting choices:
\begin{itemize}[noitemsep,topsep=1pt]
    \item Baseline: we follow the baseline configuration adopted for the anisotropic BAO analysis (\alphaiso, \alphaap) defined in \cite{KP4s2-Chen}. This consist of a configuration space analysis of post-reconstruction catalogs, using the \recsym\ convention for reconstruction \cite{KP4s4-Paillas}, 
    spline-based broadband parameterization, and Gaussian priors on redshift-space distortion parameters ($\Sigma_s$, $\Sigma_\parallel$, $\Sigma_\perp$).
    \item 1D fit: as a test, we run the baseline analysis except that we perform an isotropic BAO analysis (\alphaiso).
    \item Pre-recon: as a test, we run the baseline analysis on pre-reconstruction data as opposed to the post-reconstruction data.
    \item Power spectrum: as a test, we run the post-reconstruction analysis but in Fourier space.
    \item Polynomial broadband: as a test, instead of using the spline parameterization for the broadband, we employ a polynomial.
    \item Flat priors: as a test, instead of using informative priors on redshift-space distortion parameters, we use flat priors.
\end{itemize}\

\begin{figure}[hbt!]
    \centering
    \includegraphics[width=\textwidth]{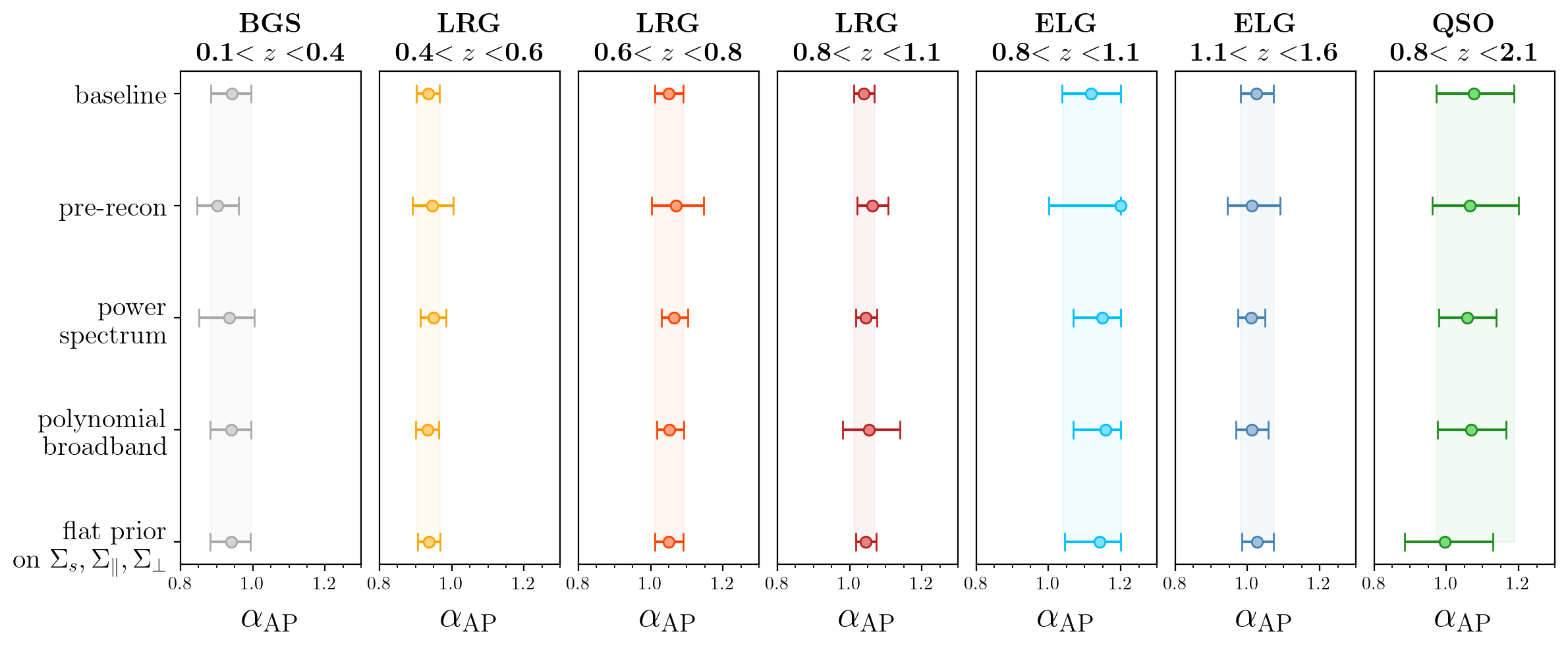}
    \includegraphics[width=\textwidth, clip=True, trim={0 0 0 35}]{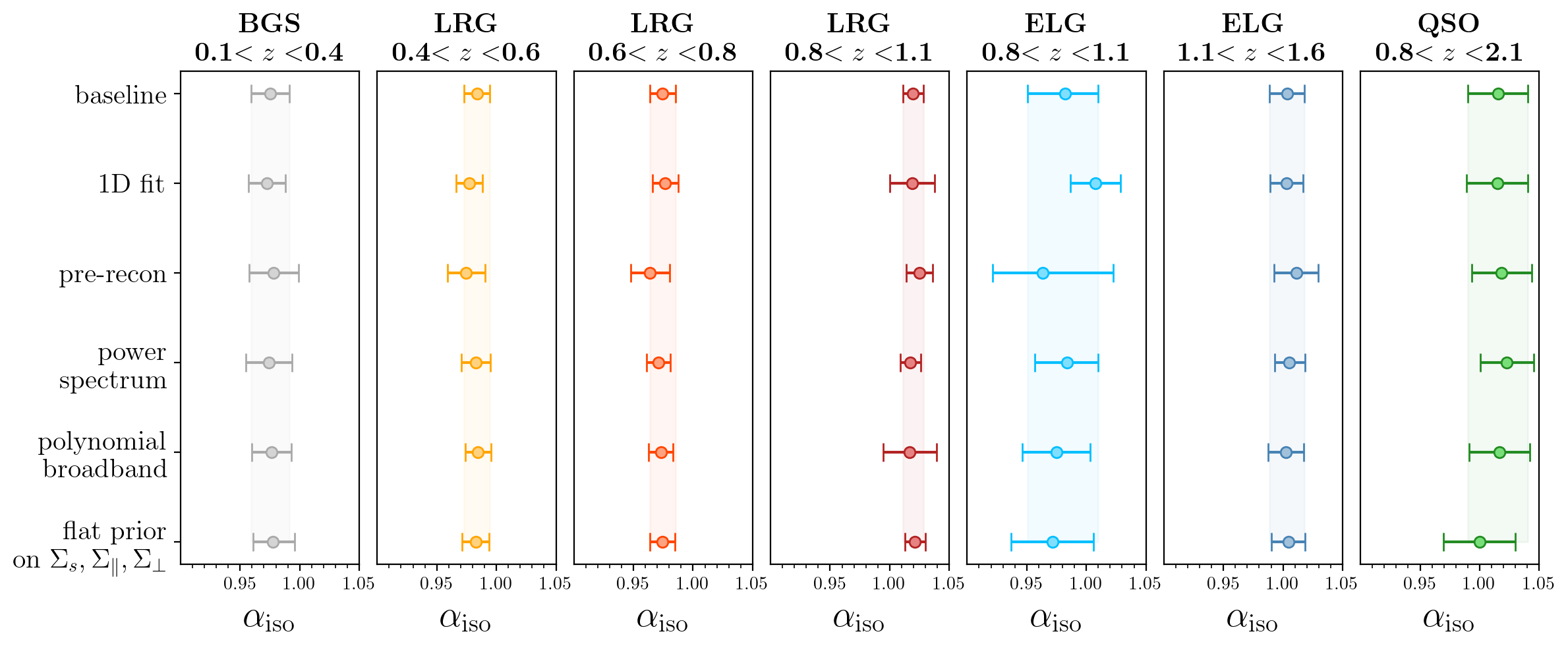}
    \caption{
    Comparison of BAO fitting parameter (\alphaap, \alphaiso) measurements from the double-blinded catalog, using the baseline analysis model and various fitting choices across the four galaxy tracers (BGS, LRG, ELG, and QSO tracers) for different redshift ranges. The top panel shows the whisker plots for \alphaap, while the bottom panel shows those for \alphaiso. We see that the various fitting choices are consistent with the baseline.
    }
    \label{fig:bao_whiskers_combined}
\end{figure}

\noindent Overall, our comparative analysis across these configurations yields consistency with the baseline results, notwithstanding minor variations discussed below. \cref{fig:bao_whiskers_combined} demonstrates these effects, showcasing the parameter constraints for each galaxy tracer under different analysis settings.

The 1D fit closely mirrors baseline findings, with deviations within \(1\sigma\) for most tracers. We see that reconstruction markedly enhances precision across most tracers, illustrating the utility of the technique in sharpening parameter estimates. The outlier of this trend is the tracer QSO which tends to be shot-noise limited and therefore does not benefit strongly from sharpening techniques \cite{eBOSS:2020gbb}.

Analysis in Fourier space yields results compatible with those from configuration space, with deviations remaining within \(1\sigma\) for most tracers. This consistency extends to the comparison between spline-based and polynomial broadband parameterizations; the former, a new spline-basis parameterization proposed by \cite{KP4s2-Chen} while the latter used in BOSS \cite{Alam2017} and eBOSS \cite{eboss2020}.

Lastly, employing flat priors for BAO damping parameters is largely consistent with the baseline, using Gaussian priors, affirming the robustness of our analytical approach to prior effects. 

We have demonstrated that the double-blind tests are robust against variations in the fitting procedures. Similarly, the \textit{single-blind} analysis has been thoroughly examined in the companion DESI paper (see, for example, Fig. 15 in~\cite{KP4s4-Paillas}), to which we refer without duplicating the results here to avoid redundancy. Together, these two tests—single-blind and double-blind—confirm that the choice of fitting procedure is not influenced by the blinding technique.

\subsubsection{Comparing Posteriors} 

Once we establish that our baseline analysis is robust against changes in various choices, we proceed with another test: a comparison of parameter estimates from the fiducial blinded catalog vs. the double-blinded one. Given that we know the blinding parameter used for blinding the double-blinded catalog, we can generate \textit{shifted fiducial} estimates, whereby we shift the inferred values from the fiducial catalog. This mimics a posterior-level blinding, achieved by simply multiplying the fiducial posterior ($X$) by the expected shift ($a_{\mathrm{shift}}$) of the parameters due to blinding, leading to a fiducial shifted posterior, $Y_{\mathrm{shifted}}= a_{\mathrm{shift}}\ X$; we calculate $a_{\mathrm{shift}}$ using the cosmology used for (second) blinding and its relation with the BAO fitting parameters (as presented in \cref{eq:aperp_apar}).

\begin{figure}[hbt!]
    \vspace*{-2em}
    \centering
    \begin{minipage}{0.23\paperwidth}
        \includegraphics[width=\textwidth]{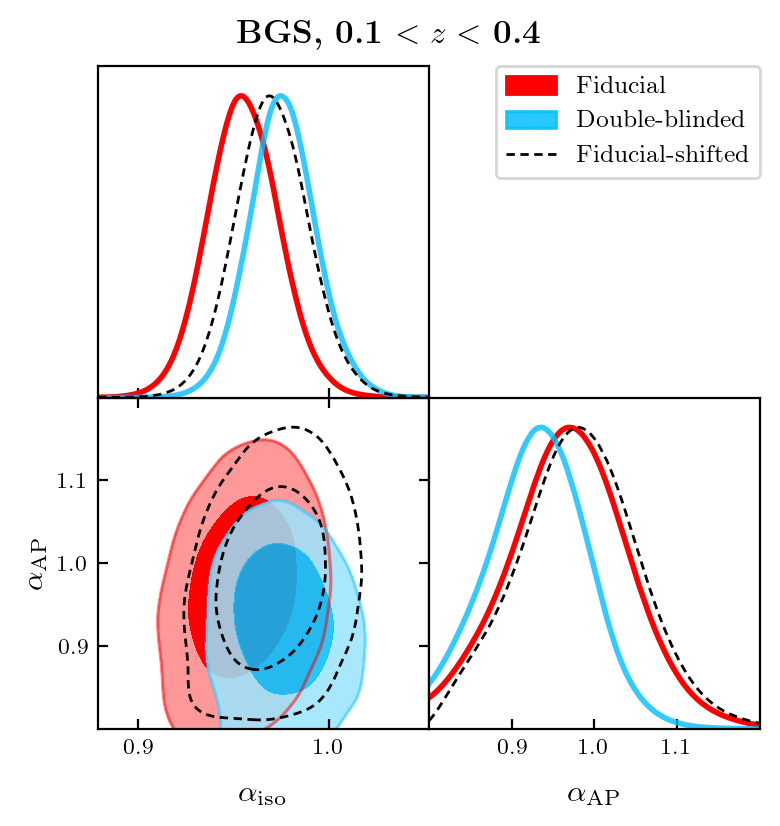}
    \end{minipage}
    \vspace*{-0.25em}
    
    \begin{minipage}{0.23\paperwidth}
        \includegraphics[width=\textwidth]{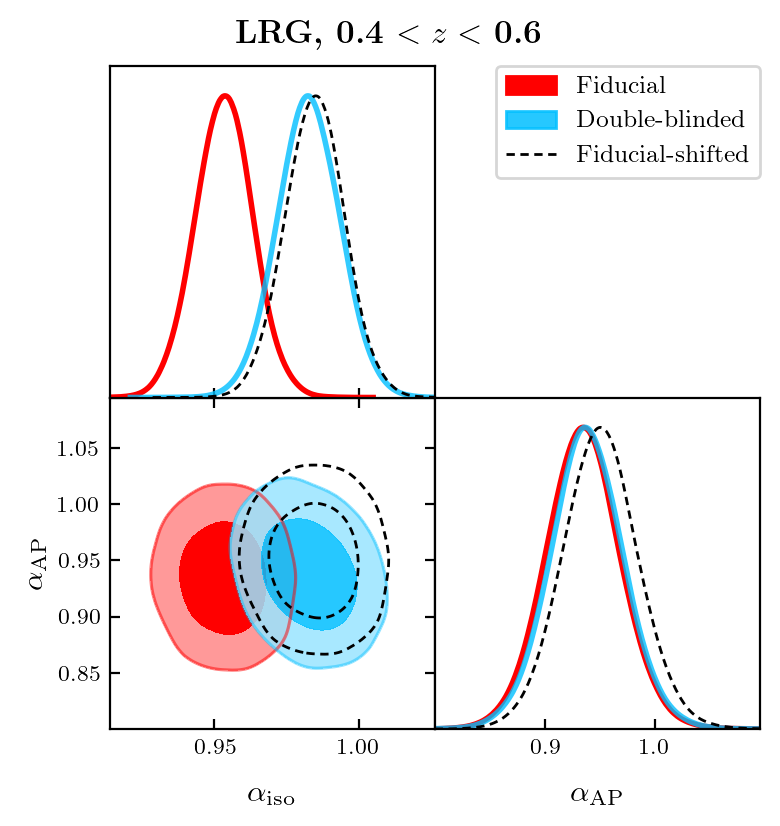}
    \end{minipage}
    \hspace*{-0.5em}
    \begin{minipage}{0.23\paperwidth}
        \includegraphics[width=\textwidth]{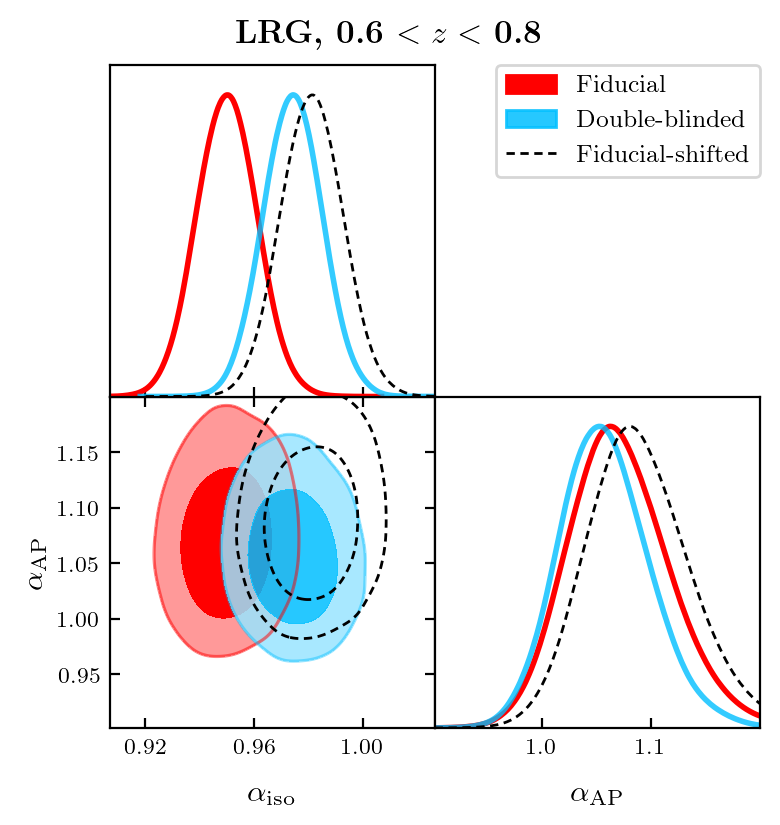}
    \end{minipage}
    \hspace*{-0.5em}
    \begin{minipage}{0.23\paperwidth}
        \includegraphics[width=\textwidth]{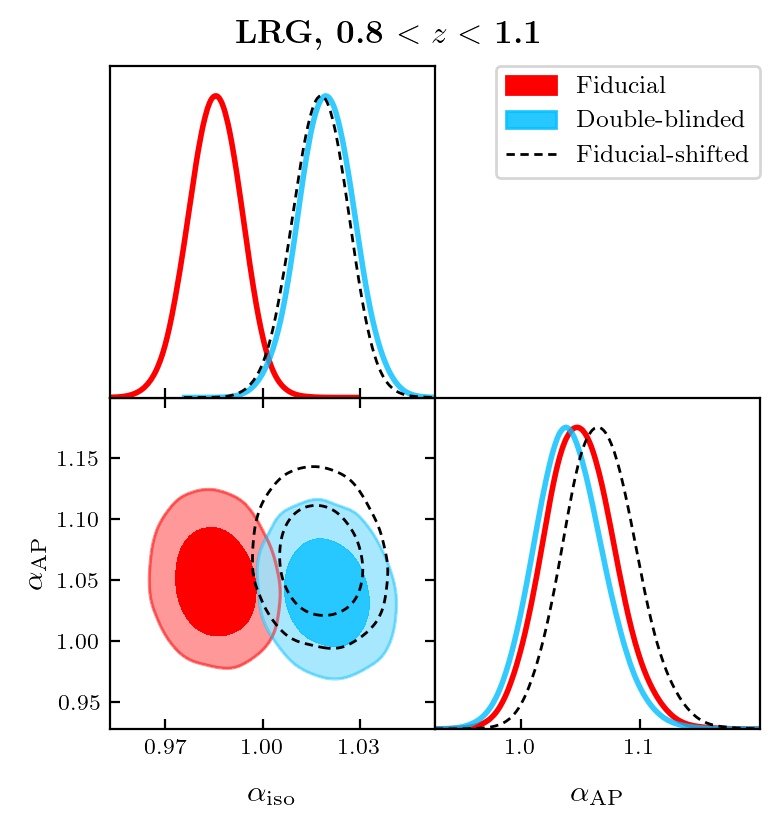}
    \end{minipage}
    \vspace*{-0.25em}
    
    \begin{minipage}{0.23\paperwidth}
        \includegraphics[width=\textwidth]{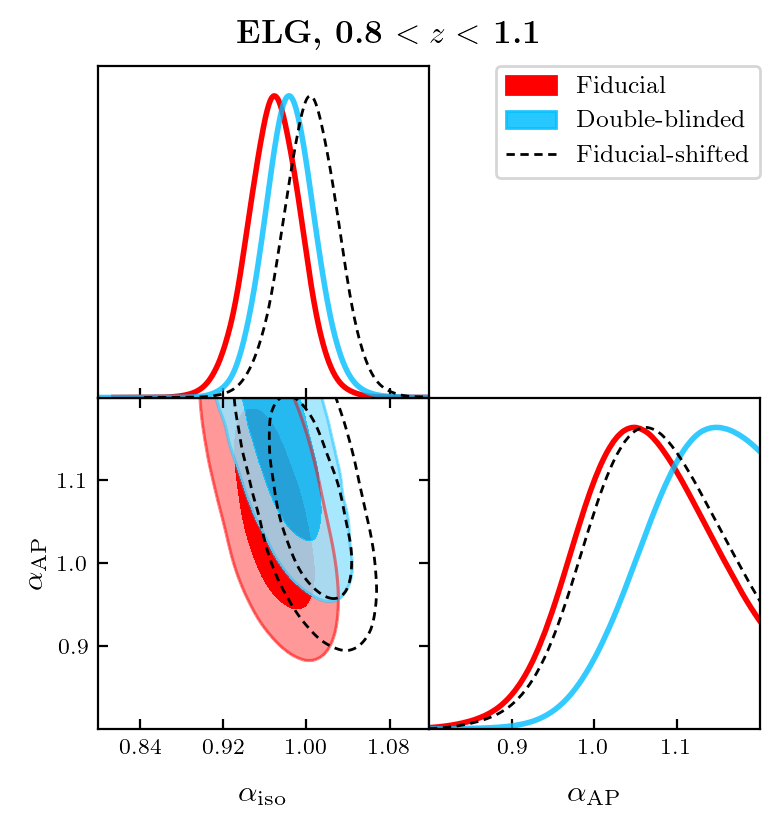}
    \end{minipage}
    \hspace*{-0.5em}
    \begin{minipage}{0.23\paperwidth}
        \includegraphics[width=\textwidth]{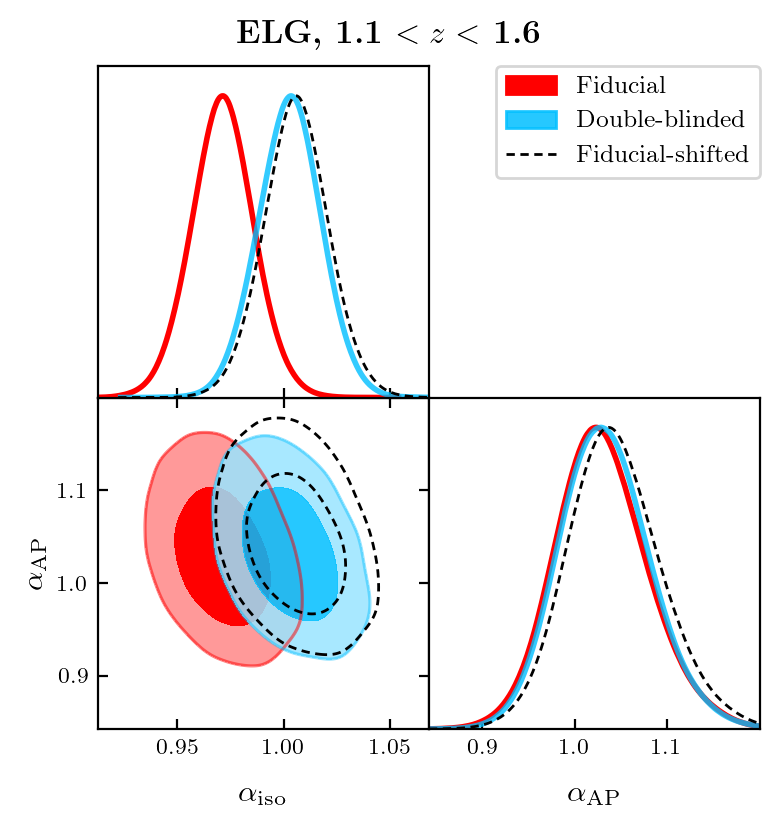}
    \end{minipage}
    \vspace*{-0.25em}
    
    \begin{minipage}{0.23\paperwidth}
         \includegraphics[width=\textwidth]{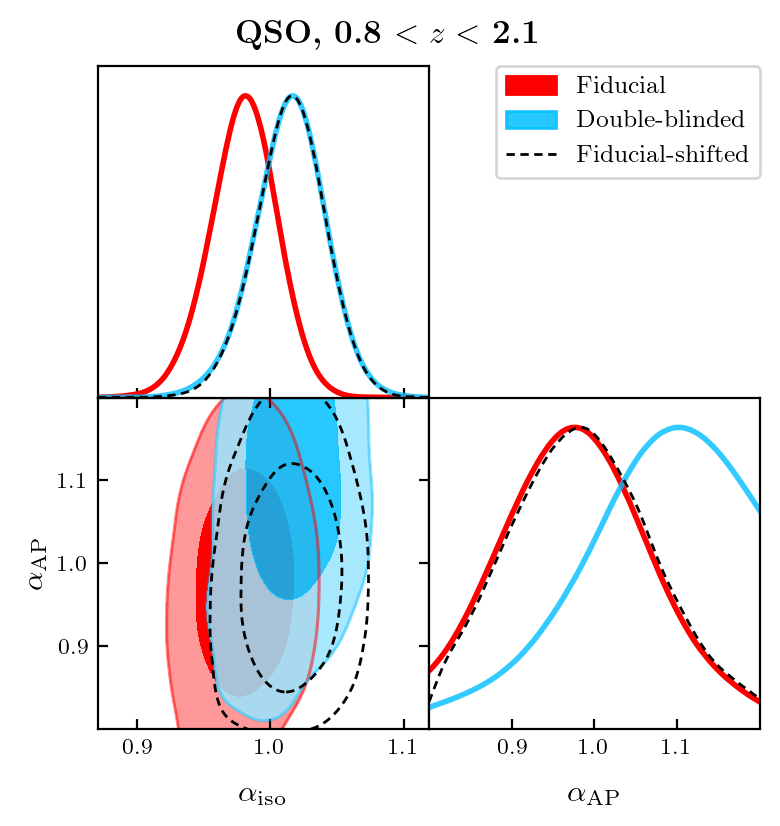}
    \end{minipage}
    \vspace*{-0.25em}
    \caption{
    \footnotesize{
    Post-reconstruction anisotropic BAO fits different tracer samples for fiducial blinded and double-blinded catalogs. Each subplot presents the 68\% and 95\% confidence level marginalized posteriors for the isotropic (\alphaiso) and anisotropic (\alphaap) scaling parameters. The red contours denote the posteriors based on the fiducial blinded catalog, while the blue contours represent those from the double-blinded catalog. The dashed contours are produced by applying the same offset to the fiducial posteriors that was used in the blinding process - a check to understand how the data-blinding prescription affects the posteriors. For the cases where the dashed-black posteriors overlap with blue ones, the blinding scheme essentially has the same effect as it would if we have blinded at the posterior level. However, we see that in most cases, the two posteriors are not exactly the same, meaning that our blinding is more complicated than a posterior-level blinding (as is indeed the case). It is also reassuring that the double-blinded posteriors do not match the fiducial ones, indicating that our blinding scheme is effective at blinding for the parameters of interest.
    }
    \label{fig:allfour_grouped}
    }
\end{figure}

\cref{fig:allfour_grouped} shows the posteriors for the four tracers, across various redshift bins. We see that the fiducial-shifted curves (black lines) do not perfectly match with those from double-blinded (blue), highlighting the distinction between a catalog-level blinding vs. a posterior-level one, i.e., the catalog-level blinding acts on the recovered parameters broadly as expected, but is more complex and thus not identical to simply shifting the posteriors.
This figure also demonstrates that the catalog-level blinding works in that the inferred parameters from the double-blinded catalog (blue) do not match those from the fiducial one (red).

\afterpage{\FloatBarrier}

\begin{figure}[!hbt]
    \centering
    \includegraphics[width=\textwidth]{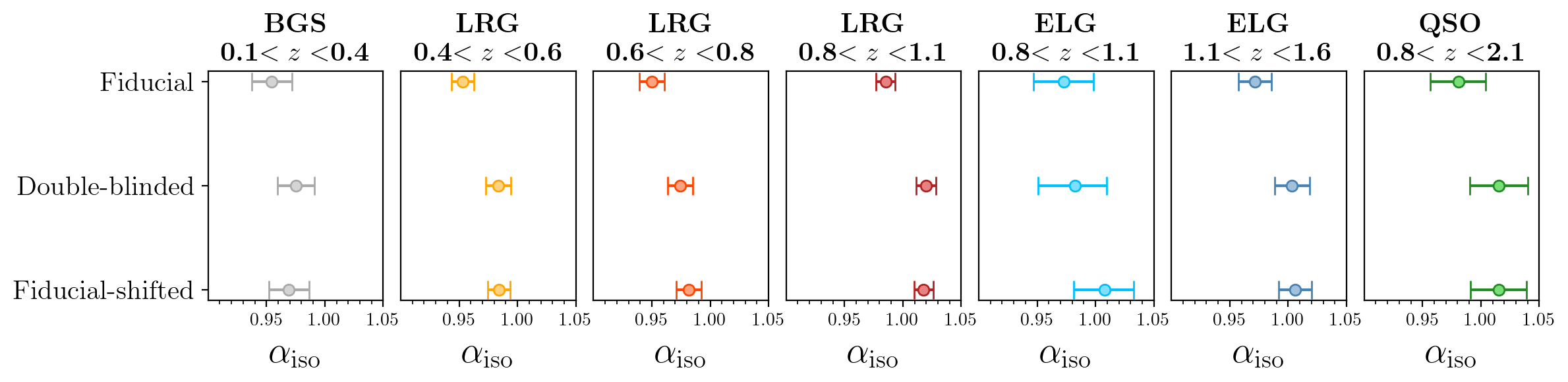}
    \hspace*{-1em}
    \includegraphics[width=0.98
\textwidth, clip=True, trim={0 0 0 35}]{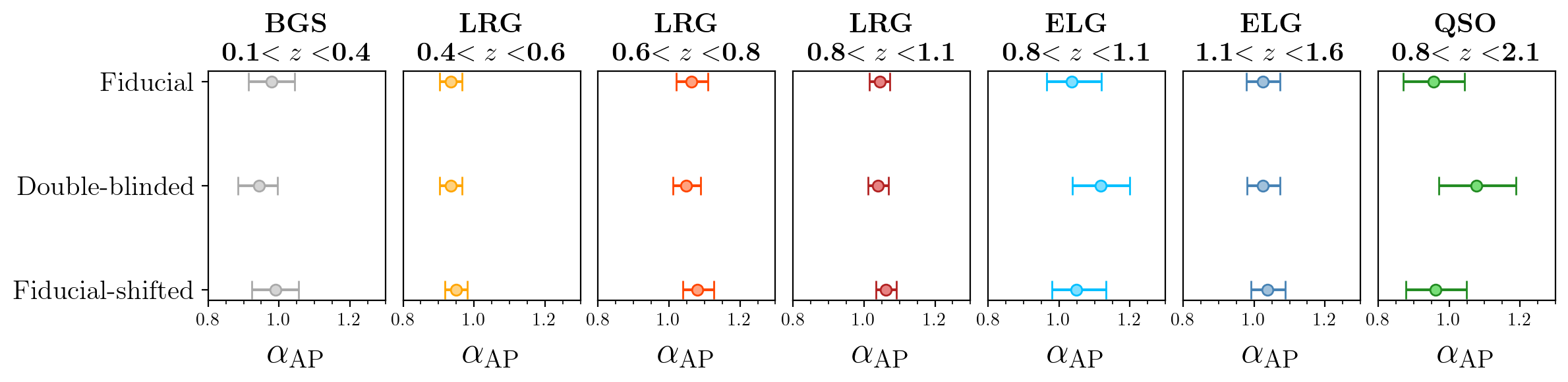}
    \caption{
    Comparison of BAO fitting parameter (\alphaiso, \alphaap) measurements from the fiducial blinded and double-blinded catalogs, alongside fiducial-shifted which mimics posterior-level blinding. As in \cref{fig:bao_whiskers_combined}, we consider all four tracers and plot the two parameters in the two rows. We see that fiducial-shifted estimates do not always follow those from the double-blinded catalogs, reinforcing the distinct impacts of a catalog-level blinding as opposed to a posterior-level one, while demonstrating that our blinding scheme effectively masks the parameters of interest.
    }
    \label{fig:bao_whiskers_iso_ap}
\end{figure}

As a summary, \cref{fig:bao_whiskers_iso_ap} shows the whisker plots for the best-fit values for the two BAO parameters. We see that the fiducial-shifted estimates do not always follow those from the double-blinded catalogs, reinforcing the distinct impacts of a catalog-level blinding as opposed to a posterior-level one; here, the difference between the fiducial cosmology used for the (double) blinding vs. the true (blinded) cosmology underlying the fiducial blinded catalog plays a role. Nevertheless, we find that our blinding scheme is effective at blinding the underlying cosmology, as it should.


\subsection{Concluding Remarks on Validation on Real Data}

The tests presented above provide a robust framework for evaluating the impact of blinding on our ability to extract cosmological information. Notably, the consistent results across different fitting methodologies affirm the resilience of our blinding scheme against analytical variations. This exercise also reinforces our confidence in the blinding process and the reliability of subsequent cosmological interpretations. As a result, we set the stage for applying similar methodologies to future data releases from DESI and other large-scale structure surveys. 

\section{CONCLUSIONS\label{sec:conclusion}}

In modern observational cosmology, it is crucial to employ blinding methods to safeguard the results against experimenter bias. In this work, we presented and validated a comprehensive blinding scheme for the DESI DR1 analysis, aimed at mitigating experimenter bias and ensuring the integrity of our cosmological parameter estimation. Through a series of rigorous tests on both mock datasets and actual blinded data, we demonstrated the effectiveness of our blinding strategy in preserving the statistical properties of the data while preventing inadvertent unblinding or bias introduction by the researchers.

Our validation process spanned a variety of tracers and included two compression techniques, i.e. BAO and ShapeFit. In particular, we performed a BAO analysis on all dark-time DESI tracers (see \cref{sec:validation-bao} for the first redshift bin LRG, i.e., [0.4, 0.6] and \cref{appendix:sup} for the LRG ([0.6, 0.8], [0.8, 1.1]), ELG and QSO samples) for 16 different blinding configurations of the {\tt abacus-1} mock. Furthermore, we performed the BAO analysis both in configuration and Fourier space, either pre- or post-reconstruction. In all cases, we found exquisite agreement (better than $1-\sigma$) of the BAO scaling parameters with the expectation. We applied the ShapeFit methodology on the same set of 16 blinded LRG (and ELG) mocks in \cref{sec:validation-shapefit} (and \cref{appendix:sup}) and found all ShapeFit parameters to agree with the expectation within $1-\sigma$. Finally, in \cref{sec:validation-wdata}, we performed a series of tests of the BAO pipeline on blinded and double-blinded data for all DESI tracers, finding our baseline choice to be robust against choosing variations such as pre-recon, power spectrum, and 1D fits or adopting a different number of broadband terms or different priors on the BAO damping parameters; we were able to largely recover the fiducial blinded posterior, by shifting the double-blinded posterior. To our knowledge, this is the first time a blinding scheme was explicitly tested with such a doubled layer.

The development and successful validation of this blinding scheme marked a significant step in ensuring that our analysis of DESI DR1 is free from experimenter bias. Furthermore, the methodologies and insights gained from this work offer valuable lessons for future DESI data releases as well as other large-scale structure surveys.

We note that there are analyses that are beyond the scope of current work, including especially the full-shape modeling of the power spectrum; \cite{DESI2024.V.KP5} will address this, including blinding validation with full-shape modeling. As we look ahead, the validated blinding scheme will serve as a crucial component of our analysis toolkit, enhancing the credibility of our findings and strengthening the foundation of cosmological research. Future studies will benefit from this foundational work, providing a stepping stone for the application of rigorous scientific methodologies in the exploration of our Universe.

\section{DATA AVAILABILITY}

The data used in this analysis will be made public as part of DESI Data Release 1. Details can be found in \url{https://data.desi.lbl.gov/doc/releases/}.

\section*{Acknowledgements}

UA acknowledges support by the Leinweber Center for Theoretical Physics at the University of Michigan Postdoctoral Research Fellowship and DOE grant DE-FG02-95ER40899.
HA acknowledges support by the Leinweber Postdoctoral Research Fellowship and DOE grant DE-SC009193.
SB acknowledges funding from the European Research Council (ERC) under the European Union’s Horizon 2020 research and innovation program (grant agreement 853291).

This material is based upon work supported by the U.S. Department of Energy (DOE), Office of Science, Office of High-Energy Physics, under Contract No. DE–AC02–05CH11231, and by the National Energy Research Scientific Computing Center, a DOE Office of Science User Facility under the same contract. Additional support for DESI was provided by the U.S. National Science Foundation (NSF), Division of Astronomical Sciences under Contract No. AST-0950945 to the NSF’s National Optical-Infrared Astronomy Research Laboratory; the Science and Technology Facilities Council of the United Kingdom; the Gordon and Betty Moore Foundation; the Heising-Simons Foundation; the French Alternative Energies and Atomic Energy Commission (CEA); the National Council of Humanities, Science and Technology of Mexico (CONAHCYT); the Ministry of Science and Innovation of Spain (MICINN), and by the DESI Member Institutions: \url{https://www.desi.lbl.gov/collaborating-institutions}. Any opinions, findings, and conclusions or recommendations expressed in this material are those of the author(s) and do not necessarily reflect the views of the U. S. National Science Foundation, the U. S. Department of Energy, or any of the listed funding agencies.

The authors are honored to be permitted to conduct scientific research on Iolkam Du’ag (Kitt Peak), a mountain with particular significance to the Tohono O’odham Nation.


\bibliographystyle{JHEP}
\bibliography{references, DESI2024_bib}

\newpage
\appendix
\counterwithin{figure}{section}
\section{Redshift Range for Blinding Parameter Space\label{appendix:zrange_blinding}}

\begin{figure}[hbt!]
    \centering
    \includegraphics[width=0.75\textwidth]{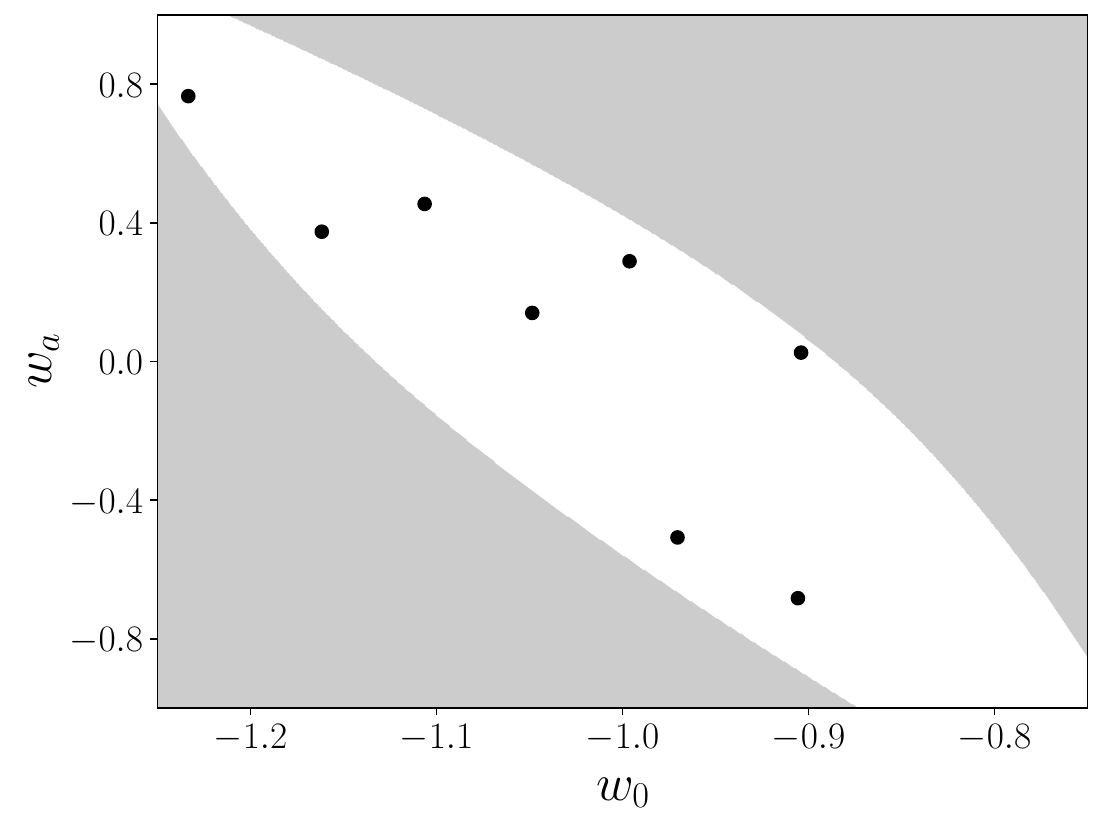}
    \caption{Same as \cref{fig:w0-wa_plane_zeffcombined_8realisations_4paper_0.4-2.1} but using the redshift range $0.1<z<2.1$ to define the allowed parameter space. We see no differences.}
    \label{fig:w0-wa_plane_zeffcombined_8realisations_4paper_0.1-2.1}
\end{figure}

At the very early stages of the analysis, the redshift range chosen to define the region allowed for $w_0$, $w_a$ (the white region in \cref{fig:w0-wa_plane_zeffcombined_8realisations_4paper_0.4-2.1}) was thought to encompass all dark-time tracers. Given that LRGs span redshifts 0.4 to 1.1, ELGs 0.8 to 1.6, and QSOs 0.8 to 2.1, we chose $0.4<z<2.1$ as our default redshift range. At a later stage, we also included the BGS in the blinding pipeline, which spans redshift 0.1 to 0.4. However, we decided not to modify the redshift range used to create \cref{fig:w0-wa_plane_zeffcombined_8realisations_4paper_0.4-2.1} because we found that including the redshift range $0.1<z<0.4$ had very little impact. In \cref{fig:w0-wa_plane_zeffcombined_8realisations_4paper_0.1-2.1} we show the same plot as in \cref{fig:w0-wa_plane_zeffcombined_8realisations_4paper_0.4-2.1} but using $0.1<z<2.1$ instead of $0.4<z<2.1$. We find that the two figures are identical.

\newpage
\section{Supplemental Plots\label{appendix:sup}}
\begin{figure}[hbt!]
    \centering
    \includegraphics[width=0.495\textwidth, clip=True, trim={0 25 0 0}]{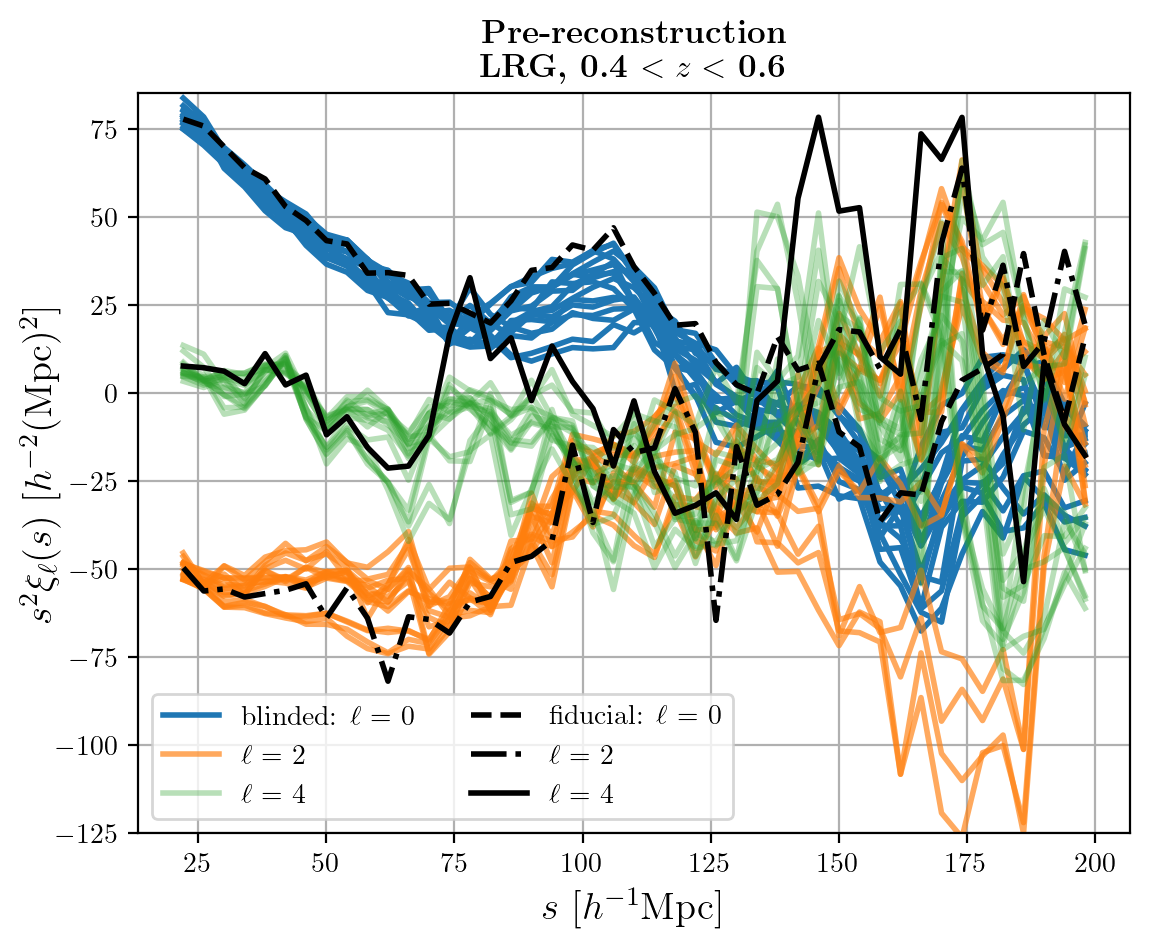}
    \hfill
    \includegraphics[width=0.495\textwidth, clip=True, trim={0 25 0 0}]{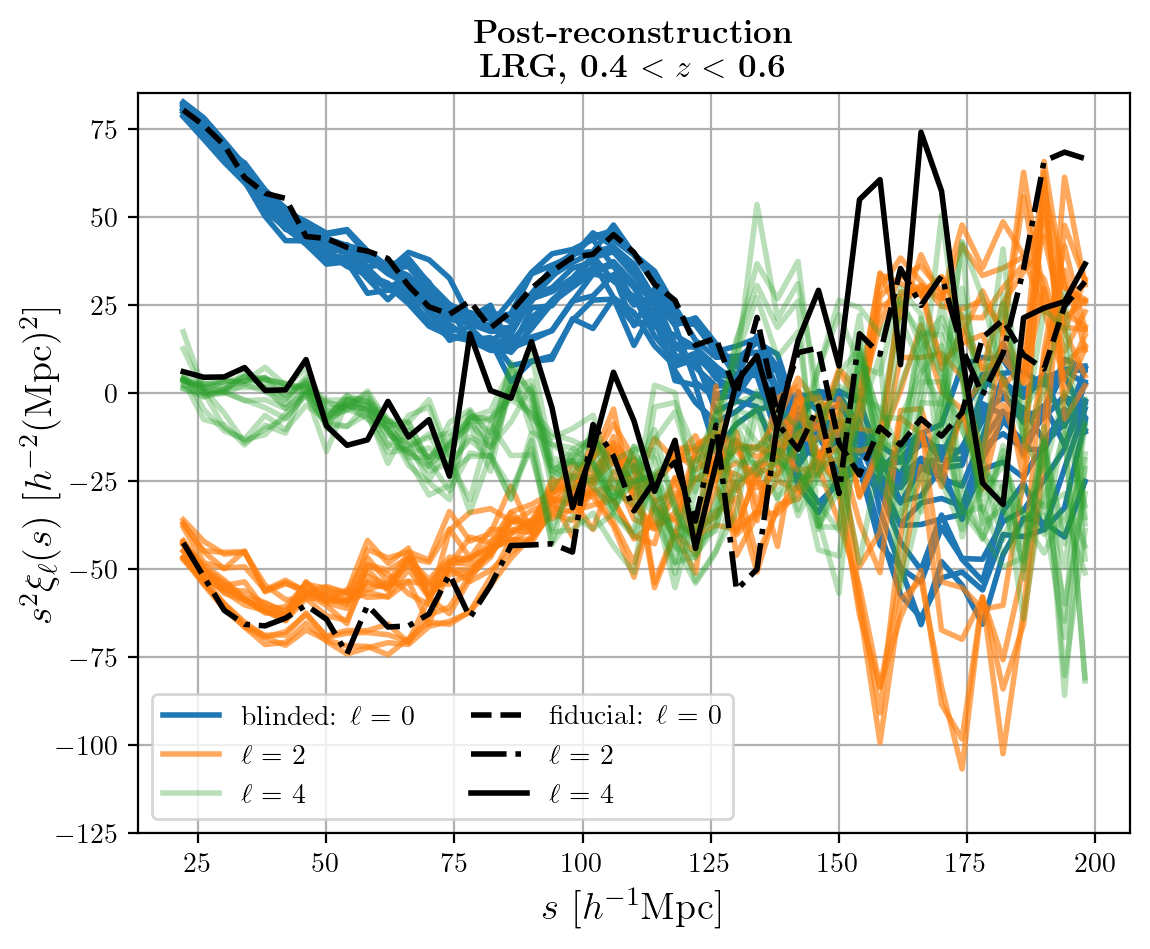}
    \hfill
    \includegraphics[width=0.495\textwidth, clip=True, trim={0 25 0 16}]{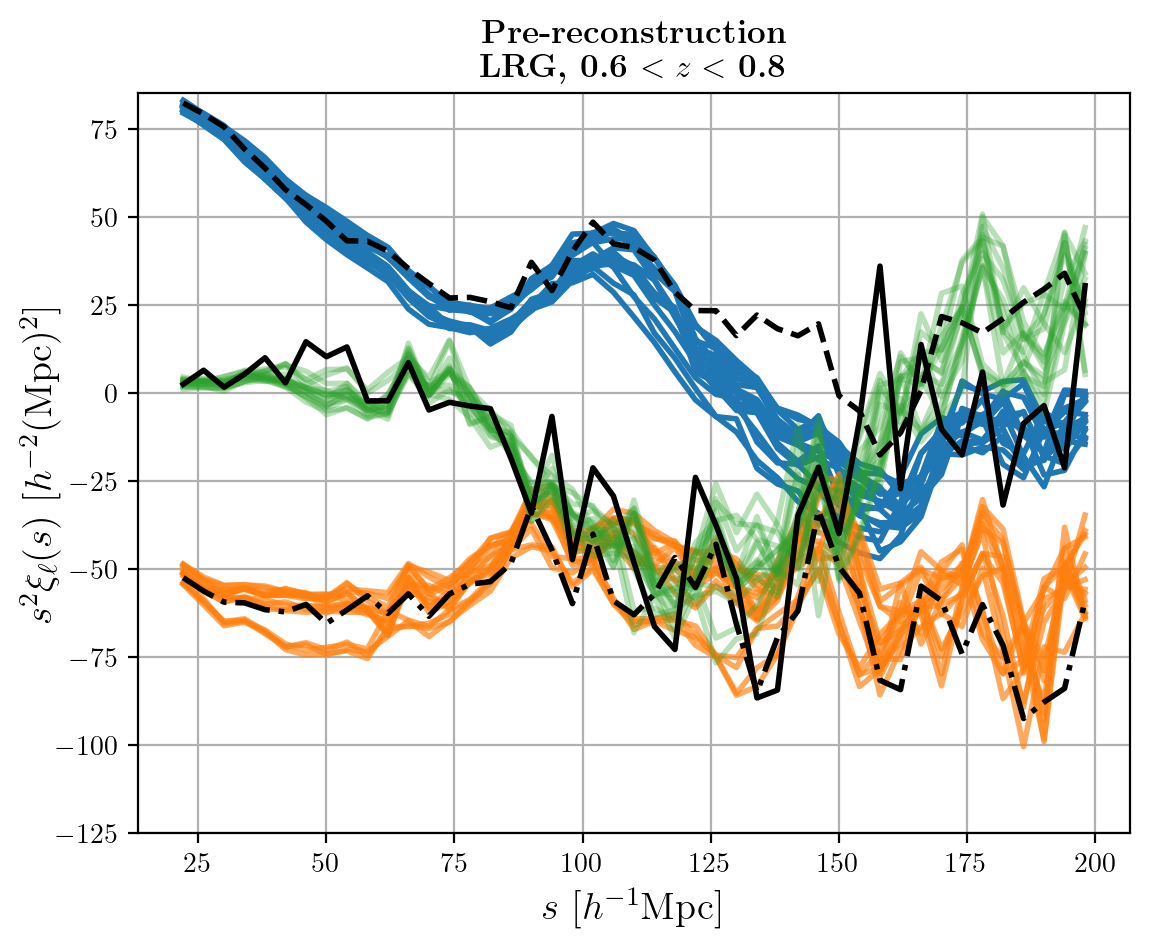}
    \hfill
    \includegraphics[width=0.495\textwidth, clip=True, trim={0 25 0 16}]{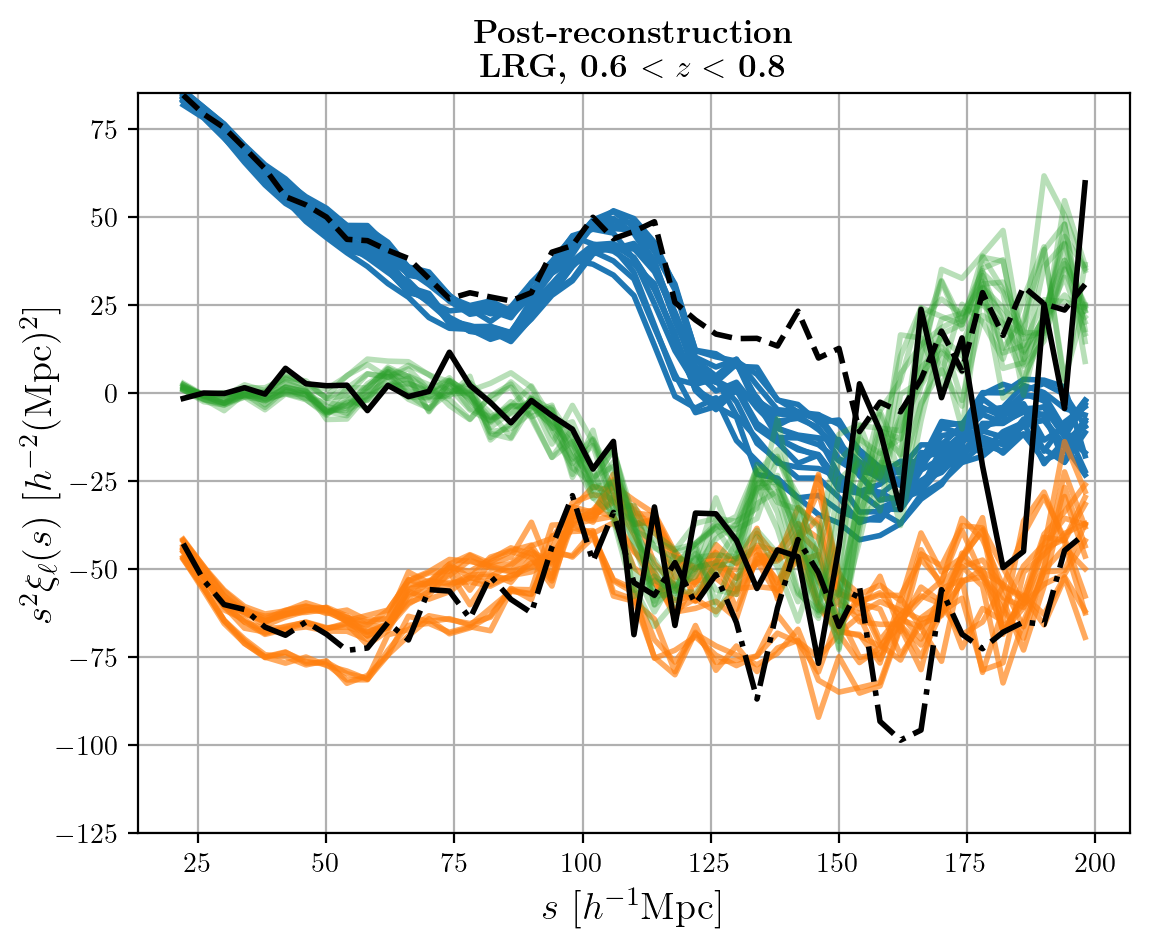}
    \hfill
    \includegraphics[width=0.495\textwidth, clip=True, trim={0 0 0 16}]{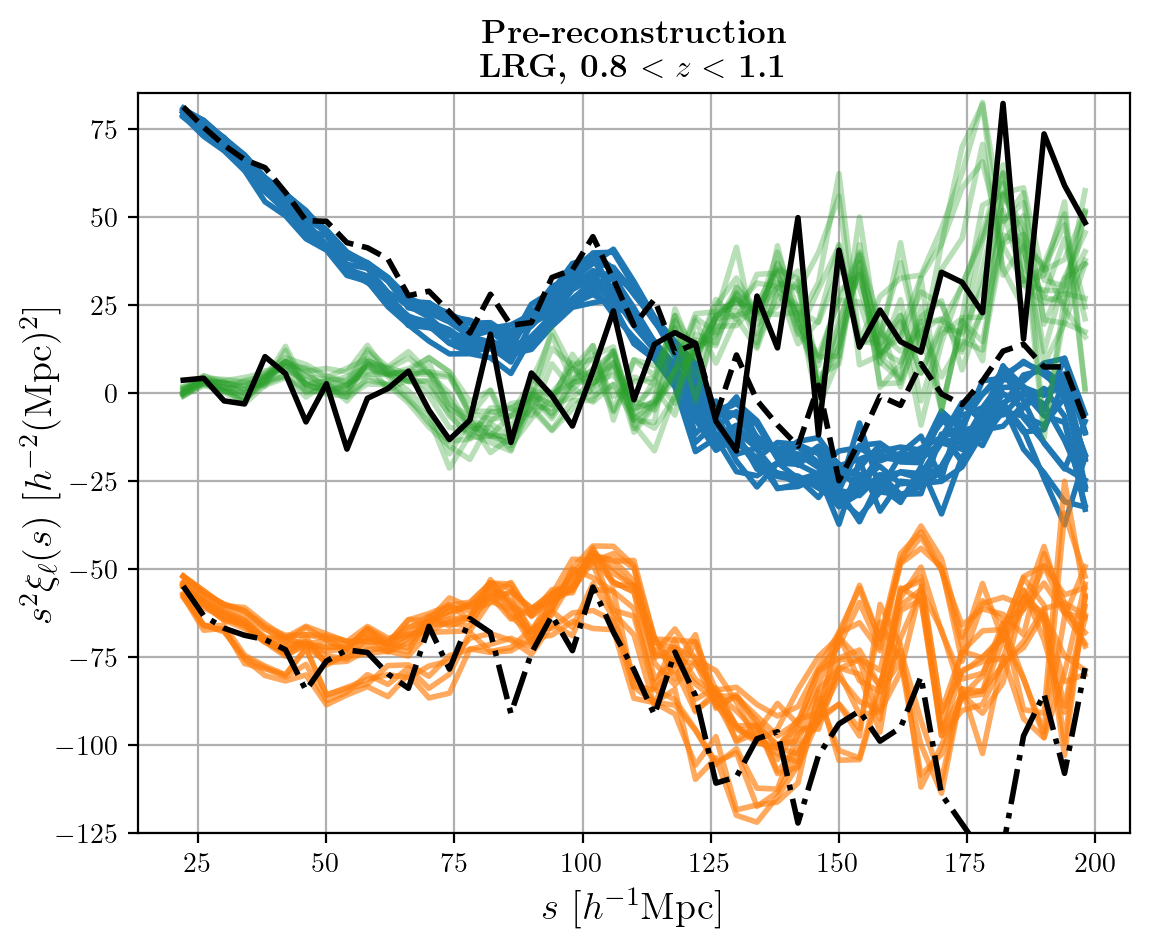}
    \hfill
    \includegraphics[width=0.495\textwidth, clip=True, trim={0 0 0 16}]{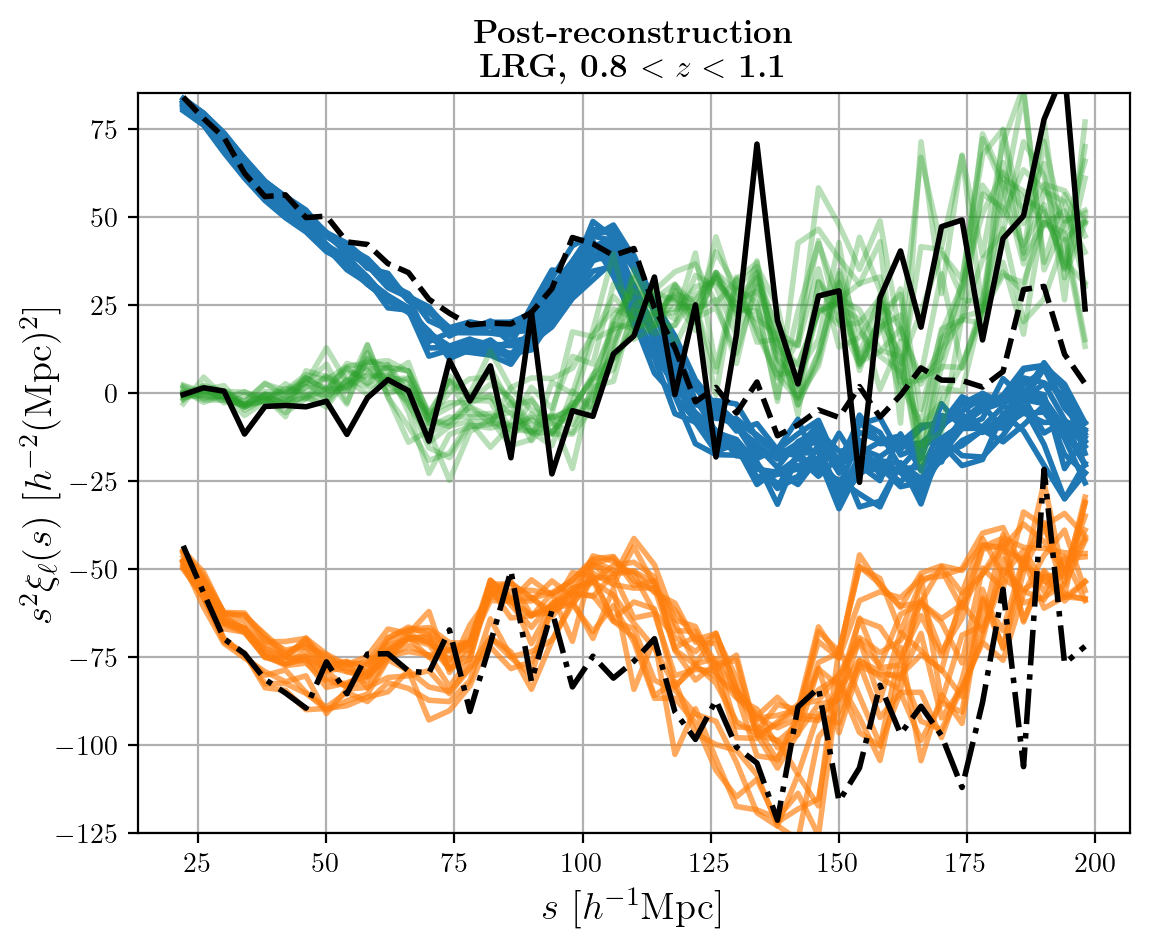}
    \caption{Correlation function multipoles for LRG samples from an \abacus\ mock catalog, presented for three redshift bins; pre-reconstruction multipoles (left) and post-reconstruction ones (right). The black lines show the fiducial simulation which serves as the baseline for generating 16 different blinded cosmological configurations; these are depicted by the colored lines, with the three colors showing the three multipoles (blue for monopole ($\ell$ = 0); orange for quadrupole ($\ell$ = 2); green for hexadecapole ($\ell$ = 4)); all panels use the same legend. These configurations are used throughout the paper in order to assess the impacts of different blinding cosmological configurations on our inference.
    }
    
    \label{fig:correlation_multipoles_combined}
\end{figure}
\begin{figure}[hbt!]
    \centering
    \includegraphics[width=0.495\textwidth, clip=True, trim={0 25 0 0}]{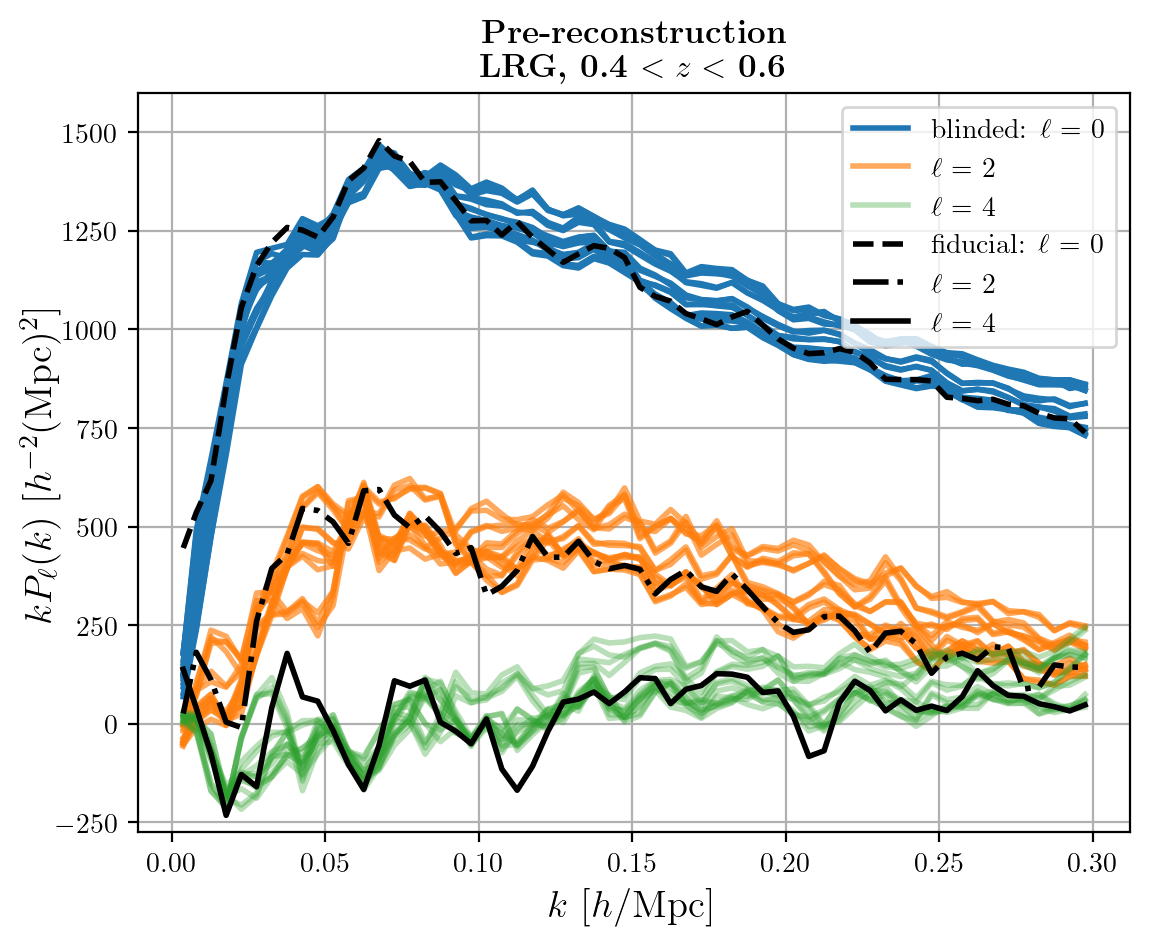}
    \hfill
    \includegraphics[width=0.495\textwidth, clip=True, trim={0 25 0 0}]{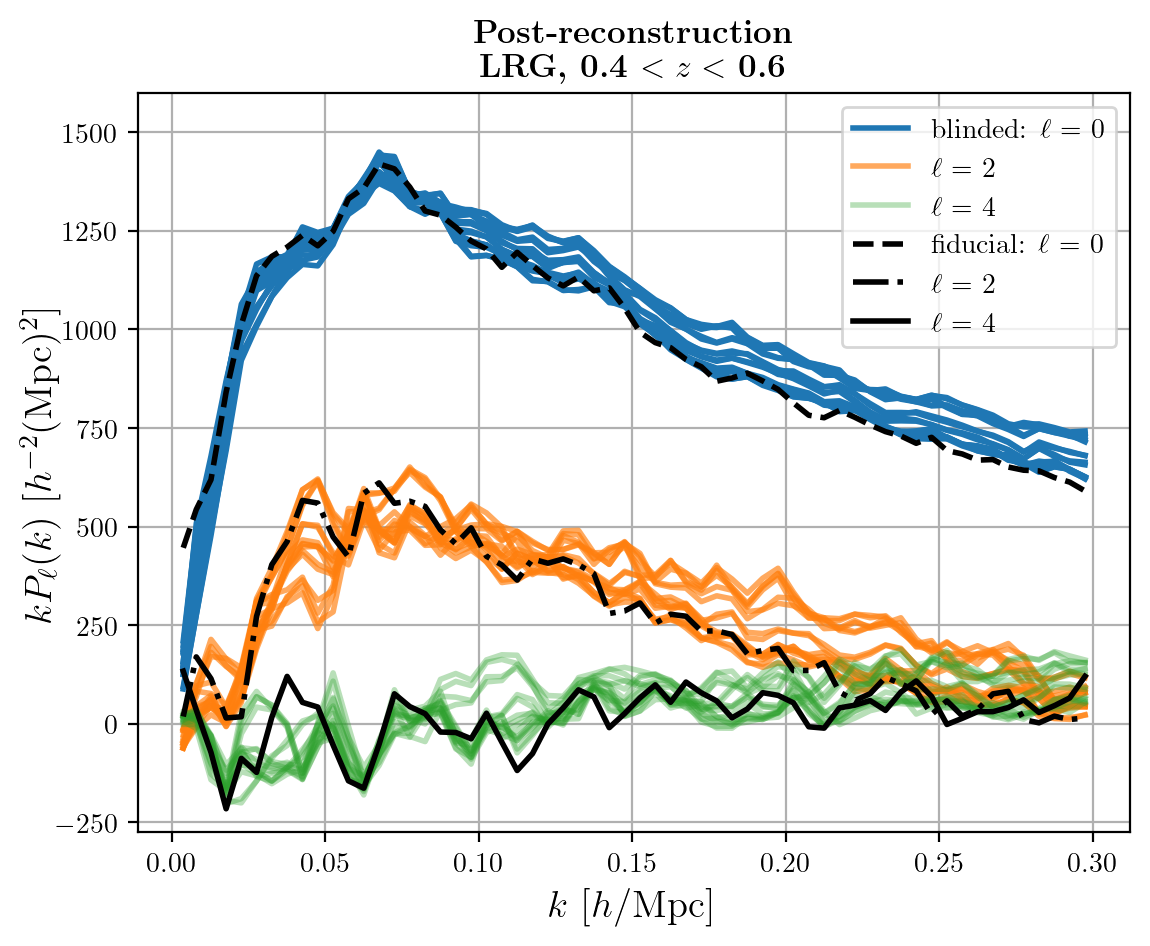}
    \hfill
    \includegraphics[width=0.495\textwidth, clip=True, trim={0 25 0 16}]{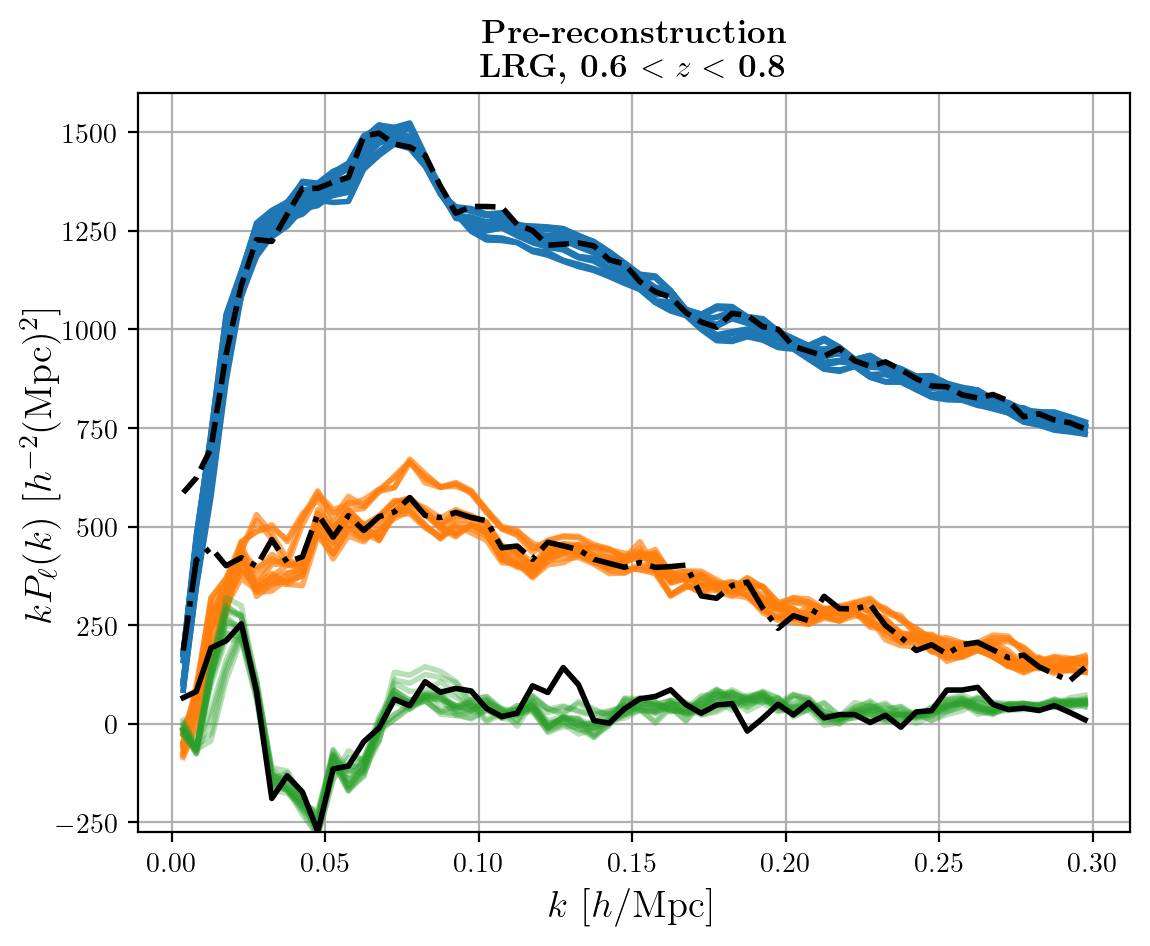}
    \hfill
    \includegraphics[width=0.495\textwidth, clip=True, trim={0 25 0 16}]{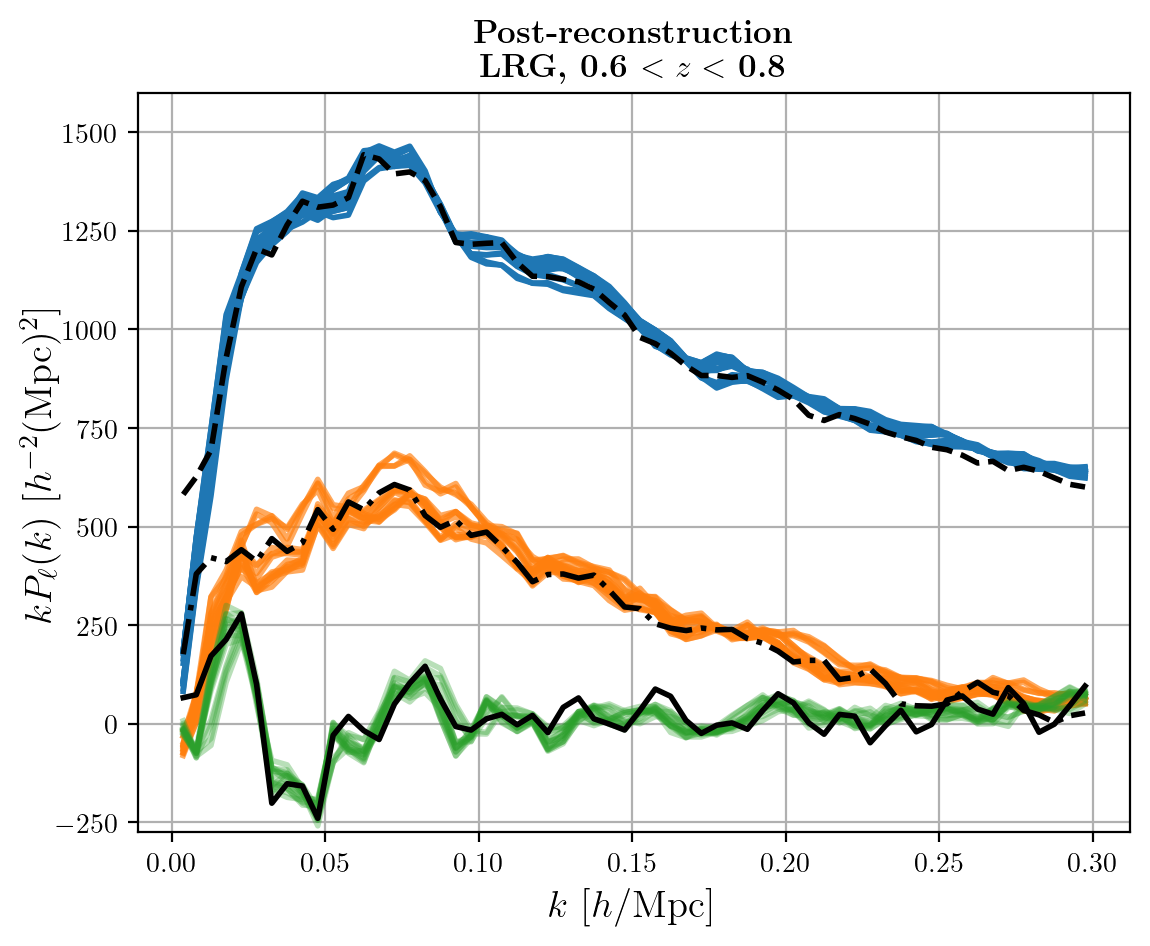}
    \hfill
    \includegraphics[width=0.495\textwidth, clip=True, trim={0 0 0 16}]{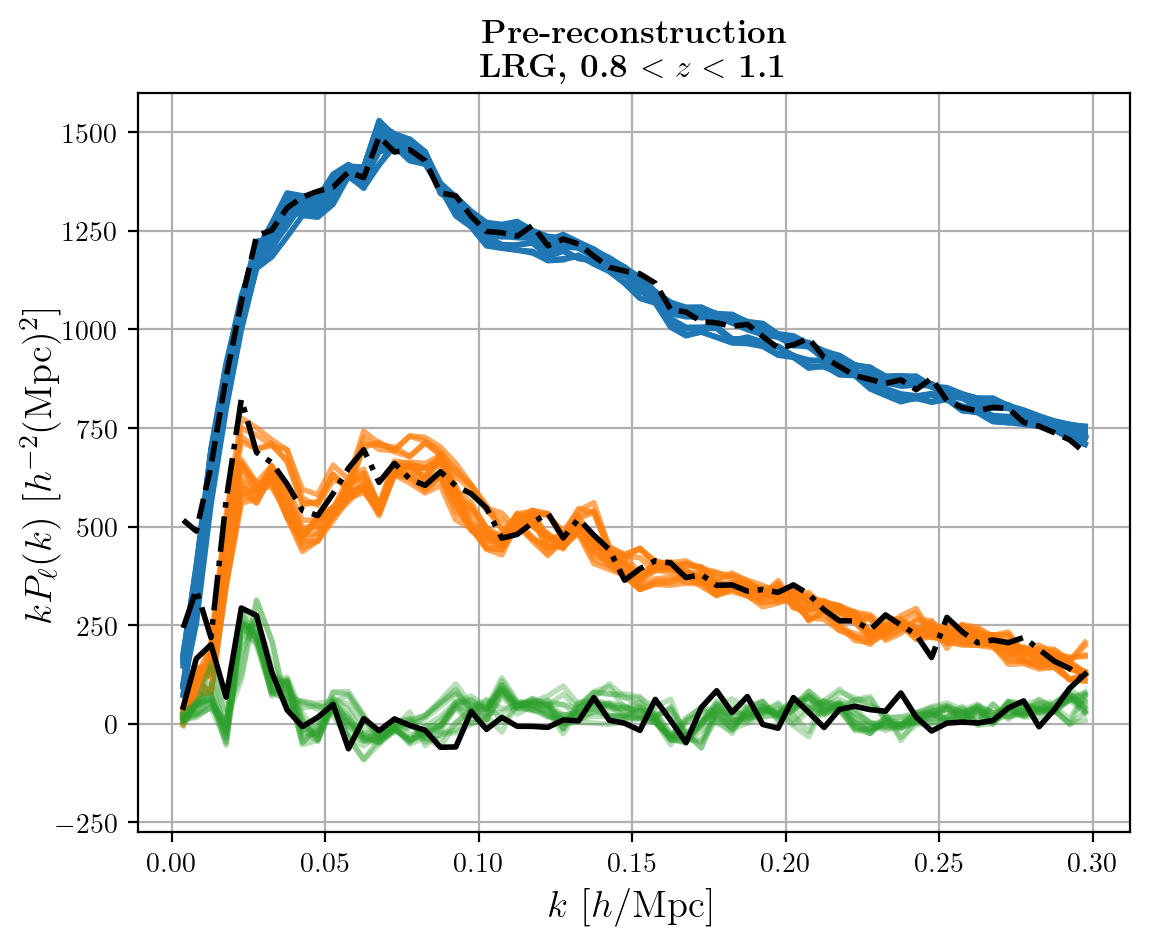}
    \hfill
    \includegraphics[width=0.495\textwidth, clip=True, trim={0 0 0 16}]{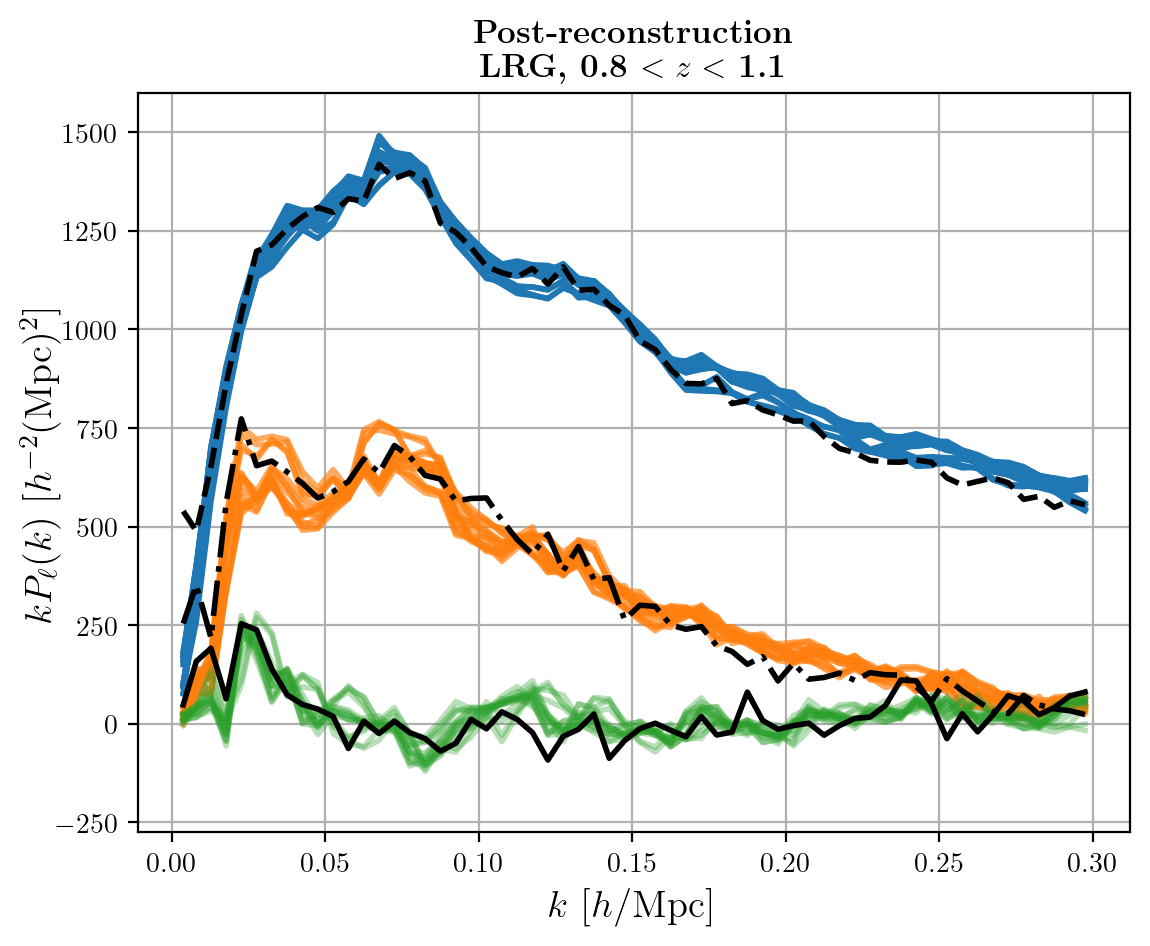}
    \caption{Power spectrum multipoles for the same LRG sample as in \cref{fig:correlation_multipoles_combined}. The rows are for different redshift bins;  pre-reconstruction multipoles (left) and post-reconstruction ones (right); see \cref{fig:correlation_multipoles_combined} caption for the rest of the details.
    }
    \label{fig:power_multipoles}
\end{figure}

\afterpage{\FloatBarrier}
\begin{figure}[hbt!]
    \centering
    \vspace*{-0.25em}
    \includegraphics[width=0.495\textwidth, clip=True, trim={0 25 0 20}]{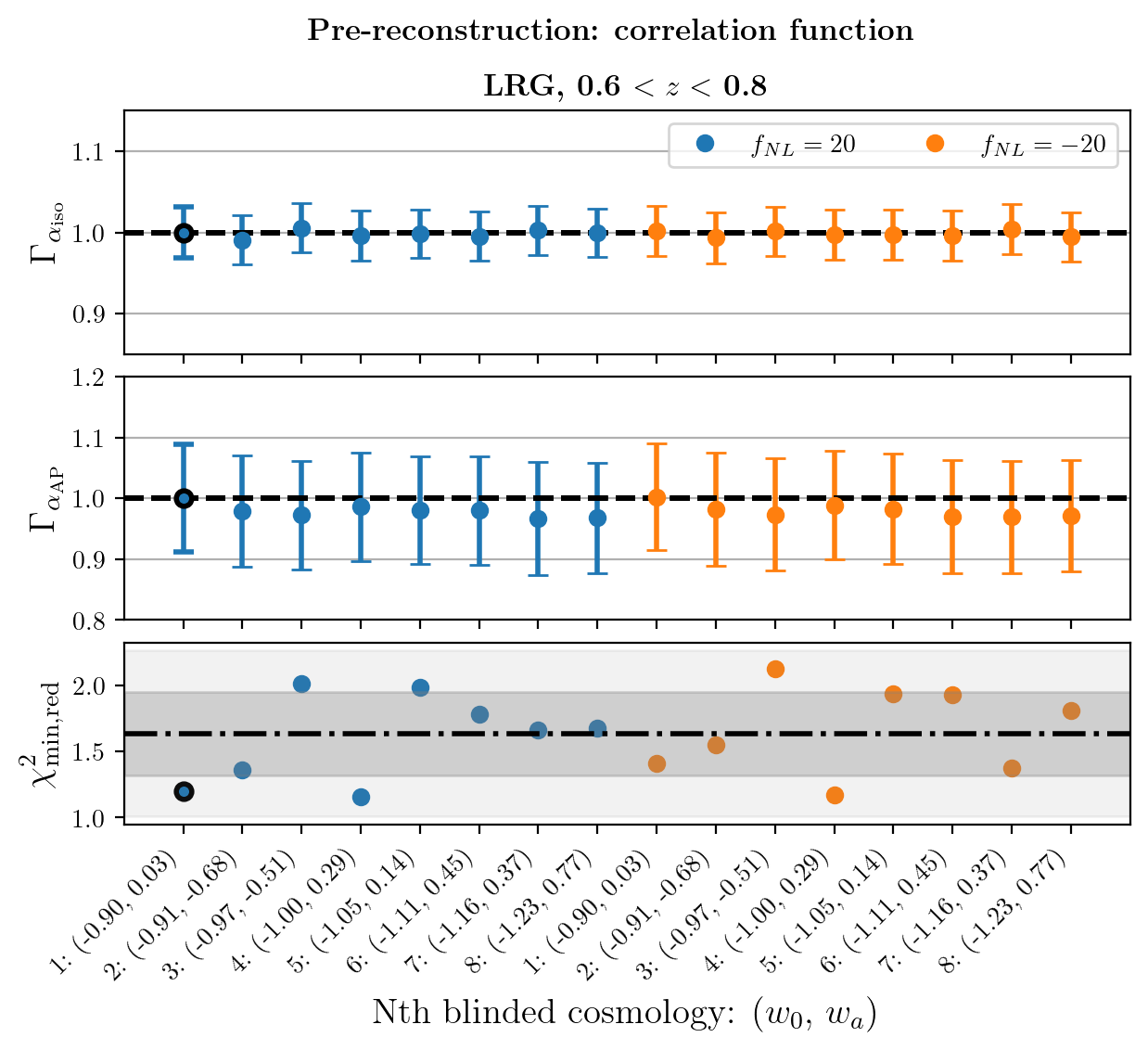}
    \includegraphics[width=0.495\textwidth, clip=True, trim={0 25 0 20}]{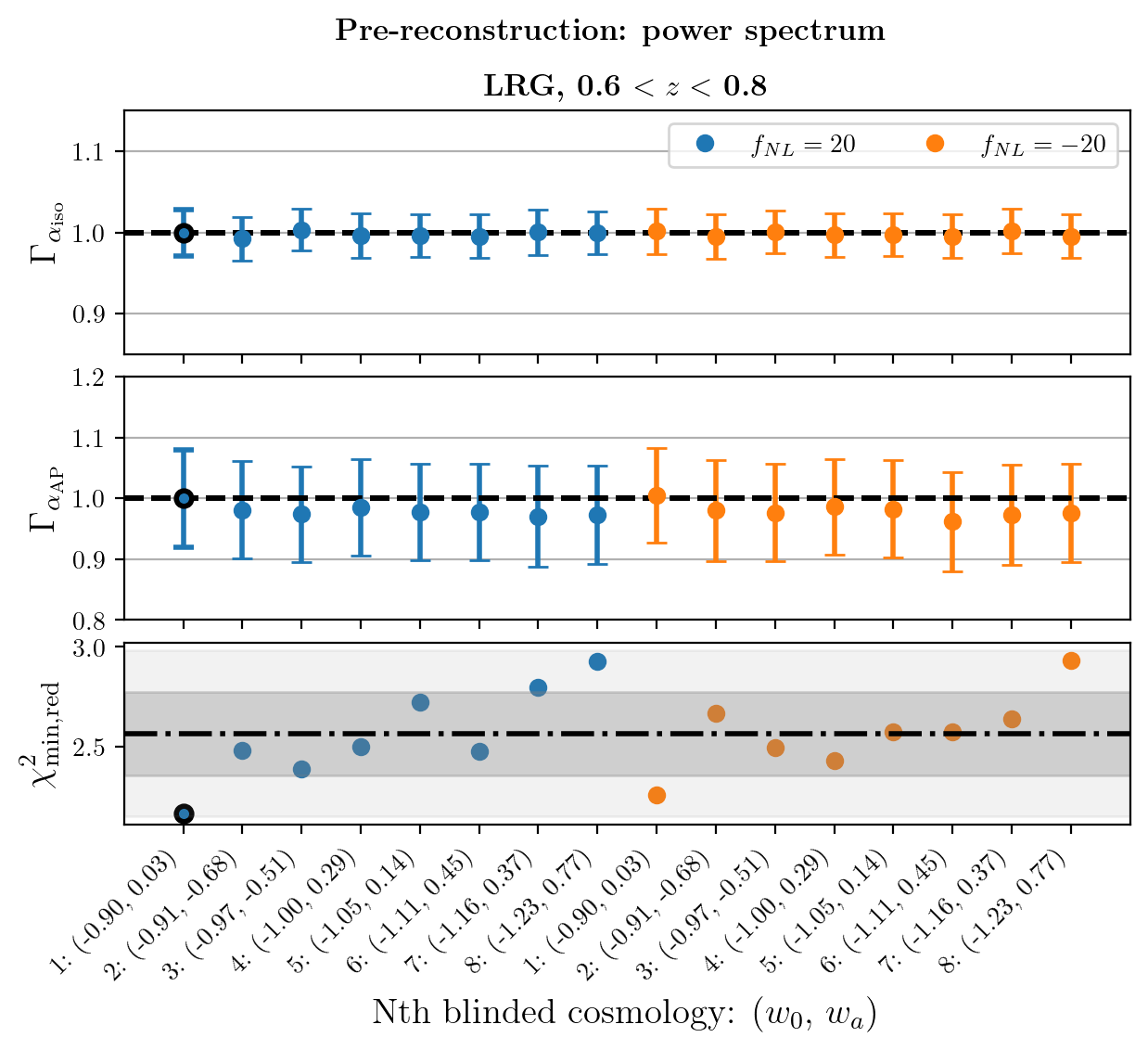}
    \includegraphics[width=0.495\textwidth, clip=True, trim={0 0 0 20}]{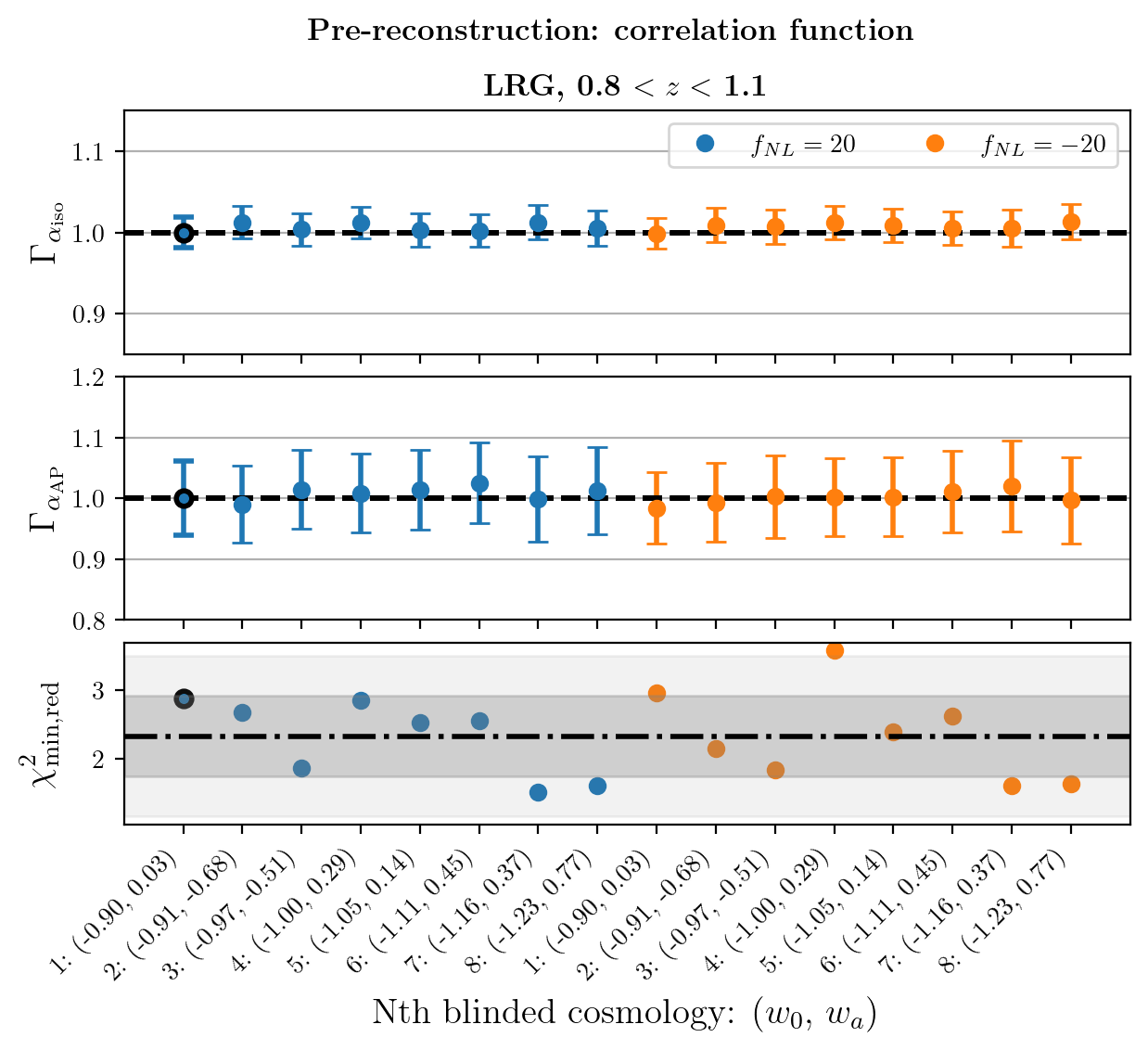}
    \includegraphics[width=0.495\textwidth, clip=True, trim={0 0 0 20}]{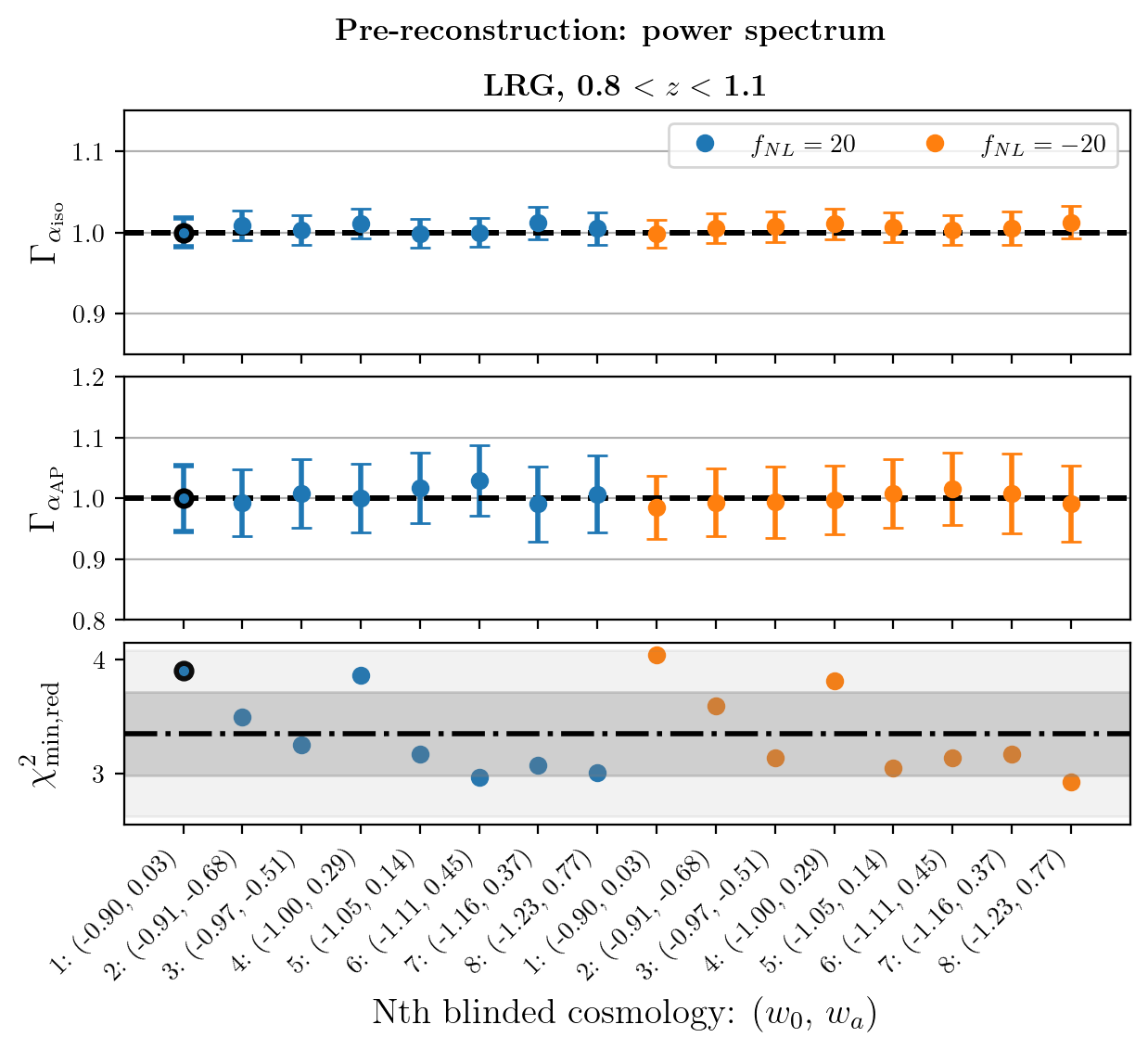}
    \caption{
    \footnotesize{
    Pre-reconstruction anisotropic BAO fits using the correlation function (left column) and the power spectrum (right column) for LRG samples for the two redshift bins (each row) from 16 different blinded mock catalogs with \wowa\ choices identified by indices 1-8 and two \fnl\ values by blue and orange, respectively. The top two subplots in each panel plot $\Gamma_i$, defined as the ratio of measured vs expected ratios of the $i$th parameter from each \texttt{sim} vs a reference \texttt{sim} (identified with black marker-edge); here $i$ = \alphaiso, \alphaap, where measured values are from the analysis pipeline while expected ones are from the theoretical connection with the respective \wowa; error bars capture the measurement uncertainties while propagating the errors. This statistic allows comparing all the sims against a reference sim. The bottom subplot in each panel displays the reduced \chitwo\ values, with shaded areas representing $1\sigma$ and $2\sigma$ regions;
    the $\sigma$-limits are obtained as the standard deviation from the mean of the \chitwo\ distribution of the 16 \chitwo\ values. This confirms the consistency and reliability of BAO measurements under various blinding shifts given the small variations.
    }
    }
    \label{fig:appendix_pre_recon_bao_fits_LRG}

\end{figure}

\begin{figure}[hbt!]
    \centering
    \includegraphics[width=0.495\textwidth, clip=True, trim={0 25 0 20}]{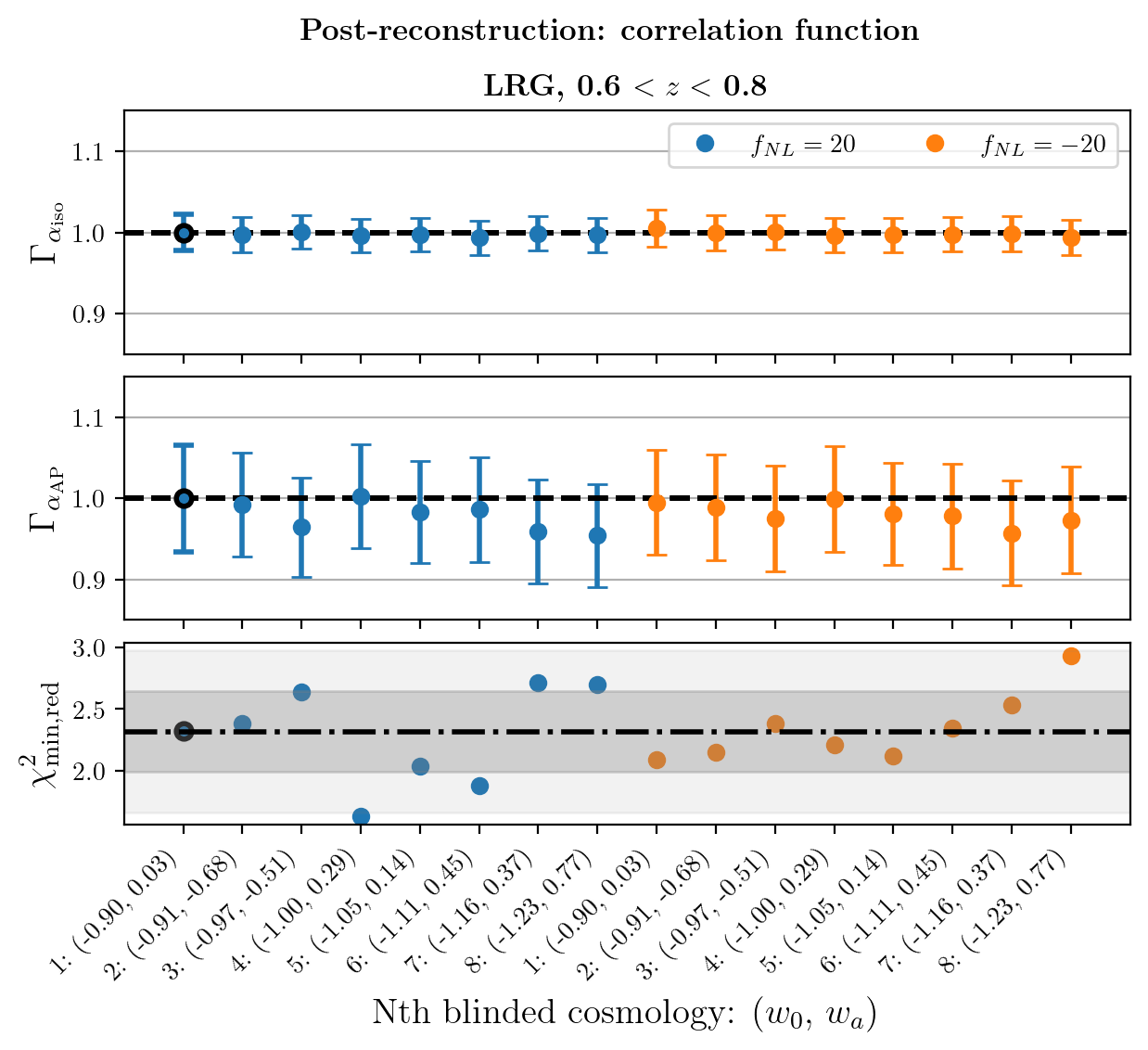}
    \includegraphics[width=0.495\textwidth, clip=True, trim={0 25 0 20}]{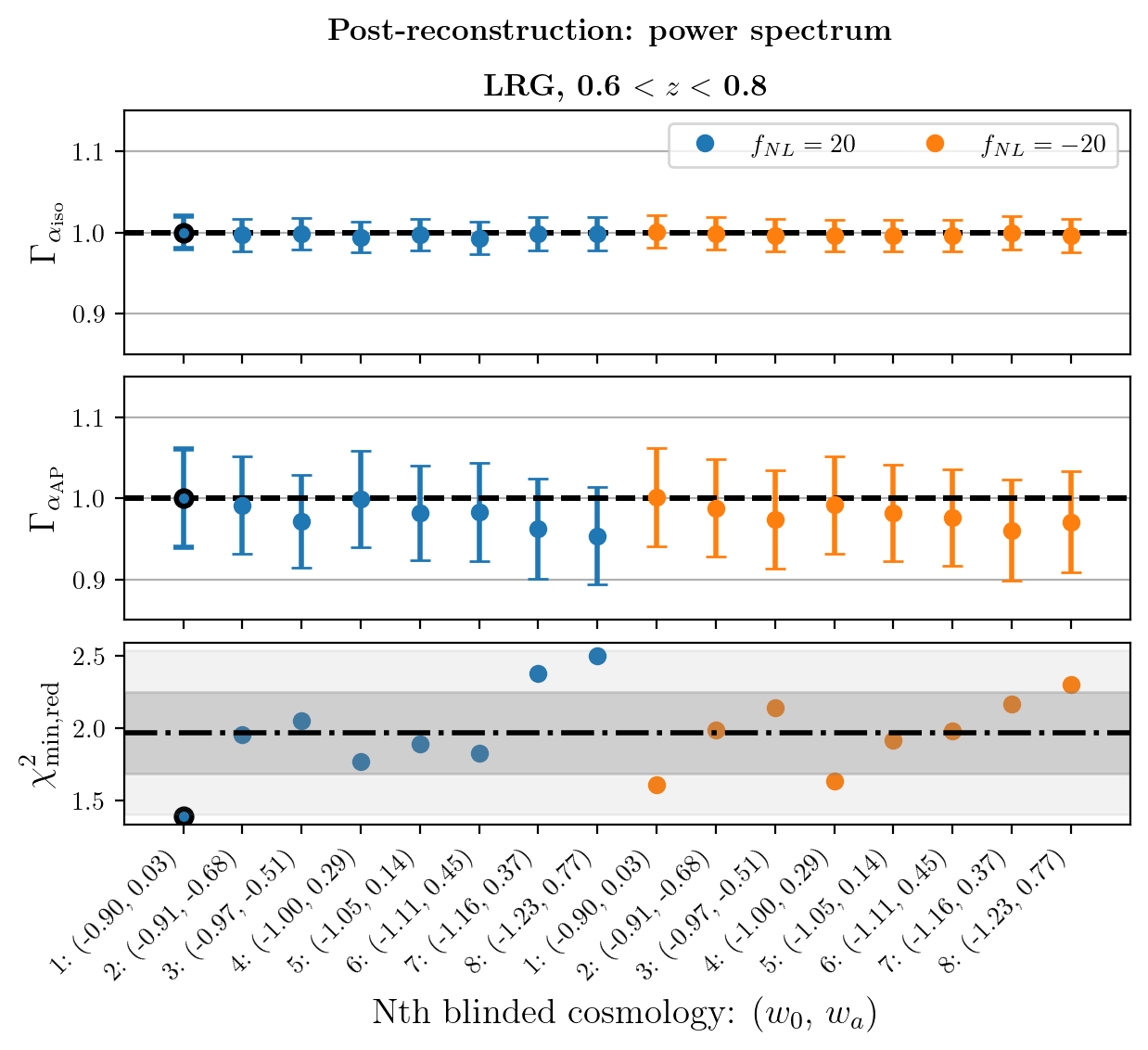}
    \includegraphics[width=0.495\textwidth, clip=True, trim={0 0 0 20}]{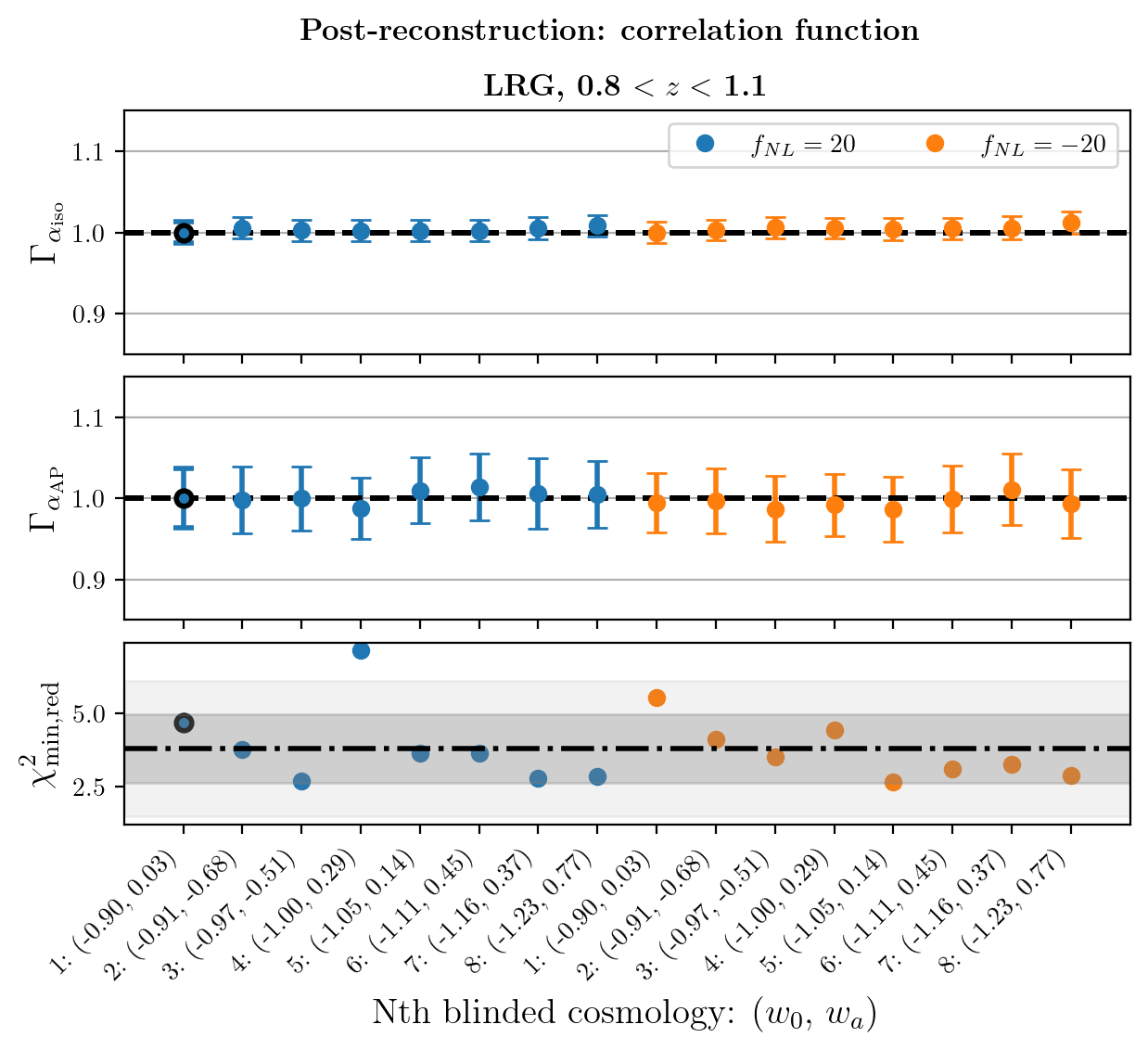}
    \includegraphics[width=0.495\textwidth, clip=True, trim={0 0 0 20}]{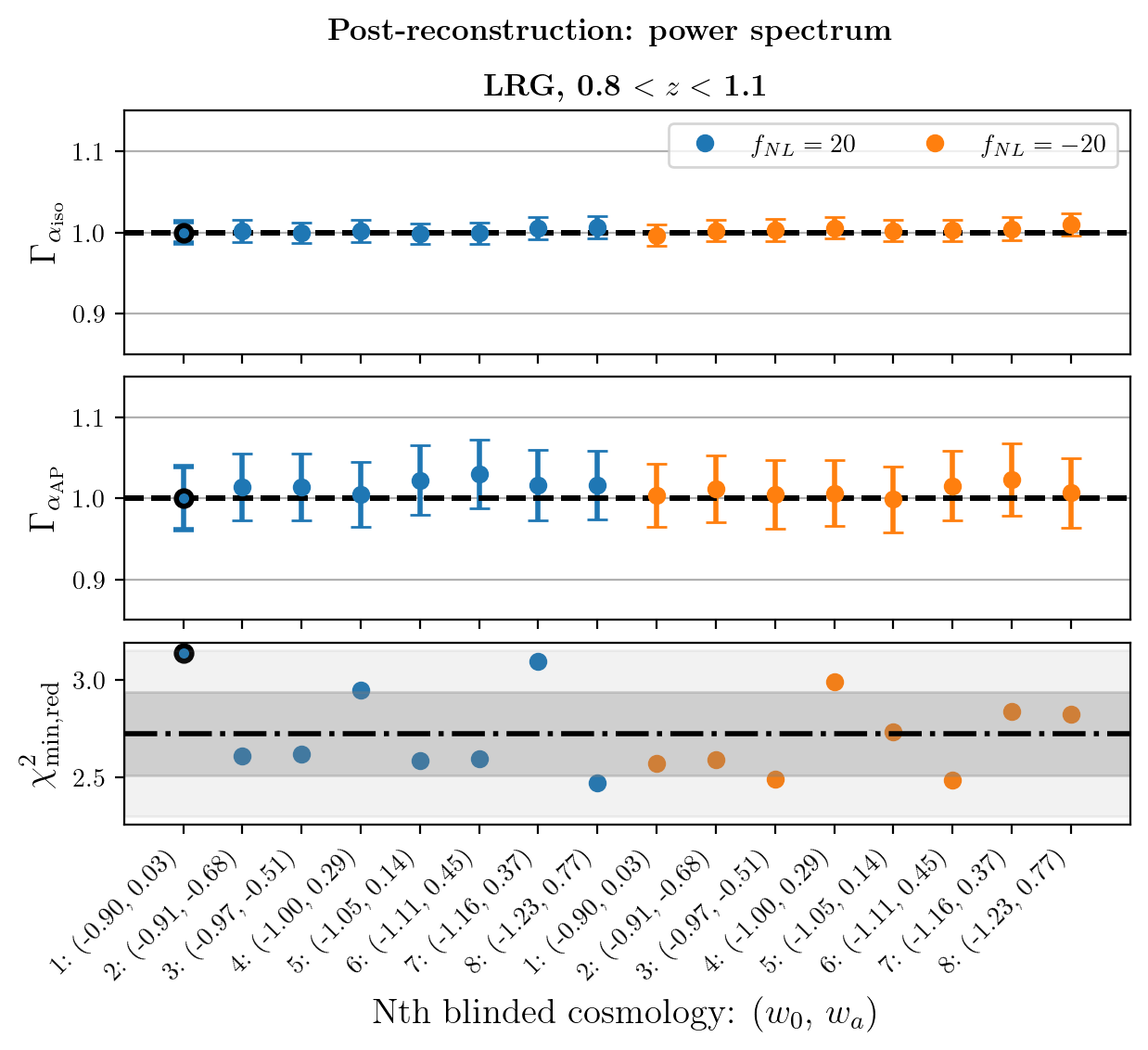}
    \caption{Post-reconstruction anisotropic BAO fits for LRG samples for the two redshift bins (each row) following the structure in \cref{fig:pre_recon_bao_fits_LRG}. Here, too, we see that while our $\Gamma$ statistic varies around the expected value of unity, the reduced \chitwo\ indicates good fits.}
    \label{fig:post_recon_bao_fits_LRG}
\end{figure}

\begin{figure}[hbt!]
    \centering
    \includegraphics[width=0.495\textwidth, clip=True, trim={0 25 0 0}]{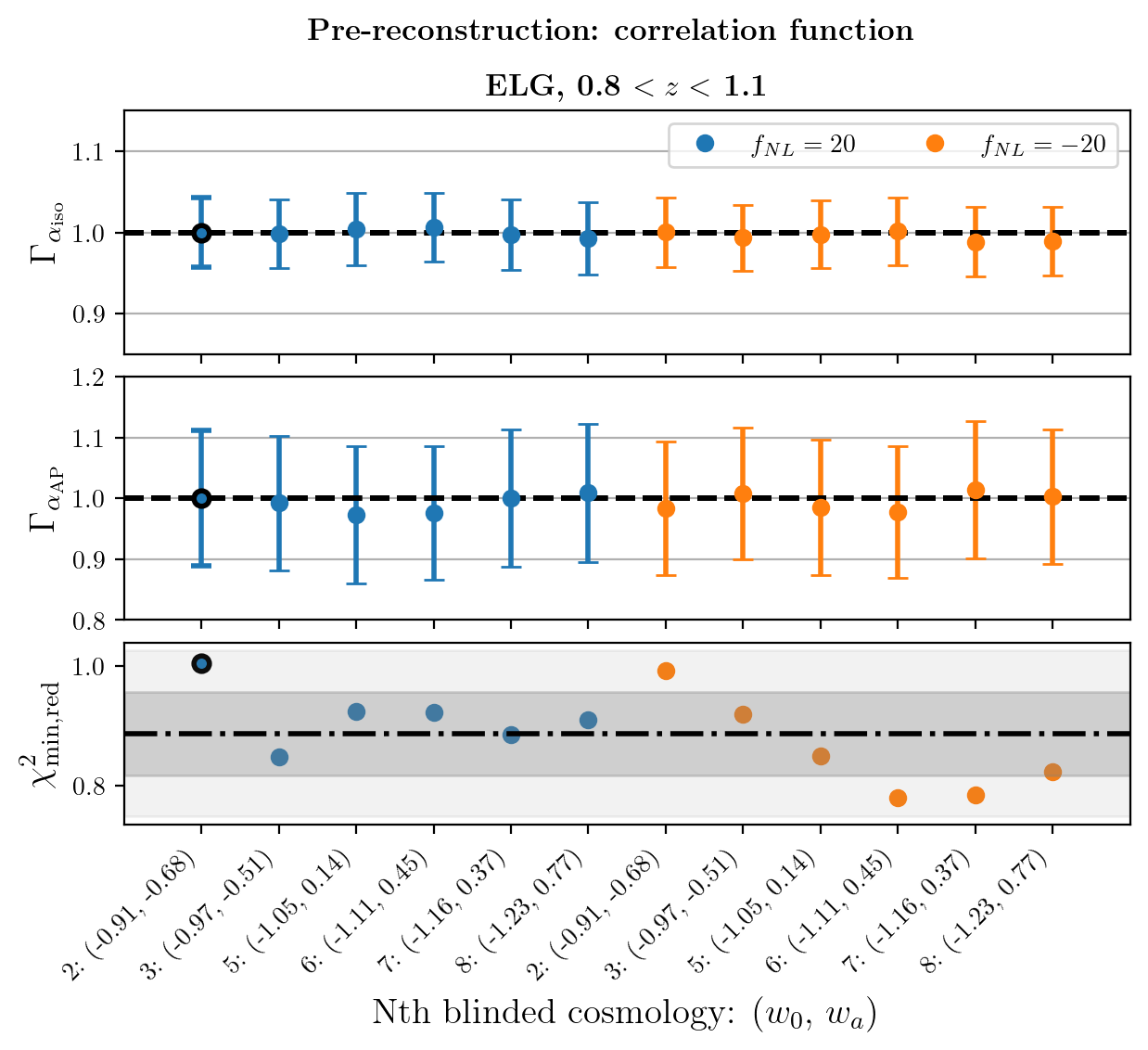}
    \includegraphics[width=0.495\textwidth, clip=True, trim={0 25 0 0}]{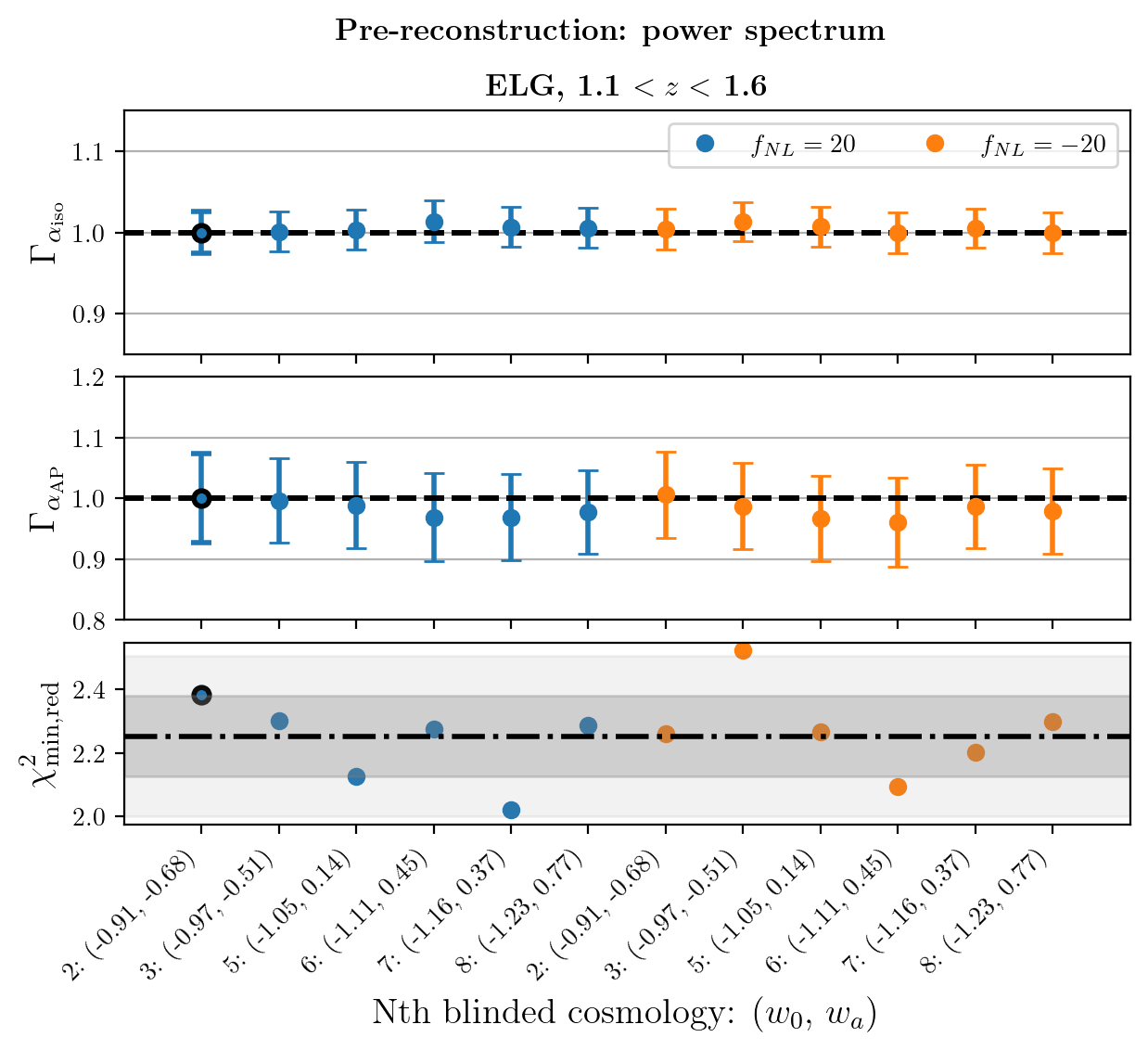}
    \includegraphics[width=0.495\textwidth, clip=True, trim={0 0 0 20}]{figures/bao_abacusmock1_blindconfigs_correlation_ELG_GCcomb_z0.8-1.1.png}
    \includegraphics[width=0.495\textwidth, clip=True, trim={0 0 0 20}]{figures/bao_abacusmock1_blindconfigs_power_ELG_GCcomb_z1.1-1.6.png}
    \caption{
    Pre-reconstruction anisotropic BAO fits for ELG samples for two different redshifts, following the structure in \cref{fig:pre_recon_bao_fits_LRG}. Here too, we see that the measured vs. expected ratios of the BAO fitting parameters vary a little across the sims and the reduced \chitwo\ indicates good fits. Note that for this tracer, we drop two of the blinded cosmologies (\wowa\ pairs 1,4) since these ELG catalogs had few randoms and therefore did not deliver reliable clustering measurements.}
    \label{fig:pre_recon_bao_fits_ELG}
\end{figure}

\begin{figure}[hbt!]
    \centering
    \includegraphics[width=0.495\textwidth, clip=True, trim={0 25 0 0}]{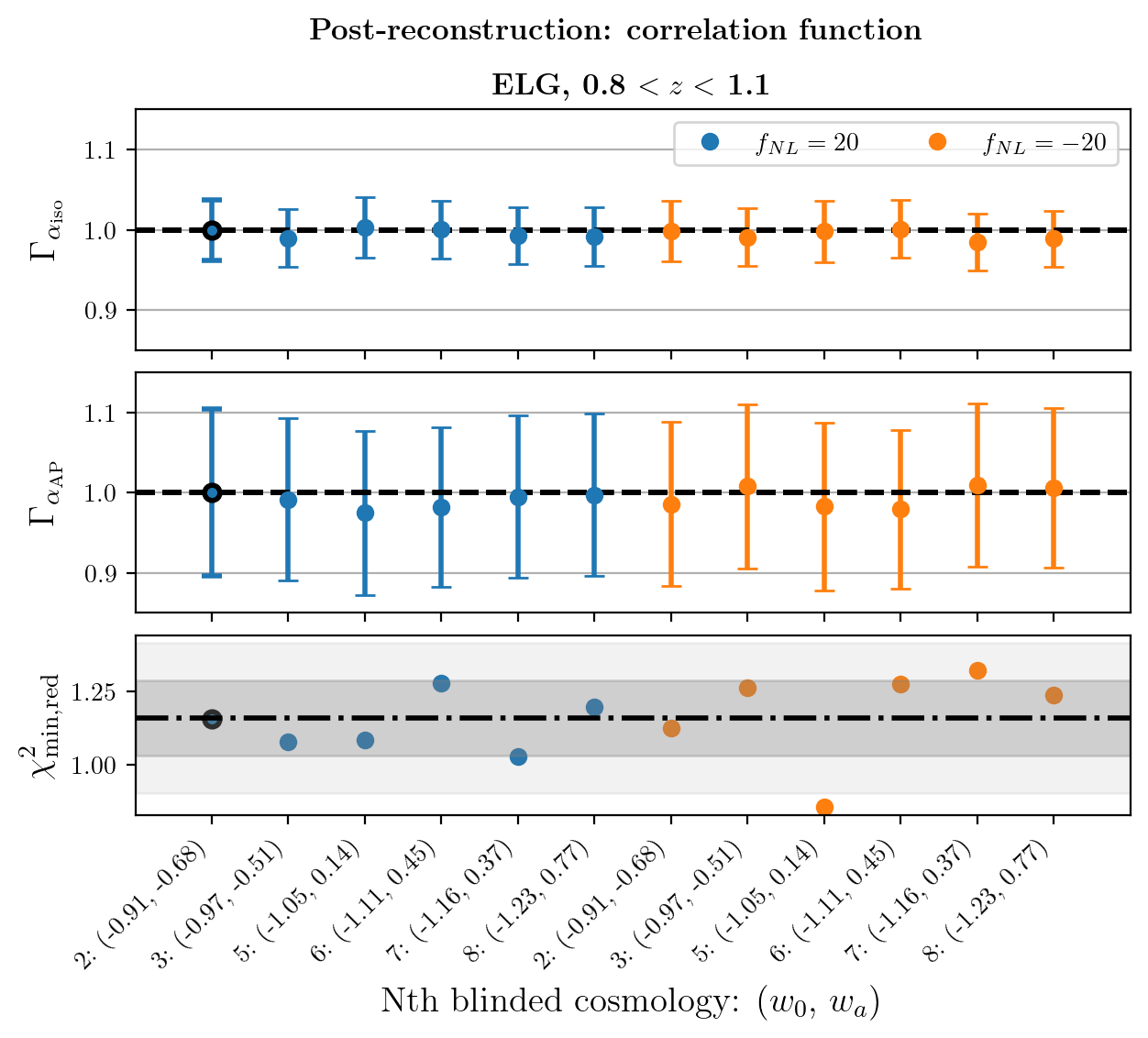}
    \includegraphics[width=0.495\textwidth, clip=True, trim={0 25 0 0}]{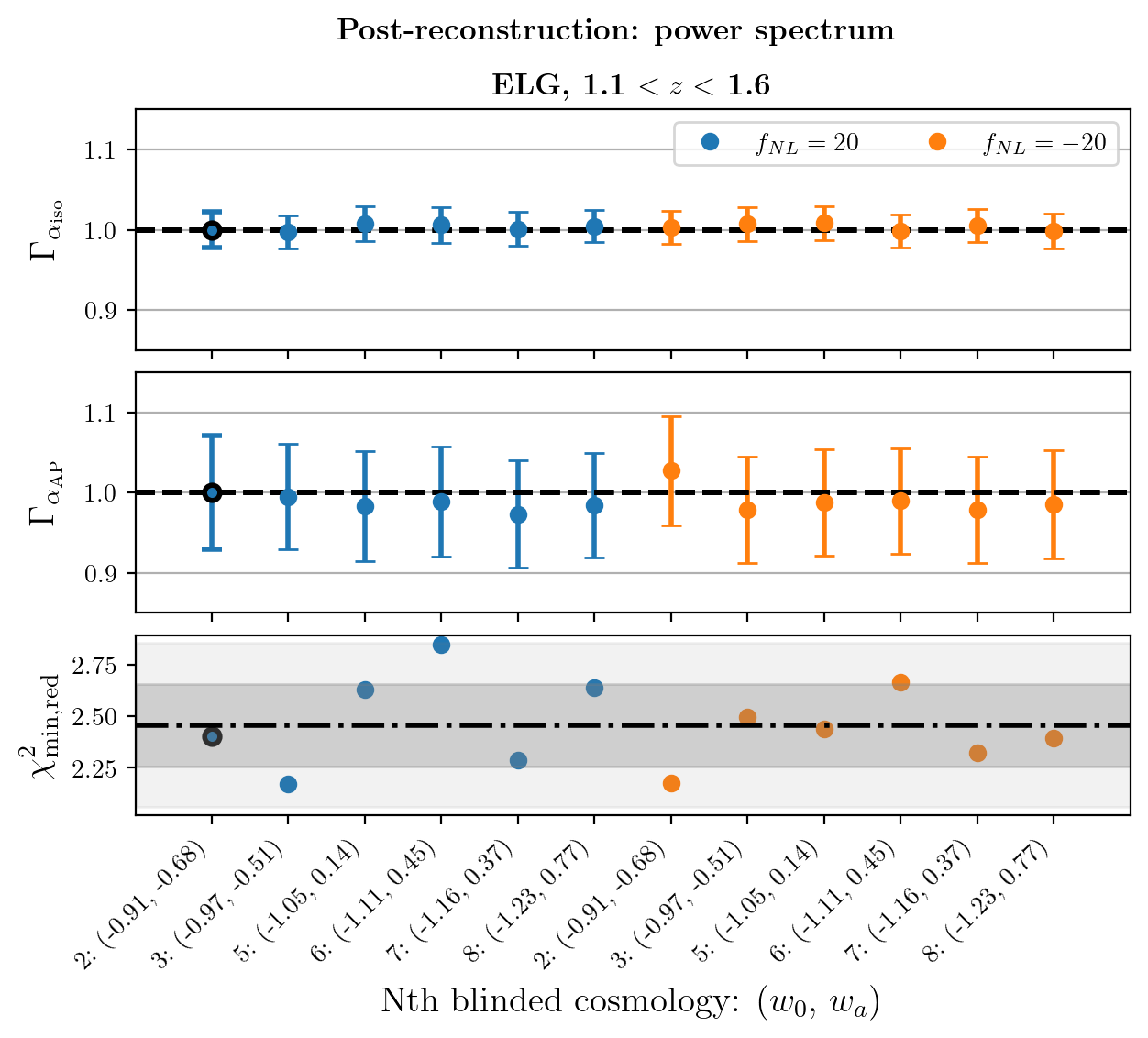}
    \includegraphics[width=0.495\textwidth, clip=True, trim={0 0 0 20}]{figures/bao_recon_abacusmock1_blindconfigs_correlation_ELG_GCcomb_z0.8-1.1.png}
    \includegraphics[width=0.495\textwidth, clip=True, trim={0 0 0 20}]{figures/bao_recon_abacusmock1_blindconfigs_power_ELG_GCcomb_z1.1-1.6.png}
    \caption{
    Post-reconstruction anisotropic BAO fits for ELG samples, following the structure in \cref{fig:pre_recon_bao_fits_LRG}. Here too, we see that the measured vs. expected ratios of the BAO fitting parameters vary a little across the sims and the reduced \chitwo\ indicates good fits. As mentioned in \cref{fig:pre_recon_bao_fits_ELG}, we drop two of the blinded cosmologies (\wowa\ pairs 1,4).}
    \label{fig:post_recon_bao_fits_ELG}
\end{figure}

\begin{figure}[hbt!]
    \centering
    \includegraphics[width=0.495\textwidth]{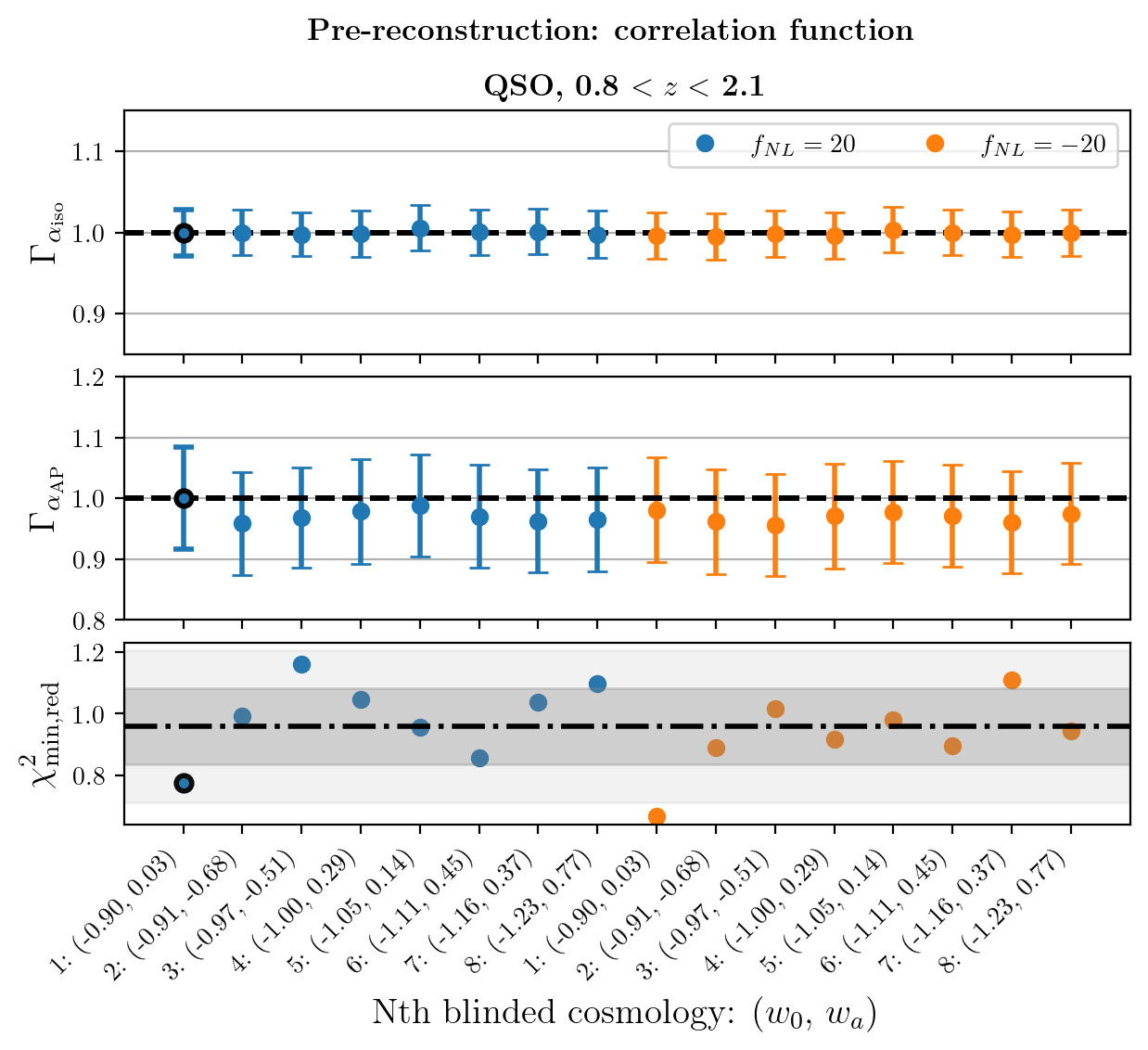}
    \includegraphics[width=0.495\textwidth]{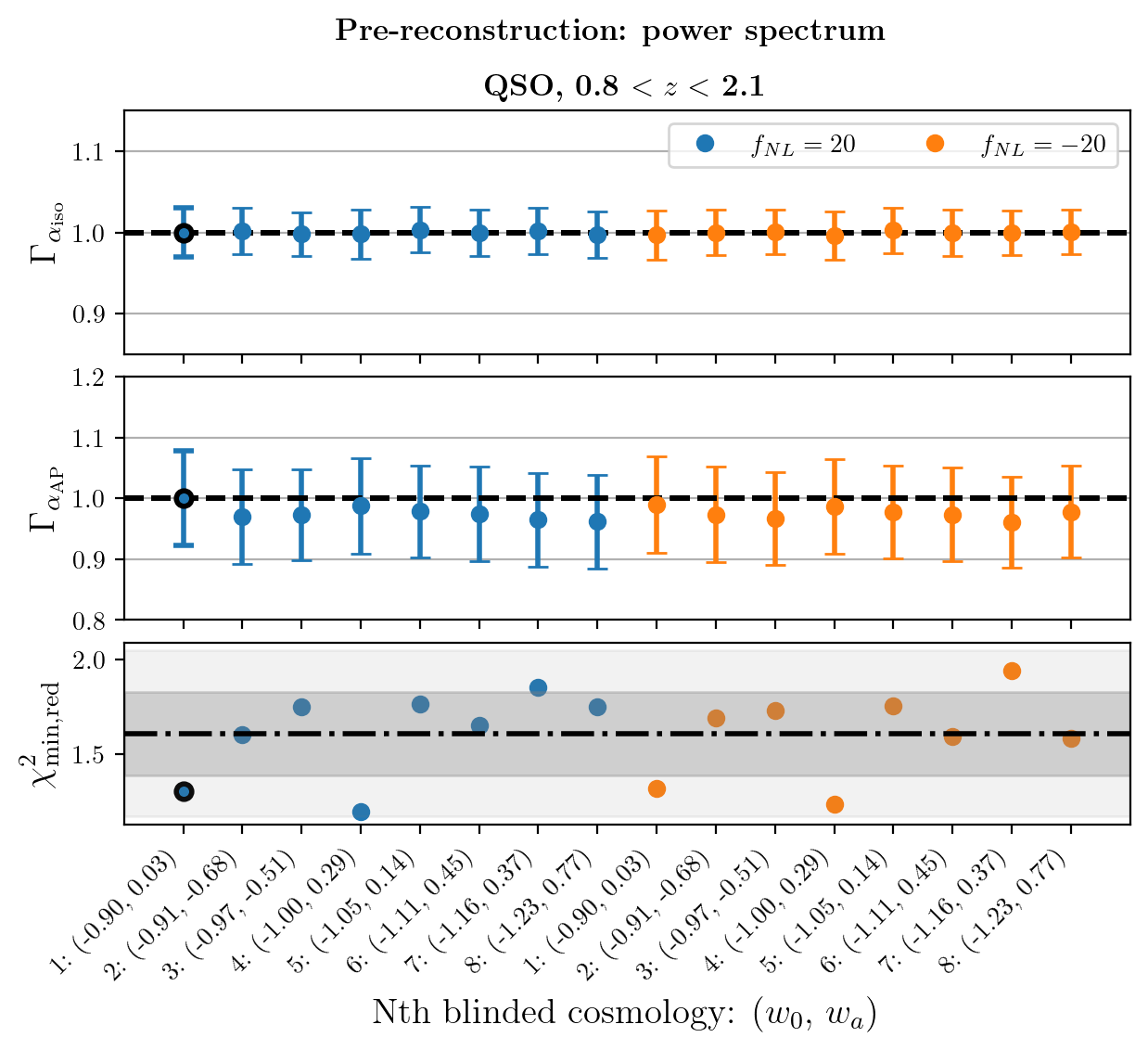}
    \caption{
    Pre-reconstruction anisotropic BAO fits for a QSO sample, following the structure in
    \cref{fig:pre_recon_bao_fits_LRG}. Here too, we see that the measured vs. expected ratios of the BAO fitting parameters vary a little across the sims while the reduced \chitwo\ indicates good fits.
    }
    \label{fig:pre_recon_bao_fits_QSO}
\end{figure}

\begin{figure}[hbt!]
    \centering
    \includegraphics[width=0.495\textwidth]{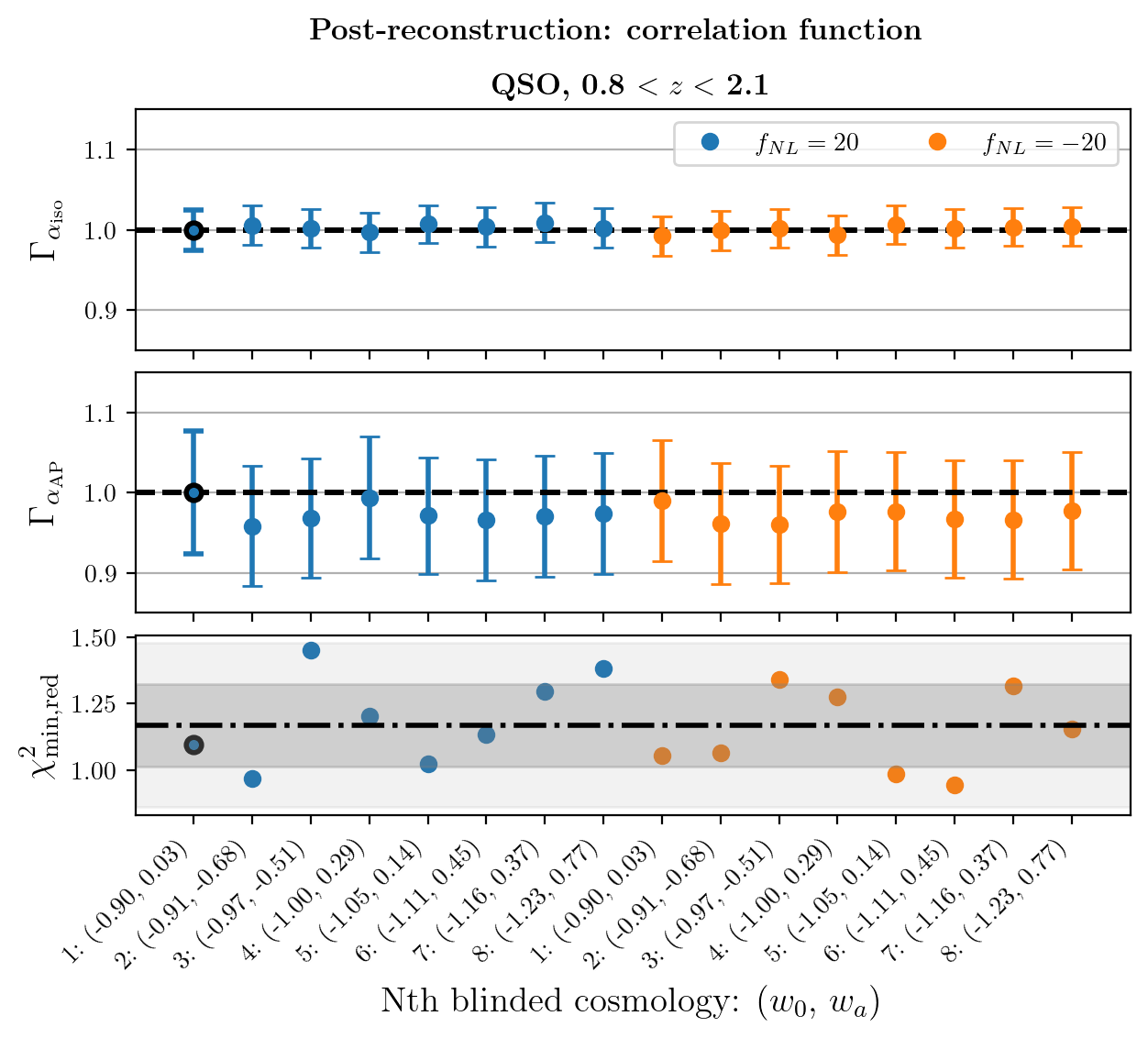}
    \includegraphics[width=0.495\textwidth]{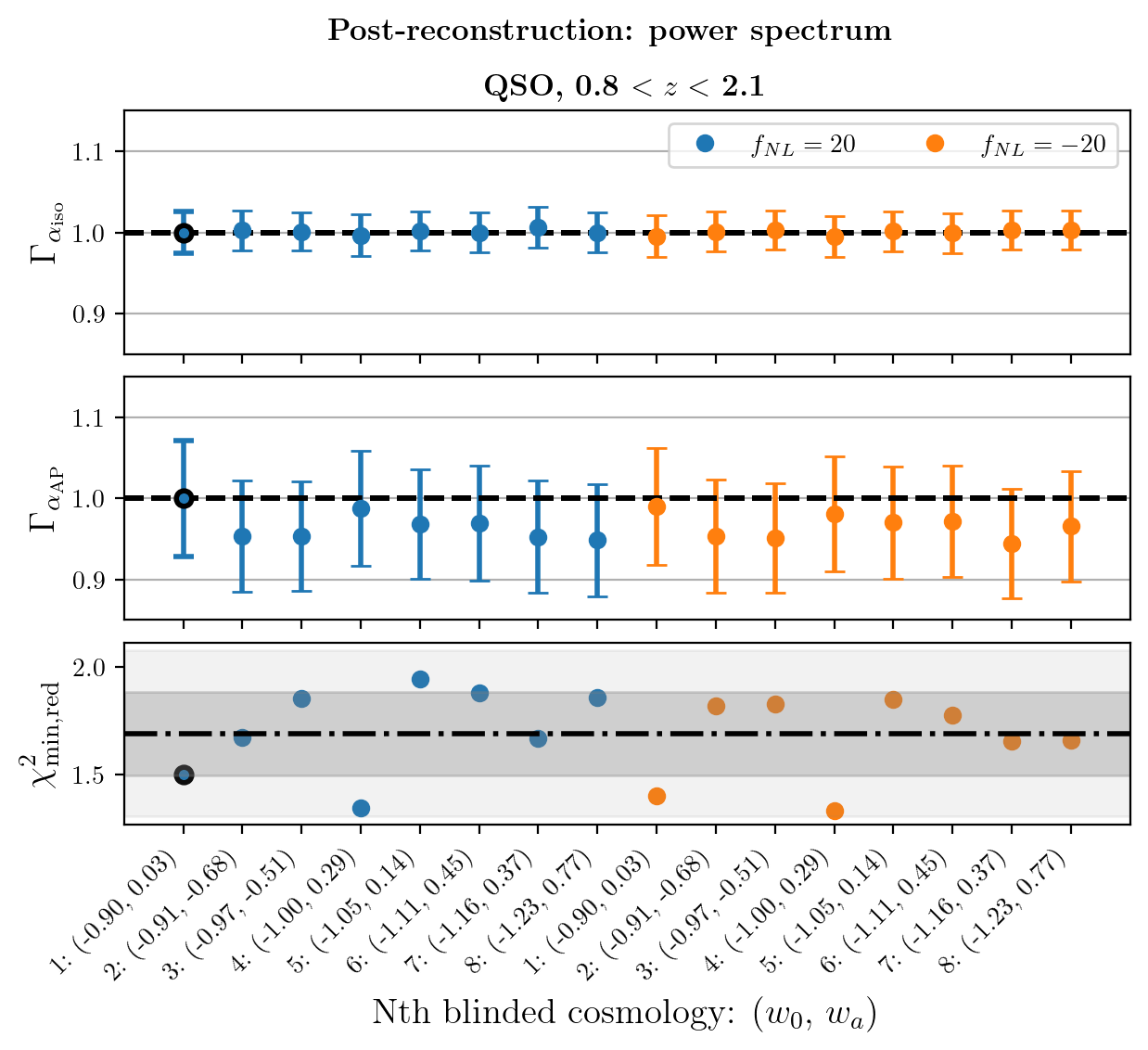}
    \caption{Post-reconstruction anisotropic BAO fits for a QSO sample, following the structure in
    \cref{fig:pre_recon_bao_fits_LRG}. Here too, we see that the measured vs. expected ratios of the BAO fitting parameters vary a little across the sims and the reduced \chitwo\ indicates good fits.}
    \label{fig:post_recon_bao_fits_QSO}
\end{figure}

\begin{figure}[hbt!]
    \centering
    \includegraphics[width=0.36\paperwidth, clip=True, trim={25 0 0 0}]{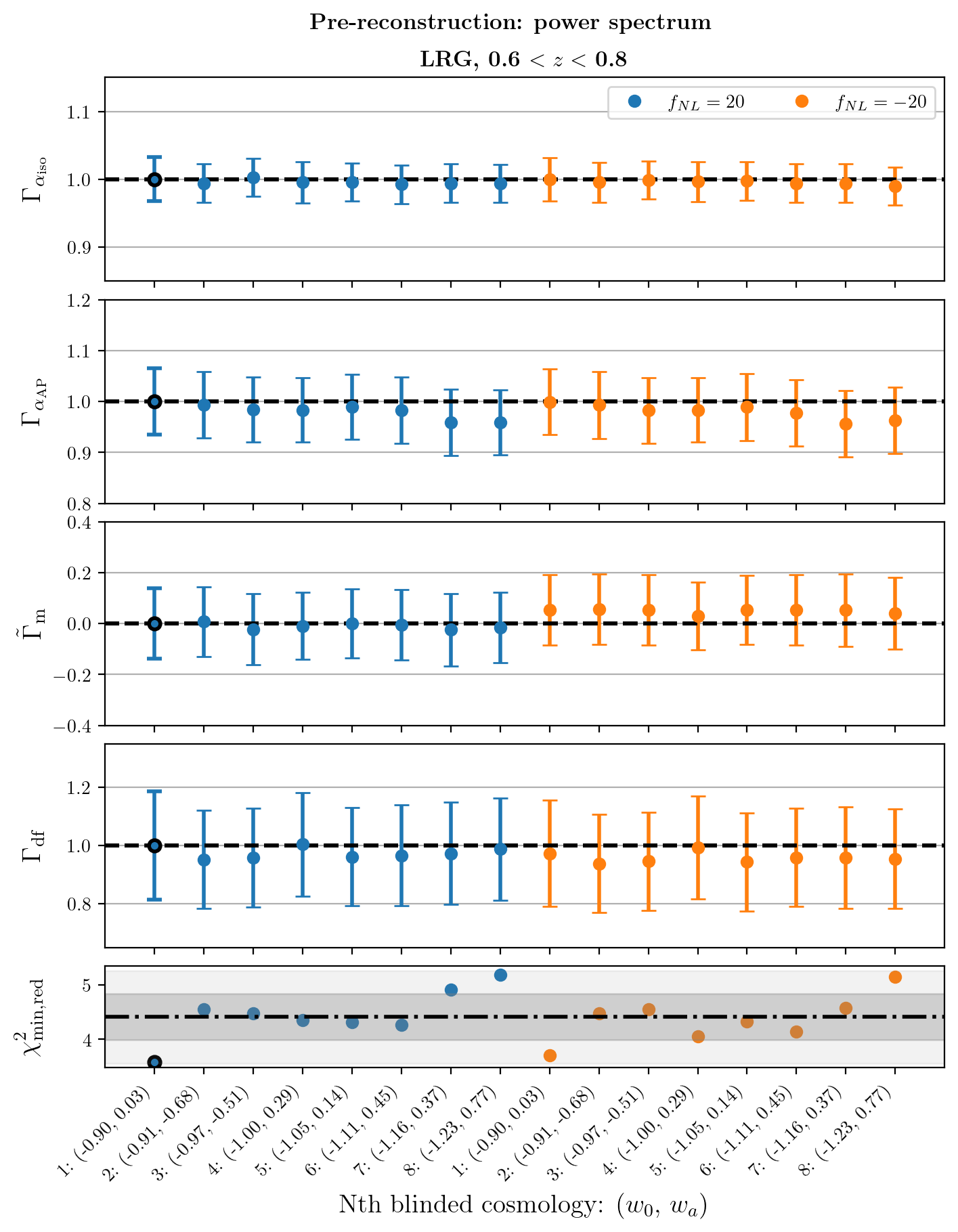}
    \includegraphics[width=0.36\paperwidth, clip=True, trim={25 0 0 0}]{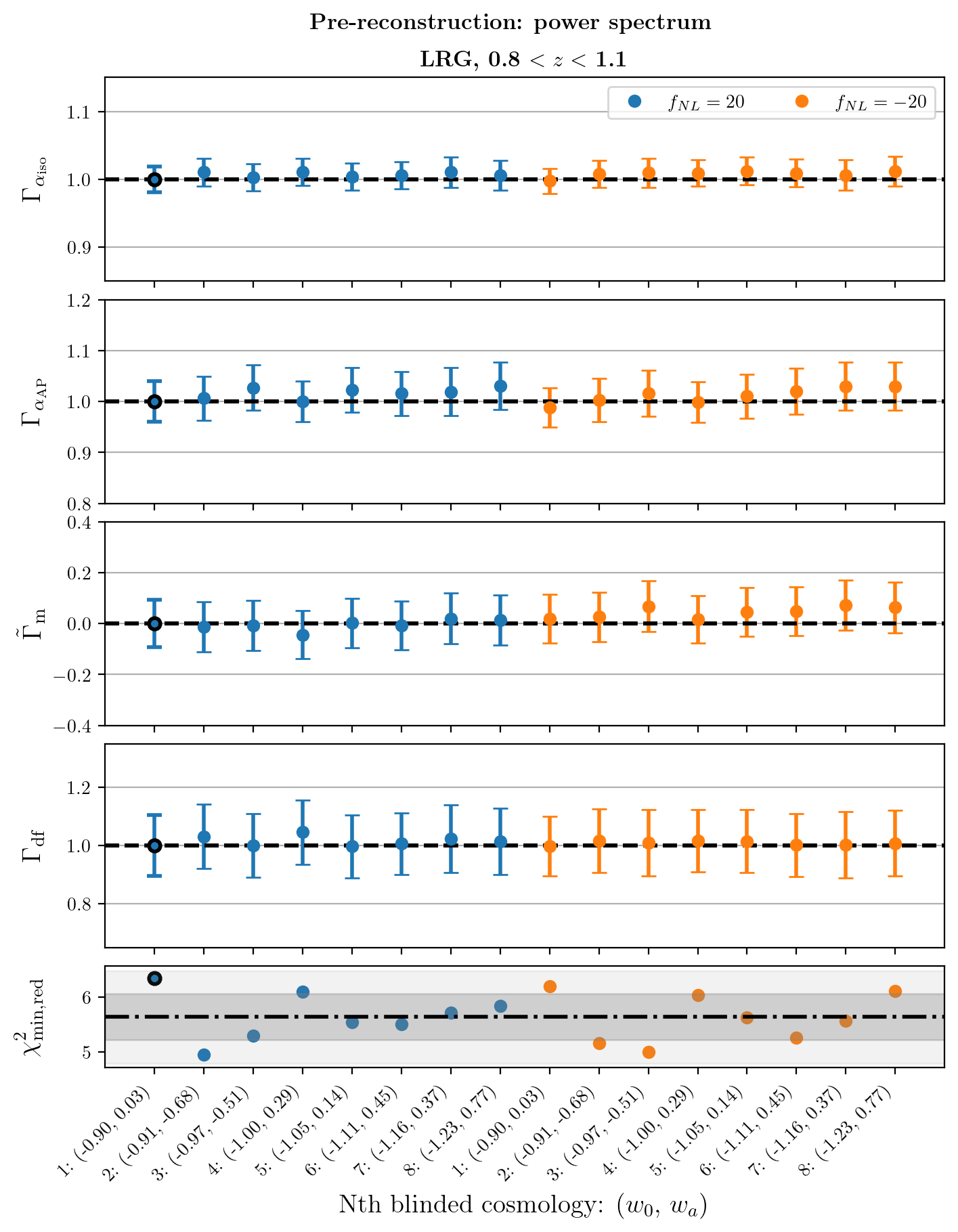}
    \caption{
    ShapeFit fits using LRG samples for the two redshift bins (each column) from 16 different blinded mock catalogs. Various details here are the same as in \cref{fig:pre_recon_bao_fits_LRG}, except that $i$ = \alphaiso, \alphaap, $df$, \dm\ in $\Gamma_i$ while $\tilde\Gamma_i$ is the same as $\Gamma_i$ but comparing differences as opposed to ratios between measured and expected (since expected is 0). As for BAO fits, we see that the ratios (differences) are close to 1 (0) and the \chitwo\ variations are within 1-2$\sigma$, demonstrating the robustness of the fits.
    }
    \label{fig:shapefit_fits_LRG23}
\end{figure}

\begin{figure}[hbt!]
    \centering
    \includegraphics[width=0.495\textwidth]{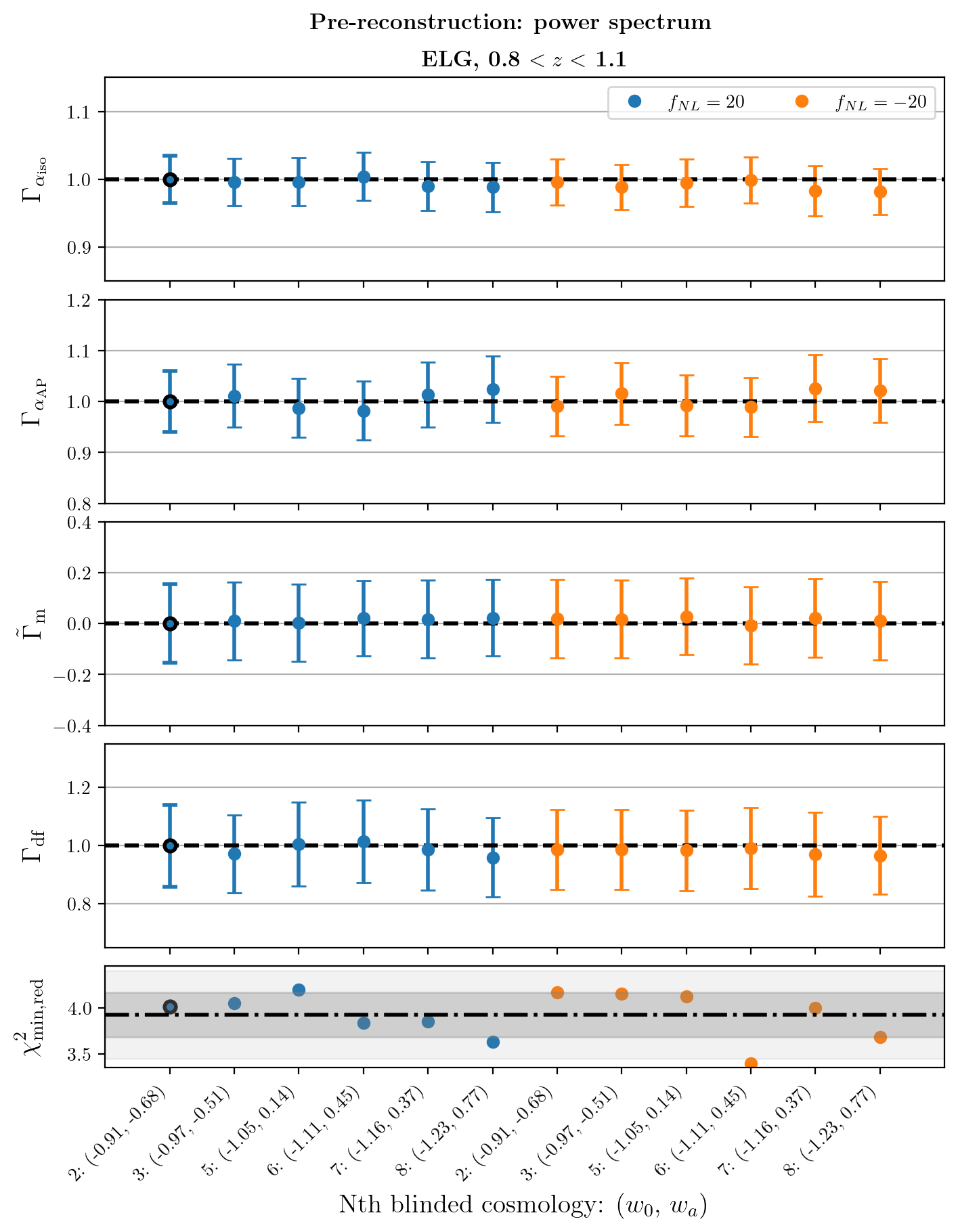}
    \includegraphics[width=0.495\textwidth]{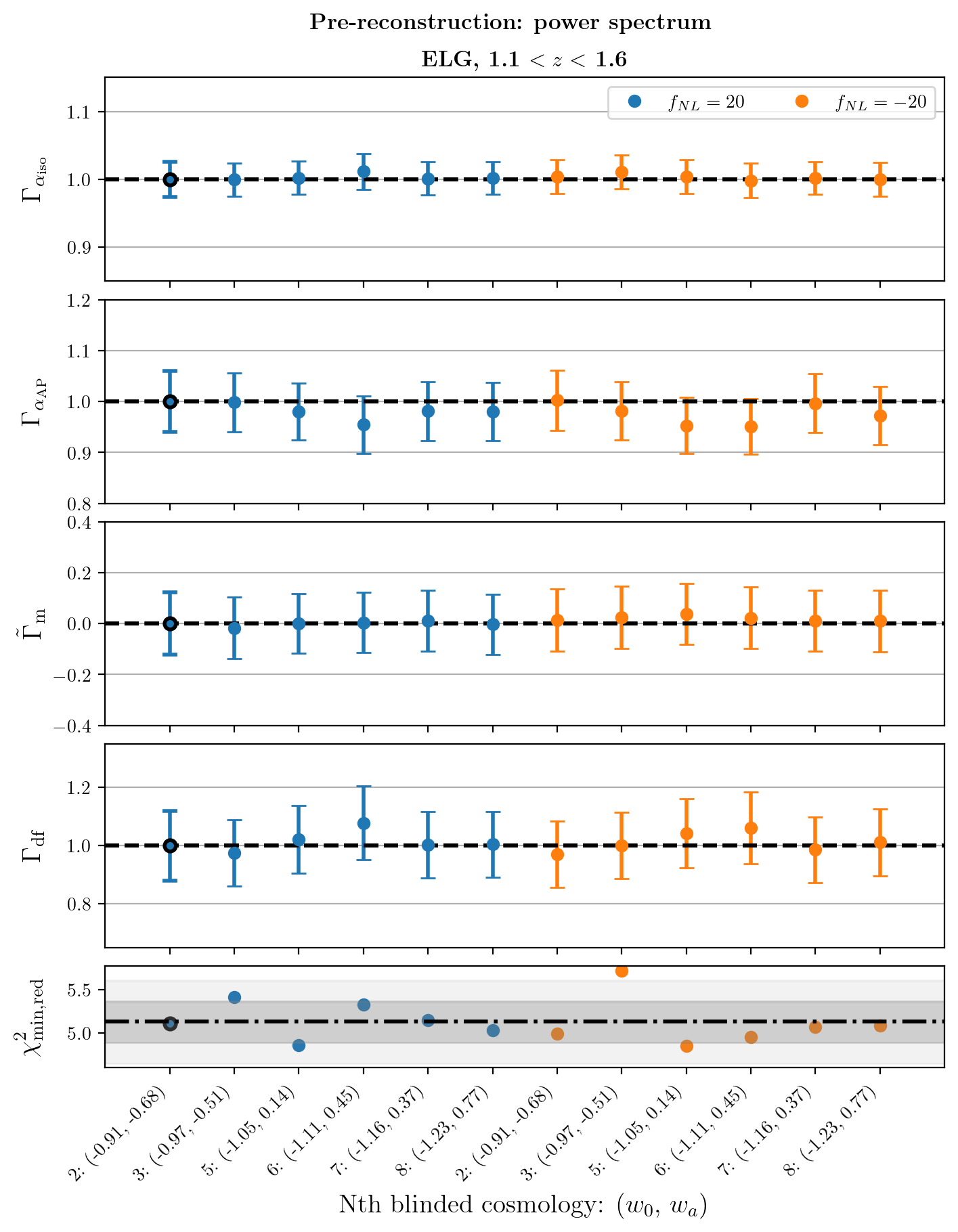}
    \caption{ShapeFit fits for ELG samples, following the structure in \cref{fig:shapefit_fits_LRG1}. Here, too, we see that while the measured vs. expected ratios/differences of the various parameters vary a little across the sims, the reduced \chitwo\  variations are within 1-2$\sigma$. As mentioned in \cref{fig:pre_recon_bao_fits_ELG}, two sims are dropped for this tracer.}
    \label{fig:shapefit_fits_ELG}
\end{figure}

\afterpage{\FloatBarrier}

\section{Author Affiliations}
\label{sec:affiliations}

\begin{hangparas}{.5cm}{1}

$^{1}${Leinweber Center for Theoretical Physics, University of Michigan, 450 Church Street, Ann Arbor, Michigan 48109-1040, USA}

$^{2}${Department of Physics, University of Michigan, Ann Arbor, MI 48109, USA}

$^{3}${Laboratoire de Physique Subatomique et de Cosmologie, 53 Avenue des Martyrs, 38000 Grenoble, France}

$^{4}${Center for Cosmology and AstroParticle Physics, The Ohio State University, 191 West Woodruff Avenue, Columbus, OH 43210, USA}

$^{5}${Department of Astronomy, The Ohio State University, 4055 McPherson Laboratory, 140 W 18th Avenue, Columbus, OH 43210, USA}

$^{6}${The Ohio State University, Columbus, 43210 OH, USA}

$^{7}${Institute for Astronomy, University of Edinburgh, Royal Observatory, Blackford Hill, Edinburgh EH9 3HJ, UK}

$^{8}${IRFU, CEA, Universit\'{e} Paris-Saclay, F-91191 Gif-sur-Yvette, France}

$^{9}${Lawrence Berkeley National Laboratory, 1 Cyclotron Road, Berkeley, CA 94720, USA}

$^{10}${Physics Dept., Boston University, 590 Commonwealth Avenue, Boston, MA 02215, USA}

$^{11}${Department of Physics \& Astronomy, University College London, Gower Street, London, WC1E 6BT, UK}

$^{12}${Department of Astronomy and Astrophysics, University of Chicago, 5640 South Ellis Avenue, Chicago, IL 60637, USA}

$^{13}${Fermi National Accelerator Laboratory, PO Box 500, Batavia, IL 60510, USA}

$^{14}${Institute for Computational Cosmology, Department of Physics, Durham University, South Road, Durham DH1 3LE, UK}

$^{15}${Instituto de F\'{\i}sica, Universidad Nacional Aut\'{o}noma de M\'{e}xico,  Cd. de M\'{e}xico  C.P. 04510,  M\'{e}xico}

$^{16}${NSF NOIRLab, 950 N. Cherry Ave., Tucson, AZ 85719, USA}

$^{17}${Kavli Institute for Particle Astrophysics and Cosmology, Stanford University, Menlo Park, CA 94305, USA}

$^{18}${SLAC National Accelerator Laboratory, Menlo Park, CA 94305, USA}

$^{19}${Departamento de F\'isica, Universidad de los Andes, Cra. 1 No. 18A-10, Edificio Ip, CP 111711, Bogot\'a, Colombia}

$^{20}${Observatorio Astron\'omico, Universidad de los Andes, Cra. 1 No. 18A-10, Edificio H, CP 111711 Bogot\'a, Colombia}

$^{21}${Institut d'Estudis Espacials de Catalunya (IEEC), 08034 Barcelona, Spain}

$^{22}${Institute of Cosmology and Gravitation, University of Portsmouth, Dennis Sciama Building, Portsmouth, PO1 3FX, UK}

$^{23}${Institute of Space Sciences, ICE-CSIC, Campus UAB, Carrer de Can Magrans s/n, 08913 Bellaterra, Barcelona, Spain}

$^{24}${Departament de F\'{\i}sica Qu\`{a}ntica i Astrof\'{\i}sica, Universitat de Barcelona, Mart\'{\i} i Franqu\`{e}s 1, E08028 Barcelona, Spain}

$^{25}${Institut de Ci\`encies del Cosmos (ICCUB), Universitat de Barcelona (UB), c. Mart\'i i Franqu\`es, 1, 08028 Barcelona, Spain.}

$^{26}${Department of Astrophysical Sciences, Princeton University, Princeton NJ 08544, USA}

$^{27}${Department of Physics, The Ohio State University, 191 West Woodruff Avenue, Columbus, OH 43210, USA}

$^{28}${School of Mathematics and Physics, University of Queensland, 4072, Australia}

$^{29}${Sorbonne Universit\'{e}, CNRS/IN2P3, Laboratoire de Physique Nucl\'{e}aire et de Hautes Energies (LPNHE), FR-75005 Paris, France}

$^{30}${Departament de F\'{i}sica, Serra H\'{u}nter, Universitat Aut\`{o}noma de Barcelona, 08193 Bellaterra (Barcelona), Spain}

$^{31}${Institut de F\'{i}sica d’Altes Energies (IFAE), The Barcelona Institute of Science and Technology, Campus UAB, 08193 Bellaterra Barcelona, Spain}

$^{32}${Instituci\'{o} Catalana de Recerca i Estudis Avan\c{c}ats, Passeig de Llu\'{\i}s Companys, 23, 08010 Barcelona, Spain}

$^{33}${Department of Physics and Astronomy, Siena College, 515 Loudon Road, Loudonville, NY 12211, USA}

$^{34}${Department of Physics and Astronomy, University of Sussex, Brighton BN1 9QH, U.K}

$^{35}${Department of Physics \& Astronomy, University  of Wyoming, 1000 E. University, Dept.~3905, Laramie, WY 82071, USA}

$^{36}${Department of Physics \& Astronomy and Pittsburgh Particle Physics, Astrophysics, and Cosmology Center (PITT PACC), University of Pittsburgh, 3941 O'Hara Street, Pittsburgh, PA 15260, USA}

$^{37}${National Astronomical Observatories, Chinese Academy of Sciences, A20 Datun Rd., Chaoyang District, Beijing, 100012, P.R. China}

$^{38}${Departamento de F\'{i}sica, Universidad de Guanajuato - DCI, C.P. 37150, Leon, Guanajuato, M\'{e}xico}

$^{39}${Instituto Avanzado de Cosmolog\'{\i}a A.~C., San Marcos 11 - Atenas 202. Magdalena Contreras, 10720. Ciudad de M\'{e}xico, M\'{e}xico}

$^{40}${Department of Physics and Astronomy, University of Waterloo, 200 University Ave W, Waterloo, ON N2L 3G1, Canada}

$^{41}${Perimeter Institute for Theoretical Physics, 31 Caroline St. North, Waterloo, ON N2L 2Y5, Canada}

$^{42}${Waterloo Centre for Astrophysics, University of Waterloo, 200 University Ave W, Waterloo, ON N2L 3G1, Canada}

$^{43}${Space Sciences Laboratory, University of California, Berkeley, 7 Gauss Way, Berkeley, CA  94720, USA}

$^{44}${University of California, Berkeley, 110 Sproul Hall \#5800 Berkeley, CA 94720, USA}

$^{45}${Instituto de Astrof\'{i}sica de Andaluc\'{i}a (CSIC), Glorieta de la Astronom\'{i}a, s/n, E-18008 Granada, Spain}

$^{46}${Center for Astrophysics $|$ Harvard \& Smithsonian, 60 Garden Street, Cambridge, MA 02138, USA}

$^{47}${Department of Physics, Kansas State University, 116 Cardwell Hall, Manhattan, KS 66506, USA}

$^{48}${Department of Physics and Astronomy, Sejong University, Seoul, 143-747, Korea}

$^{49}${CIEMAT, Avenida Complutense 40, E-28040 Madrid, Spain}

$^{50}${Department of Physics \& Astronomy, Ohio University, Athens, OH 45701, USA}

\end{hangparas}

\end{document}